\documentclass{aa}  

\usepackage{colortbl}
\usepackage{graphicx}
\usepackage{txfonts}
\usepackage[]{caption}
\usepackage{subfigure}
\usepackage{multirow}
\usepackage{comment}
\usepackage{ulem}
\definecolor{pink}{rgb}{1.0, 0, 0.8}
\begin{document} 

\title{Apsidal motion in massive eccentric binaries in NGC\,6231}
%\title{Apsidal motion in the massive eccentric binary HD\,152219}
\subtitle{The case of HD\,152219}
\author{S.\ Rosu\inst{1}\fnmsep\thanks{Research Fellow FRS-FNRS (Belgium).} \and G.\ Rauw\inst{1} \and M.\ Farnir\inst{2}\and M.-A.\ Dupret\inst{1} \and A. Noels\inst{1}}
\mail{sophie.rosu@uliege.be}
\institute{Space sciences, Technologies and Astrophysics Research (STAR) Institute, Universit\'e de Li\`ege, All\'ee du 6 Ao\^ut, 19c, B\^at B5c, 4000 Li\`ege, Belgium \and Coventry CV4 7AL, United Kingdom }
 % \email{sophie.rosu@uliege.be}} 
\date{}

\abstract{The measurement of the apsidal motion in close eccentric massive binary systems provides essential information to probe the internal structure of the stars that compose the system.}{Following the determination of the fundamental stellar and binary parameters, we make use of the tidally induced apsidal motion to infer constraints on the internal structure of the stars composing the binary system HD\,152219. }{The extensive set of spectroscopic, photometric, and radial velocity observations allows us to constrain the fundamental parameters of the stars together with the rate of apsidal motion of the system. Stellar structure and evolution models are further built with the \texttt{Cl\'es} code testing different prescriptions for the internal mixing occurring inside the stars. The effect of stellar rotation axis misalignment with respect to the normal to the orbital plane on our interpretation of the apsidal motion in terms of internal structure constants is investigated.}{Made of an O9.5\,III primary star ($M_1=18.64 \pm 0.47$\,$M_\odot$, $R_1=9.40^{+0.14}_{-0.15}\,R_\odot$, $T_\text{eff,1}=30\,900 \pm 1000$\,K, $L_\text{bol,1}=(7.26 \pm 0.97) \times 10^4\,L_\odot$) and a B1-2\,V-III secondary star ($M_2=7.70 \pm 0.12$\,$M_\odot$, $R_2=3.69\pm 0.06\,R_\odot$, $T_\text{eff,2}=21\,697\pm 1000$\,K, $L_\text{bol,2}=(2.73 \pm 0.51) \times 10^3 \,L_\odot$), the binary system HD\,152219 displays apsidal motion at a rate $(1.198 \pm 0.300)^\circ\,\text{yr}^{-1}$. The weighted-average mean of the internal structure constant of the binary system is inferred: $\bar{k}_2 = 0.00173 \pm 0.00052$. For the \texttt{Cl\'es} models to reproduce the $k_2$-value of the primary star, a significant enhanced mixing is required, notably through the turbulent mixing, but at the cost that other stellar parameters cannot be reproduced simultaneously. }{The difficulty to reproduce the $k_2$-value simultaneously with the stellar parameters as well as the incompatibility between the age estimates of the primary and secondary stars are indications that some physics of the stellar interior are still not completely understood.} 
% 5 {} token are mandatory

\maketitle
%
%-------------------------------------------------------------------

\section{Introduction\label{sect:introduction}}
The young open cluster NGC\,6231, at the core of the Sco~OB1 association \citep{Rei08,Kuh17}, hosts a substantial population of O-type stars, a large fraction of which were shown to be binary systems \citep[][and references therein]{San08}. Based on the properties of the population of low-mass stars, it has been shown that the cluster has an age spread of 1 -- 7\,Myr with a slight peak near 3\,Myr \citep{San07,Sung13}. Variations of the mean age across the cluster are however small and no obvious global trend (e.g.\ a radial age gradient) was found \citep{Kuh17}. The cluster core has a radius of $1.2 \pm 0.1$\,pc and the cluster has probably considerably expanded from its initial size \citep{Kuh17}. Stars more massive than 8\,$M_{\odot}$ appear more concentrated in the cluster core, whereas low and intermediate mass stars display no mass segregation \citep{Rab98,Kuh17}.

Since the rate of apsidal motion in eccentric short-period binaries depends on the evolutionary stage, and thus the age, of the binary components, it can help us constrain the star formation history of the cluster. So far, this technique was applied to two massive binaries of NGC\,6231: \citet{rauw16} determined an age of $5.8\pm 0.6$\,Myr for HD\,152218, while \citet{rosu20b} determined an age of $5.15\pm 0.13$\,Myr for HD\,152248. Here, we report our analysis of a third system, HD\,152219.

The spectroscopic binarity of HD\,152219 was first reported by \citet{Hil74} who presented an SB1 orbital solution with an orbital period of 4.16\,days and an eccentricity near 0.1. Subsequent SB1 orbital solutions were published by \citet{Lev83} and \citet{Gar01}. The systematic detection of the spectral signature of the secondary star was achieved by \citet{San06} who proposed an O9.5\,III + B1-2\,V-III spectral classification and derived the first full SB2 orbital solution. Considering their own data combined with radial velocities (RVs) from the literature, \citet{San06} obtained a dataset consisting of 79 observations spanning 13\,186\,days and derived an orbital period of 4.24028\,days. The photometric eclipses of HD\,152219 were discovered by \citet{Ote05} based on ASAS-3 data. These authors derived a photometric period of 4.24038\,days.

\citet{San06} reported apparent line profile variations of the primary star which they tentatively attributed to non-radial pulsations. However, in a subsequent study, \citet{San09} used an intensive spectroscopic monitoring of the star to show that these line profile variations are restricted to a short phase interval around primary eclipse and arise from the Rossiter-McLaughlin effect. Any additional line profile variability due to non-radial pulsations would have an amplitude below 0.5\% of the continuum level \citep{San09}.

Over recent years, an increasing number of massive stars were found to be part of multiple systems \citep{duchene13, sana12}. In the case of HD\,152219, some of the more distant neighbours might actually be cluster members without a gravitational connection to the binary system. \citet{San14} detected 6 companions with the NACO instrument. The closest component was found at an angular separation of 83.6\,mas, with a $\Delta K_s$ of 2.8\,mag. Five additional companions were found at angular separations between 2.2 and 7.2\,arcsec and with $K_s$ magnitude differences (compared to the main binary) between 4.2 and 6.9\,mag. The closest companion was however not detected during a speckle interferometric survey by \citet{Mas09}.

The present paper is organised as follows. The observational data are introduced in Sect.\,\ref{sect:observations}. In Sect.\,\ref{sect:specanalysis} we perform the spectral disentangling, reassess the spectral classification of the stars, and analyse the reconstructed spectra by means of the {\tt CMFGEN} model atmosphere code \citep{Hillier}. The RVs inferred from the spectral disentangling are combined with data from the literature in Sect.\,\ref{sect:omegadot} to establish a value for the apsidal motion rate of the system. Photometric data are analysed in Sect.\,\ref{sect:photom} by means of the {\tt Nightfall} binary star code. The link between apsidal motion and internal structure of the stars is recalled in Sect.\,\ref{sect:k2} and a weighted-average mean of the internal structure constant is inferred. Stellar structure and evolution models are computed with the \texttt{Cl\'es} code in Sect.\,\ref{sect:cles} where theoretical results are confronted to observational constraints. Finally, in Sect.\,\ref{sect:discussion}, we investigate the effects that could bias our interpretation of the apsidal motion in terms of the internal structure constant, notably the impact of a possible misalignment of the stellar rotation axis with respect to the normal to the orbital plane, and we give our conclusions in Sect.\,\ref{sect:conclusion}.

\section{Observational data \label{sect:observations}}
\subsection{Spectroscopy}
To investigate the optical spectrum of HD\,152219, we extracted 93 high-resolution \'echelle spectra, retrieved from the ESO archive. All were obtained with the Fiber-fed Extended Range Optical Spectrograph (FEROS) mounted on the European Southern Observatory (ESO) 1.5\,m or 2.2\,m telescopes in La Silla, Chile \citep{Kaufer}. These data were collected between May 1999 and June 2006 \citep{San06,San09}. The FEROS instrument has a spectral resolving power of 48\,000, and its detector is an EEV CCD with 2048\,$\times$\,4096 pixels of 15\,$\times$\,15\,$\mu$m. Thirty-seven orders cover the wavelength domain from 3650 to 9200\,\AA. Exposure times range from 300 to 1200 seconds. The data were reduced using the FEROS pipeline of the {\tt MIDAS} software. Residual cosmic rays were removed within {\tt MIDAS}, and the {\it telluric} tool within {\tt IRAF} was used along with the atlas of telluric lines of \citet{Hinkle} to remove the telluric absorptions near the H$\alpha$ and He\,{\sc i} $\lambda$\,5876 lines. The spectra were normalised with {\tt MIDAS} by fitting low-order polynomials to the continuum. The journal of the spectroscopic observations is presented in Appendix\,\ref{appendix:spectrotable}, Table~\ref{Table:spectro+RV}.

\subsection{Photometry\label{subsect:photometry}}
Photometric data of HD\,152219 exist in various public databases. Unfortunately, not all of them were useful for our analysis. Indeed the white light observations collected with the Optical Monitor Camera \citep{OMC} onboard the INTEGRAL spacecraft\footnote{http://sdc.cab.inta-csic.es/omc/} are too scarce to perform a meaningful analysis of the light curve. Likewise, the Hipparcos/Tycho data \citep{ESA97} have large uncertainties and HD\,152219 ($m_V \sim 7.57$) is heavily saturated in the All-Sky Automated Survey for Supernovae \citep[ASAS-SN,][]{ASAS-SN}\footnote{https://asas-sn.osu.edu/} data. The remaining useful data are the $V$-band data from the All Sky Automated Survey \citep[ASAS-3,][]{ASAS}\footnote{http://www.astrouw.edu.pl/$\sim$gp/asas/} that allowed \citet{Ote05} to discover the existence of photometric eclipses and the observations collected by the Transiting Exoplanet Survey Satellite \citep[TESS,][]{TESS}.

\begin{figure}[h]
\centering
\includegraphics[clip=true,trim=30 40 30 70,width=\linewidth]{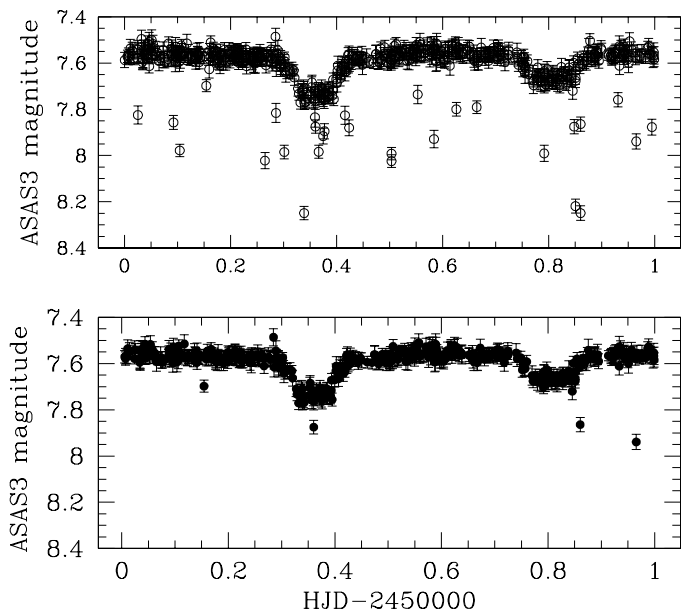}
\caption{ASAS-3 lightcurve phase-folded using the orbital period given in Table\,\ref{bestfitTable}. The top panel illustrates the initial ASAS-3 data set, whilst the bottom panel shows the data after the $1.25\,\sigma$ clipping. \label{fig:ASASlightcurve}}
\end{figure}

The ASAS-3 data were obtained at the Las Campanas Observatory with two wide-field ($8.8^{\circ} \times 8.8^{\circ}$) telescopes, each equipped with a 200/2.8 Minolta telephoto lens and a $2048 \times 2048$ pixels AP-10 CCD camera, complemented by a 25\,cm Cassegrain telescope equipped with the same type of CCD camera but covering a narrower field ($2.2^{\circ} \times 2.2^{\circ}$). The ASAS-3 archive provides aperture photometry for five different apertures varying in diameter from 2 to 6 pixels. We adopted the 4 pixel aperture data in our analysis. Following \citet{Mayer08}, we discarded the observations taken before HJD~2\,452\,455.  To further clean the data from obvious outliers, we applied a $1.25\,\sigma$ clipping to the data about the mean light curve built from normal data points computed for phase bins of 0.01. With a few exceptions, this criterion allowed us to remove all points that were off by $\geq 0.05$\,mag from the mean magnitude in the phase bins. A few remaining discrepant points were associated with phase bins containing very few data points, but were not removed afterwards. The dispersion about the mean within 0.01 phase bins (average over all phase bins containing more than 1 data point) evolved from an initial value of $4.7\,10^{-2}$\,mag to $3.0\,10^{-2}$\,mag after the $1.25\,\sigma$ clipping. In this way, we ended up with a total of 469 data points obtained between HJD~2\,452\,456 and HJD~2\,455\,112. The phase-folded ASAS-3 lightcurve is illustrated in Fig.\,\ref{fig:ASASlightcurve} adopting the orbital period given in Table\,\ref{bestfitTable}.

TESS is an all-sky survey space telescope operated by NASA that provides high-precision photometry for a huge number of stars. TESS is equipped with four 10\,cm aperture refractive cameras covering the sky in sectors of $24^{\circ} \times 96^{\circ}$. Each sector is observed during two consecutive spacecraft orbits. In the case of HD\,152219, the data were taken with camera 1 during sector 12 (i.e.\ between 21 May and 19 June 2019, hereafter TESS-12) and sector 39 (i.e. between 27 May and 24 June 2021, hereafter TESS-39). The TESS bandpass ranges from 6000\,\AA\, to 1\,$\mu$m \citep{TESS}. The CCD detectors have $15 \times 15\,\mu$m$^2$ pixels which correspond to (21\arcsec )$^2$ on the sky, and undersample the instrument PSF. Photometry is always extracted over several pixels. Bright objects saturate the central pixel, but their photometry can be recovered in the pipeline processing \citep{Jen16} from the excess charges that are spread into adjacent pixels via the blooming effect.
For TESS-12, we extracted the 2\,minutes high-cadence light curve of HD\,152219 from the Mikulski Archive for Space Telescopes (MAST) portal\footnote{http://mast.stsci.edu/}. These light curves provide background-corrected aperture photometry as well as so-called PDC photometry obtained after removal of trends that correlate with spacecraft or instrumental effects. We discarded all data points with a quality flag different from 0, as well as data taken immediately before the gap between the two orbits. For sector 39, we used the Lightkurve\footnote{https://docs.lightkurve.org/} Python software to extract light curves with a 10\,minutes cadence from the full frame images (FFI). The background level was evaluated in three different ways, either restricting ourselves to those pixels in a $50 \times 50$ pixels cutout that were below the median flux, or applying a principal component analysis including 2 or 5 components. We discarded the observations taken after 2\,459\,386 because of the large dispersion.  This results in a total of 11\,944 and 3411 data points for sectors 12 and 39, respectively, with a formal photometric accuracy ranging between 0.34 and 0.38\,mmag. 
Since HD\,152219 was located near the edge of one of the CCDs of the TESS instrument during the sector 12 observations, we also performed an extraction of the light curve using the FFI with a 30\,minutes cadence. For this purpose, we used the Lightkurve software, restricting the source region to the central (21\arcsec )$^2$ pixel. The background level was evaluated as for sector 39. The results were identical within the errors, and in good agreement with those obtained from the 2\,minutes cadence PDC photometry.

Since the region around HD\,152219 is rather crowded, both the TESS and the ASAS-3 photometry of our source are contaminated by nearby sources. Within a radius of 1\arcmin, we found two stars with a magnitude difference in the $I$-band of less than 4\,mag compared to HD\,152219 ($m_I = 7.27$). The nearest contaminator is CPD$-41^{\circ}$\,7706 (V\,945\,Sco), a $\beta$\,Cep variable located at 38.5\arcsec with $m_I = 9.26$ according to \citet{Sung98}. The second contaminator, CPD$-41^{\circ}$\,7712, is located at 54.0\arcsec and has $m_I = 8.84$ \citep{Sung98}. Assuming all the light from both sources contaminates the photometry of HD\,152219, we expect a third light fractional contribution of $\sim 0.28$. If instead, the contamination only comes from the nearest neighbour, the third light contribution would amount to $\sim 0.16$. The masks adopted for the TESS light curves extraction (as specified in the CROWDSAP keyword of the PDC files) and the full frame images suggest that the likely third light contribution due to contamination by these two sources should be about 0.09. Whilst the PDC pipeline corrects the photometry for this level of third light, this is not the case for the FFI photometry, which might thus still contain a non-zero third light contribution.

\begin{figure}[htb]
\centering
\includegraphics[width=\linewidth]{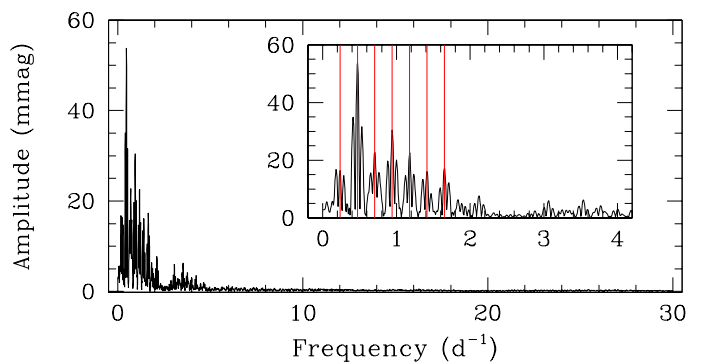}
\caption{Fourier periodogram of the TESS data from sector 12. The insert zooms on the low frequency domain. The red lines yield the orbital frequency $\nu_{\rm orb} = 0.2358$\,d$^{-1}$ and its first six harmonics.\label{fig:periodogram}}
\end{figure}

Using the Fourier method of \citet{HMM} and \citet{Gos01}, we analysed the periodogram of the TESS data to search for signals with frequencies up to 30\,d$^{-1}$ (see Fig.\,\ref{fig:periodogram}). At frequencies up to 4\,d$^{-1}$, the periodogram is largely dominated by HD\,152219's orbital frequency $\nu_{\rm orb} = 0.2358$\,d$^{-1}$ and a number of its harmonics: frequencies up to 7\,$\nu_{\rm orb}$ are clearly seen, and even 9\,$\nu_{\rm orb}$, 13\,$\nu_{\rm orb}$, and 15\,$\nu_{\rm orb}$ are clearly detected. As outlined above, the TESS light curve could be contaminated by third light from two neighbouring sources, one of them is the $\beta$~Cep variable V\,945~Sco \citep{Bal83}. Studies of this star consistently reported a signal near 14.91\,d$^{-1}$ with an amplitude near 5.5\,mmag in the $V$-band \citep{Bal83,Mei13}. Given the magnitude difference between V\,945~Sco and HD\,152219, this signal should have an amplitude in the combined light of at most 0.8\,mmag. In our periodogram of the TESS PDC photometry, the amplitude at this frequency is $\sim 0.5$\,mmag, which is consistent with the white noise level of the periodogram. Whilst we cannot totally rule out that the contaminating $\beta$\,Cep variable contributes to the non-orbital variability, its contribution must be very small, most probably comparable to or lower than the intrinsic variations of the OB-stars in HD\,152219 and any residual instrumental effects.
To reduce the amplitude of these non-orbital variations, we have constructed a mean light curve consisting of the normal points computed for phase bins of 0.01. Whilst this phase bin size does not completely remove the non-orbital variations, we note that adopting a larger bin size would smear out the details of the shape of the eclipses and thus impact the outcome of the analysis in Sect.\,\ref{sect:photom}. The TESS light curves were obtained over sufficiently short time intervals to ensure that we can assume a constant $\omega$ over the duration of each TESS sector.

\section{Spectral analysis\label{sect:specanalysis}}

\subsection{Spectral disentangling\label{subsect:spectraldisentangling}}
The FEROS spectra were used as input to our disentangling code, which is based on the method described by \citet{GL}, to establish the individual spectra of the binary components as well as their RVs at the times of the observations. We only considered those 93 spectra taken outside of the photometric eclipses to avoid biasing the reconstruction of the spectra of the binary components. The method as well as its limitations are described in details in \citet[Sect.\,3.1]{Rosu}. 

For the synthetic {\tt TLUSTY} spectra used as cross-correlation template for the RV determination, we assumed $T_\text{eff} = 30\,000$\,K, $\log g = 3.75 $, and $v \sin i =  160$\,km\,s$^{-1}$ for the primary star and $T_\text{eff} = 21\,000$\,K, $\log g = 4.00$, and $v \sin i = 160$\,km\,s$^{-1}$ for the secondary star. We performed 60 iterations to disentangle the spectra and determine the RVs. The first 30 iterations were performed with the RVs fixed to the initial input values estimated from Gaussian fits of the He\,{\sc i} $\lambda$\,4026 line. The last 30 iterations were then performed allowing the RVs to vary as we found that this number of iterations was necessary to achieve a relative error on the derived RVs of less than 1\% between the last two iterations. To check the robustness of the results of the disentangling against any bias that the initial approximation of the spectrum of star B, as a featureless continuum, could introduce, we repeated the disentangling procedure by interchanging the roles of stars A and B. The agreement was very good and the final reconstructed spectra were taken to be the mean of the two approaches.

The spectral disentangling was performed separately over a number of wavelength domains: A0[3800:3920]\,\AA, A1[3990:4400]\,\AA, A2[4300:4570]\,\AA, A3[4570:5040]\,\AA, A4[5380:5860]\,\AA, A5[5830:6000]\,\AA, A6[6400:6750]\,\AA,~and A7[7000:7100]\,\AA. As outlined in \citet{Rosu}, the presence of interstellar lines or diffuse interstellar bands close to spectral lines in some of these spectral domains (A0 and A4) affects the quality of the resulting reconstructed spectra. In these cases, the disentangling code erroneously associated some of the line flux of the non-moving interstellar medium (ISM) lines to the stars or vice versa. In addition, not enough spectral lines -- especially lines of the secondary star --  are present in the A5 and A7 domains. Both situations affect the determination of the stellar RVs. Therefore, we first processed the wavelength domains (A1, A2, A3, and A6) for which the code was able to reproduce the individual spectra and simultaneously estimate the RVs of the stars. We then computed the mean of the stellar RVs for each observation. The RVs from the individual wavelength domains were weighted according to the number of strong lines present in these domains (five lines for A1, three for A2, four for A3, and two for A6). We note that the H$\gamma$ and He\,{\sc i} $\lambda$\,4388 lines are present in both A1 and A2 domains, meaning that their weight is artificially doubled in the computation of the mean RVs. Solving this issue by lowering the weight of A1 and/or A2 gives mean RVs which differ by less than 1\%. Hence, we decided to keep the weight consistent with the number of lines in the domains. The resulting RVs of both stars are reported in Table\,\ref{Table:spectro+RV} together with their $1\sigma$ errors. We finally performed the disentangling on the four remaining domains (A0, A4, A5, and A7) with RVs fixed to these weighted averages, and, for the A0 and A4 domains only, using a version of the disentangling code designed to deal with non-moving ISM lines \citep{Mahy12}. 

\citet{Rosu} outlined several limitations of the disentangling method for massive binaries. First of all, spectral disentangling can only reconstruct features that are well sampled by the Doppler excursions during the orbital cycle. If broad lines, such as H\,{\sc i} Balmer lines, only partially deblend over the orbital cycle, this might lead to artefacts in the wings of these lines. Yet, in the case of HD\,152219, the RV amplitudes are sufficient to ensure a reliable reconstruction of the wings of the Balmer absorption lines. Furthermore, there is no indication of a significant emission in the He\,{\sc ii} $\lambda$\,4686 and H$\alpha$ lines arising from a wind interaction zone which could bias the reconstruction of these lines. We thus conclude that the reconstructed spectra of the components of HD\,152219 should be free of such biases.

\subsection{Spectral classification and absolute magnitudes\label{subsect:spectralclass}}
The reconstructed spectra of the binary components of HD\,152219 allowed reassessing the spectral classification of the binary components. 

For the primary star, we used the criterion of \citet{Conti71} complemented by \citet{Mathys88} to determine the spectral type of the star. We found that $\log W' = \log[\text{EW(He\,{\sc i} $\lambda$\,4471)/EW(He\,{\sc ii} $\lambda$\,4542)}]$ amounts to $0.61\pm 0.01$, which corresponds to spectral type O9.5. Hereafter, errors are given as $\pm1\sigma$. In addition, we used the criteria of \citet{Sota11} and \citet{Sota} based on the ratio between the strengths of several lines: the ratio He\,{\sc ii} $\lambda$\,4542/He\,{\sc i} $\lambda$\,4388 amounts to $0.67\pm0.01$, the ratio He\,{\sc ii} $\lambda$\,4200/He\,{\sc i} $\lambda$\,4144 amounts to $0.88\pm0.01$ and the ratio Si\,{\sc iii} $\lambda$\,4552/He\,{\sc ii} $\lambda$\,4542 amounts to $0.48\pm0.01$. All three ratios being lower than unity, the spectral type O9.7 is clearly excluded and the spectral type O9.5 is confirmed. We further applied the criteria of \citet{martins18} based on ratios of EWs. The ratio $\text{EW(He\,{\sc i} $\lambda$\,4144)/EW(He\,{\sc ii} $\lambda$\,4200)}$ amounts to $1.03 \pm 0.01$, suggesting a spectral type O9.2 with spectral types O9, O9.5, and O9.7 within the error bars; the ratios $\text{EW(He\,{\sc i} $\lambda$\,4388)/EW(He\,{\sc ii} $\lambda$\,4542)}$ and $\text{EW(Si\,{\sc iii} $\lambda$\,4552)/EW(He\,{\sc ii} $\lambda$\,4542)}$ amount to $1.81 \pm 0.01$ and $0.46 \pm 0.01$, respectively, both suggesting an O9.5 spectral type, excluding the O9.2 spectral type but including the O9.7 spectral type within the error bars.
To assess the luminosity class of the primary star, we used the criterion from \citet{Conti71}. 
We found that $\log W' = \log[\text{EW(Si\,{\sc iv} $\lambda$\,4089)/EW(He\,{\sc i} $\lambda$\,4143)}]$ amounts to $0.21\pm 0.01$, which corresponds to luminosity class III. The relative strength of the He\,{\sc i} $\lambda$\,4026 and Si\,{\sc iv} $\lambda$\,4089 lines, as well as of the He\,{\sc ii} $\lambda$\,4686 and He\,{\sc i} $\lambda$\,4713 lines suggests a luminosity class between V and III according to the criteria and spectral atlas of \citet{Sota11}. Similar conclusions are drawn from the criteria of \citet{martins18}: the ratios $\text{EW(He\,{\sc ii} $\lambda$\,4686)/EW(He\,{\sc i} $\lambda$\,4713)}$ and $\text{EW(Si\,{\sc iv} $\lambda$\,4089)/EW(He\,{\sc i} $\lambda$\,4026)}$ amount to $1.77 \pm 0.01$ and $0.50 \pm 0.01$, respectively, both suggesting a luminosity class III, excluding luminosity classes IV and II. We determine that the primary star is of spectral type O9.5\,III, which agrees with the determination of \citet{San06}.

For the secondary star, we used the atlases and criteria of \citet{Gray}, \citet{WP90} and \citet{liu19}. We first observed that the He\,{\sc ii} $\lambda$\,4200 line is absent in the spectra, hence excluding spectral-type O. Qualitatively, we used the ratio between the strengths of the He\,{\sc i} $\lambda$\,4471 and Mg\,{\sc ii} $\lambda$\,4481 lines, which suggests that the spectral type cannot be later than B2.5. This statement is reinforced by the fact that the Si\,{\sc ii} $\lambda\lambda$ 4128-32 lines are only marginally detected, if at all, whilst they should be clearly visible for a spectral type later than B2.5. Furthermore, the presence of C\,{\sc iii} $\lambda$\,4650, N\,{\sc iii} $\lambda$\,4097, and Si\,{\sc iv} $\lambda\lambda$\,4089, 4116 lines suggests a spectral type no later than B2. The fact that the Si\,{\sc iv} $\lambda$\,4089 line is clearly visible but weaker than the Si\,{\sc iii} $\lambda$\,4552 line suggests a spectral type B1 with an uncertainty of one spectral type. The ratio of the strengths of the Si\,{\sc iii} $\lambda$\,4552 and He\,{\sc i} $\lambda$\,4387 lines suggests a main-sequence or giant luminosity class \citep{WP90}. Furthermore, the O\,{\sc ii} $\lambda$\,4348 line is nearly absent, O\,{\sc ii} $\lambda$\,4416 is weak and the Balmer lines are broad, as expected for luminosity classes V or III. In conclusion, the classification criteria suggest that the secondary star should be a B1-2\,V-III star, again agreeing with \citet{San06}.

The brightness ratio in the visible domain was estimated based on the ratio between the EWs of the spectral lines of the  secondary\footnote{We did not apply this technique to the primary star because the uncertainties on the brightness ratio determined in this manner are on the order of the contribution of the secondary star.} star and {\tt TLUSTY} synthetic spectra of similar effective temperatures. For this purpose, we used the H$\beta$, He\,{\sc i} $\lambda\lambda$\,4026, 4144, 4471, 4921, 5016, and 5876 lines and {\tt TLUSTY} synthetic spectra of effective temperatures equal to 20\,000, 21\,000, and 22\,000\,K. The ratio $\text{EW}_{\text{\tt TLUSTY}}/\text{EW}_{sec} = (l_1+l_2)/l_2$ is equal to $10.95\pm2.07$, $10.77\pm2.02$, and $10.45\pm1.89$ for {\tt TLUSTY} synthetic spectra of 20\,000, 21\,000, and 22\,000\,K, respectively. As the three results are very close, we adopt the value for a {\tt TLUSTY} spectrum of 21\,000\,K and infer a value for $l_2/l_1$ of $0.101\pm0.021$. \\

The second {\it Gaia} data release \citep[DR2,][]{DR2} quotes a parallax of $\varpi = 0.644 \pm 0.084$\,mas, corresponding to a distance of $1507^{+231}_{-178}$\,pc \citep{bai18}. From there we derive a distance modulus of $10.89^{+0.33}_{-0.26}$. \citet{Sung98} reported mean $V$ and $B$ magnitudes of $7.552\pm0.013$ and $7.705\pm0.035$, respectively. Adopting a value of $-0.26\pm0.01$ for the intrinsic colour index $(B-V)_0$ of an O9.5 V-III star \citep{MP} and assuming the reddening factor in the V-band $R_V$ equal to $3.3\pm0.1$ \citep{Sung98}, we infer an absolute magnitude of the binary system $M_V=-4.70^{+0.29}_{-0.36}$. The brightness ratio then yields individual absolute magnitudes $M_{V,1} =-4.60^{+0.29}_{-0.36}$ and $M_{V,2} =-2.12^{+0.35}_{-0.41}$ for the primary and secondary stars, respectively. 
Comparing with the typical magnitudes reported by \citet{MP}, we find that $M_{V,1}$ is between the values of an O9.5\,V and an O9.5\,III star.

These results, along with the radii determined in \citet{San06} and in Sect.\,\ref{sect:photom} below, seem to confirm that contrary to what the luminosity class criteria suggest, the primary star would be a main-sequence star rather than a giant or subgiant. The reason is likely the fact that the surface gravity $\log g$ is not only set by the ratio of mass over radius squared, but reflects the gradient of the binary potential evaluated at the surface of the star. Likewise, comparing the magnitude obtained for the secondary star to those reported by \citet{humphreys} suggests that the luminosity classes IV and III can clearly be excluded. Instead, the secondary star's absolute magnitude is coherent within the error bars with the magnitude of a B2\,V star \citep{humphreys}.

\subsection{Projected rotational velocities\label{subsect:vsini}}
\begin{figure*}
\includegraphics[width=0.49\linewidth, trim=2cm 0cm 3cm 6.5cm, clip=true]{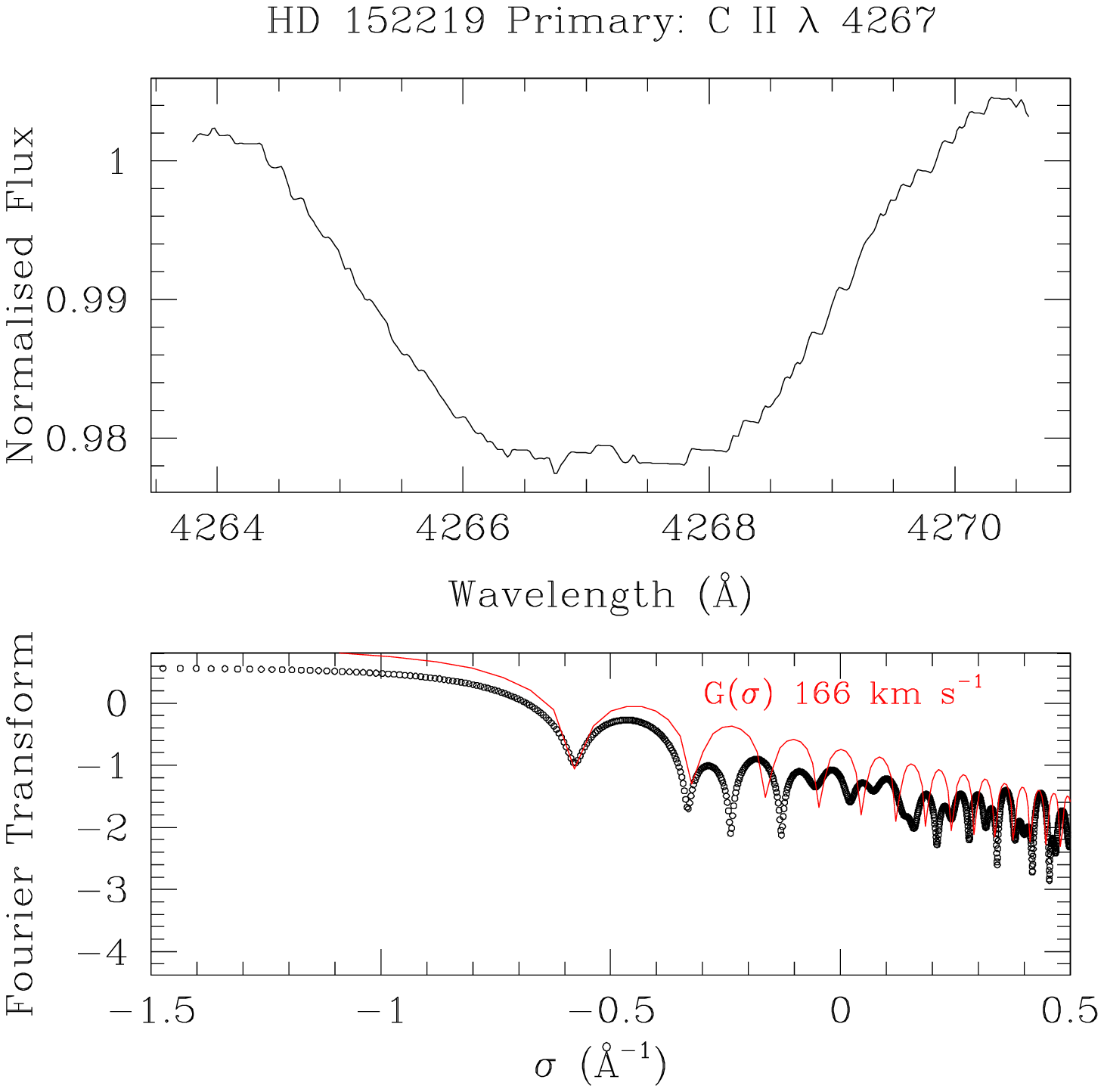}
\includegraphics[width=0.49\linewidth, trim=2cm 0cm 3cm 6.5cm, clip=true]{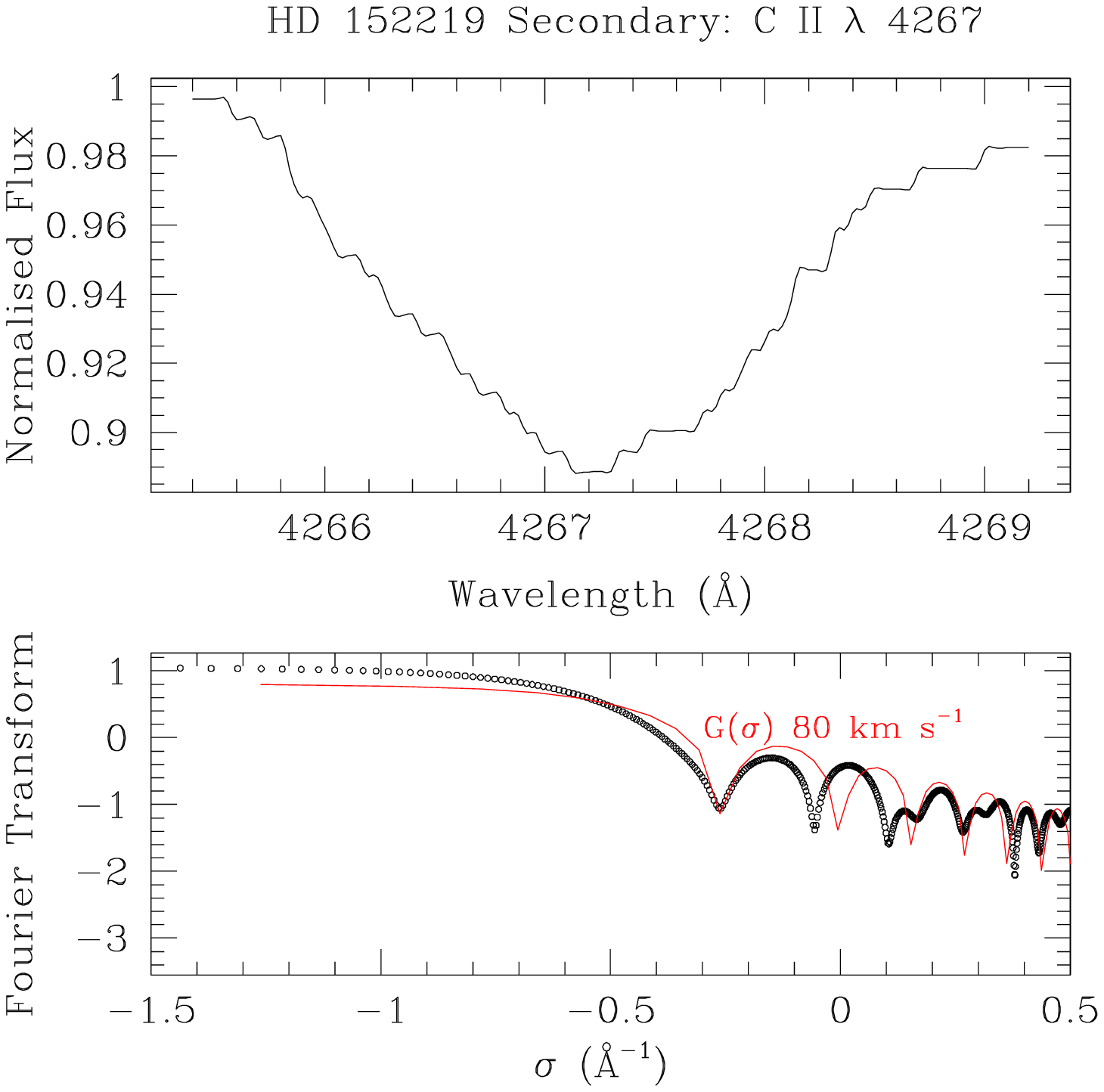}
\caption{\textit{Top row:} C\,{\sc ii} $\lambda$\,4267 line profiles of the separated spectra obtained after application of the brightness ratio for the primary (\textit{left panel}) and secondary (\textit{right panel}) stars. \textit{Bottom row:} Fourier transform of those lines (in black) and best-match rotational profile (in red) for the primary (\textit{left panel}) and secondary (\textit{right panel}) stars.\label{fig:vsini}}
\end{figure*}

The projected rotational velocities of both stars were derived using the Fourier transform method \citep{Simon-Diaz,Gray08}, which has the advantage of being able to separate the effect of rotation from other broadening mechanisms such as macroturbulence. We applied this method to a set of well-isolated spectral lines, which are therefore expected to be free of blends. The results are presented in Table\,\ref{vsiniTable}, and the Fourier transforms are illustrated for the C\,{\sc ii} $\lambda$\,4267 line in Fig.\,\ref{fig:vsini} for the primary and secondary stars. It is commonly recommended to use metal lines to compute the projected rotational velocity. The H\,{\sc i}, He\,{\sc i}, and He\,{\sc ii} line profiles are indeed more affected by non-rotational broadening mechanisms such as Stark broadening, which alter the position of the first zero in the Fourier transform~\citep{Simon-Diaz}. However, when the projected rotational velocity of the star exceeds 100 km\,s$^{-1}$, the effect of Stark broadening on the He\,{\sc i} lines is expected to be small and these lines can then be used as a complement to the metal lines. The results presented in Table\,\ref{vsiniTable} show that the mean $v \sin i_\text{rot}$ computed on metal lines alone or on all the lines agree very well. We adopt a mean $v \sin i_\text{rot}$ of $166\pm10$ km\,s$^{-1}$ for the primary star and $95\pm9$ km\,s$^{-1}$ for the secondary star. For the primary star, our results agree with the values obtained by \citet{Lev83} and \citet{San06}, but are lower than those obtained by \citet{contiebbets} (see Table\,\ref{vsiniTable}).

\begin{table}[h!]
\caption{Best-fit projected rotational velocities as derived from the disentangled spectra of HD\,152219 and comparison with previous determinations.}
\centering
\begin{tabular}{l l l}
\hline\hline
Line & \multicolumn{2}{c}{$v\sin i_{\text{rot}}$ (km\,s$^{-1}$)} \\
& Primary & Secondary \\
\hline
Si\,{\sc iv} $\lambda$\,4089 & 163 & ... \\ 
Si\,{\sc iv} $\lambda$\,4212 & 135 & ... \\ 
O\,{\sc ii} $\lambda$\,4254 & 166 & 104 \\ 
C\,{\sc ii} $\lambda$\,4267 & 166 & 80 \\ 
O\,{\sc iii} $\lambda$\,4277 & 169 & ... \\ 
N\,{\sc iii} $\lambda$\,4379 & 156 & ... \\ 
O\,{\sc ii} $\lambda$\,4610 & 148 & ... \\ 
N\,{\sc iii} $\lambda$\,4641 & 161 & ... \\ 
O\,{\sc iii} $\lambda$\,5592 & 165 & ... \\ 
He\,{\sc i} $\lambda$\,3820 & 184 & 104 \\ 
He\,{\sc i} $\lambda$\,4026 & 170 & ... \\ 
He\,{\sc i} $\lambda$\,4143 & 172 & ... \\ 
He\,{\sc i} $\lambda$\,4387 & 172 & ... \\ 
He\,{\sc i} $\lambda$\,4471 & 168 & ... \\ 
He\,{\sc i} $\lambda$\,4713 & 177 & ... \\ 
He\,{\sc i} $\lambda$\,4922 & 167 & ... \\ 
He\,{\sc i} $\lambda$\,5016 & 170 & 88 \\
He\,{\sc i} $\lambda$\,5875 & 170 & 97 \\
He\,{\sc i} $\lambda$\,6678 & 170 & ... \\
He\,{\sc i} $\lambda$\,7065 & 166 & 98 \\
\hline
Mean (metal lines) & $159\pm11$ &$92\pm17$ \\ 
Mean (He\,{\sc i} lines) & $171\pm5$ &$97\pm7$ \\ 
Mean (all lines) & $166\pm10$ & $95\pm9$ \\
\hline
\citet{contiebbets} & 250 & ... \\ 
\citet{Lev83} & $160$ & ...\\
\citet{San06}  & & \\
~~~~~ He\,{\sc i} lines & $190\pm10$ & ... \\
~~~~~ Si\,{\sc iv} $\lambda$\,4089 & $150\pm10$ & ... \\
~~~~~ O\,{\sc iii} $\lambda$\,5592 & $175\pm10$ & ... \\
~~~~~ He\,{\sc ii} $\lambda$\,4542 & $130\pm10$ & ... \\
~~~~~ He\,{\sc ii} $\lambda$\,4686 & $160\pm10$ & ... \\
\hline
\end{tabular}
\tablefoot{The values quoted by \citet{contiebbets} were obtained by visual comparison with standards and the values quoted by \citet{Lev83} were obtained by visual comparison with the standards given by \citet{Slettebak}. The values of \citet{San06} were estimated through a comparison of the measured Full Widths at Half Maximum (FWHMs) of the various lines with those measured, for a range of $v\sin i$, on model spectra of equivalent effective temperature and gravity computed using the codes {\tt TLUSTY} and {\tt SYNSPEC}.}  
\label{vsiniTable}
\end{table}

\subsection{Wind terminal velocity\label{subsect:vinf}}
\begin{figure}
\includegraphics[width=\linewidth]{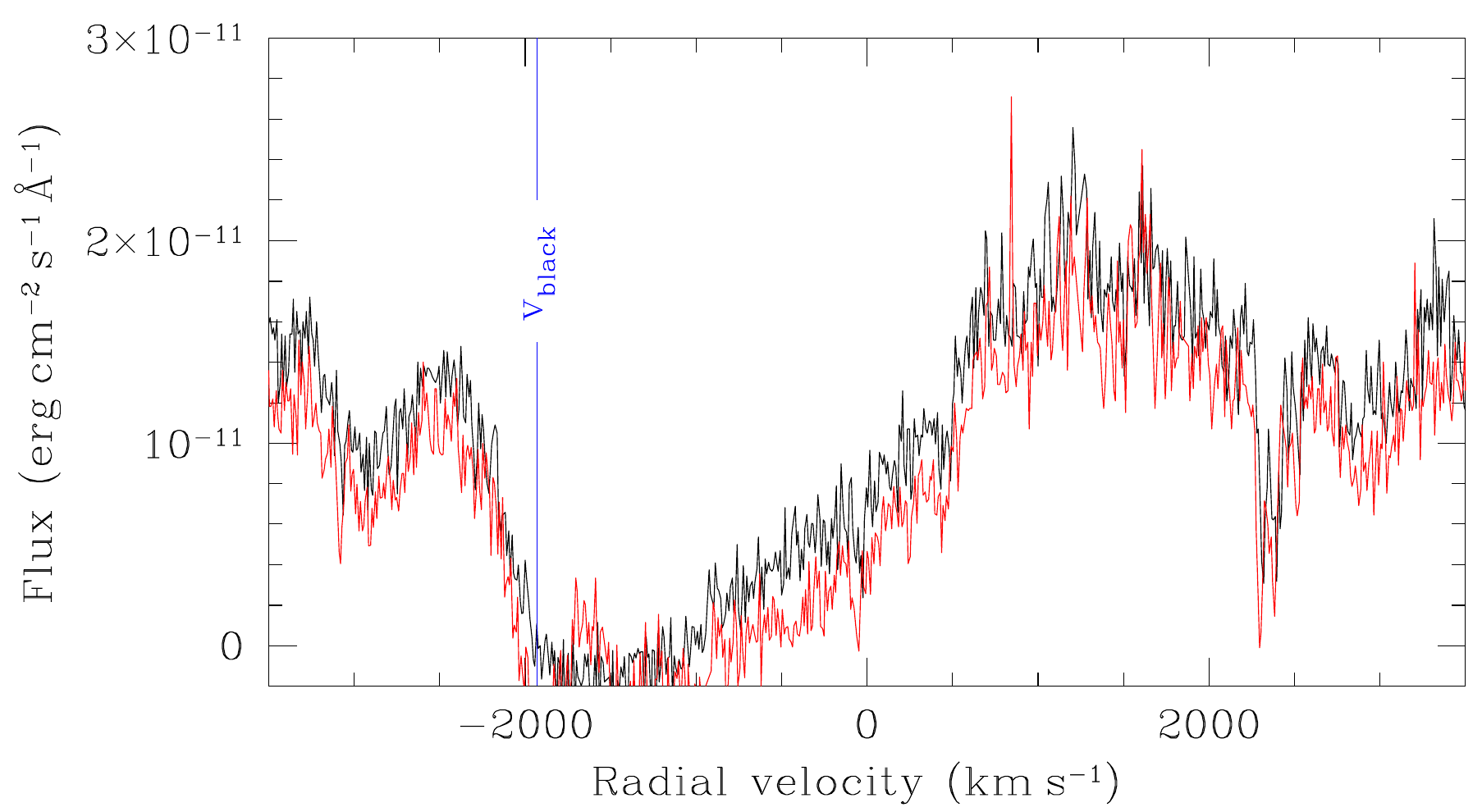}
\caption{Determination of $v_{\rm black}$ on the absorption trough of C\,{\sc iv} $\lambda\lambda$ 1548, 1551 as observed in the IUE spectra of HD\,152219. The black and red curves correspond to observations SWP~56939 and SWP~56953, respectively.\label{fig:iue}}
\end{figure}

To estimate the wind terminal velocity $v_{\infty}$ of the primary star of HD\,152219, we extracted seven high-resolution spectra taken with the short wavelength primary (SWP) camera of the International Ultraviolet Explorer ({\it IUE}) satellite. The most prominent wind feature in these spectra is clearly the saturated C\,{\sc iv} $\lambda\lambda$\,1548,1551 P-Cygni profile. We measured the velocity of the violet edge of zero residual intensity in the absorption trough $v_{\rm black}$ (see Fig.\,\ref{fig:iue}). In this way, we derived $v_{\rm black} = (1930 \pm 120)$\,km\,s$^{-1}$. The dispersion most probably arises (at least partially) from the binarity of HD\,152219. Indeed, the interaction of the primary wind with either the secondary star or the wind of the latter leads to a reduction of the primary wind velocity in some directions (and thus of $v_{\rm black}$ at some orbital phases). Since $v_{\rm black}$ was shown to be a good indicator of $v_{\infty}$ for OB stars \citep{Pri90}, we thus adopted $v_{\infty} = 1930$\,km\,s$^{-1}$.

\subsection{Model atmosphere fitting\label{subsect:cmfgen}}
The reconstructed spectra of the binary components were analysed by means of the {\tt CMFGEN} model atmosphere code \citep{Hillier} to constrain the fundamental properties of the stars. This code solves the radiative transfer equations, as well as the equations of statistical and radiative equilibrium in the comoving frame assuming the star and its wind are spherically symmetric. Line-blanketing and clumping are included in the code. For the wind, a standard $\beta$-velocity law is adopted through the relation
\begin{equation}
v(r) = v_\infty\left(1-\frac{R_*}{r}\right)^\beta,
\end{equation}
where $v_\infty$ is the terminal wind velocity, $R_*$ the radius of the star, and $r$ the radial distance from the centre of the star. The wind clumping is introduced in the models using a two-parameter exponential law for the volume filling factor
\begin{equation}
f(r) = f_1+(1-f_1) \exp\left(-\frac{v(r)}{f_2}\right),
\end{equation}
where the parameters $f_1$ and $f_2$ are the filling factor value when $r \rightarrow \infty$ and the onset velocity of clumping, respectively. Including clumping in the models leads to a reduced estimate of the mass-loss rate compared to an unclumped model. To compare with values obtained with unclumped models, the mass-loss rate derived in this paper has to be multiplied by a factor of $1/\sqrt{f_1}$~\citep{Martins11}. Regarding the microturbulent velocity $v_\text{micro}$, {\tt CMFGEN} assumes that it depends on the position $r$ as 
\begin{equation}
v_\text{micro} = v_\text{micro}^\text{min} + \left(v_\text{micro}^\text{max} - v_\text{micro}^\text{min}\right) \frac{v(r)}{v_\infty},
\end{equation}
where $v_\text{micro}^\text{min}$ and $v_\text{micro}^\text{max}$ are the minimum and maximum $v_\text{micro}$. The value of $v_\text{micro}^\text{max}$ is fixed to $0.1v_\infty$, whilst $v_\text{micro}^\text{min}$ is adjusted on metal lines. \\

The {\tt CMFGEN} spectra were first broadened by the projected rotational velocities determined in Sect.\,\ref{subsect:vsini}. The stellar and wind parameters were then adjusted following the procedure outlined by \citet{Martins11}. We proceeded slightly differently for the two stars as they have quite different spectral types.  

\subsubsection{Primary star}
The macroturbulence velocity was adjusted on the wings of the O\,{\sc iii} $\lambda$\,5592 and Balmer lines and we derived a value of $v_\text{macro}=120\pm 10$\,km\,s$^{-1}$. We adjusted the microturbulence velocity on the metal lines and inferred a value of $15\pm3$\,km\,s$^{-1}$.

The effective temperature was determined by searching for the best overall fit of the He\,{\sc i} and He\,{\sc ii} lines. This was clearly a compromise as we could not find a solution that perfectly fits all helium lines simultaneously. For instance, we found that the strength of the He\,{\sc i} $\lambda$\,4471 line is underestimated while other He\,{\sc i} lines are well-adjusted. Given the luminosity class of the primary, it seems unlikely that this discrepancy reflects the dilution effect discussed by \citet{voels89} and \citet{herrero92}. Discarding some lines that are severely affected by blends (such as He\,{\sc i} $\lambda$\,4121 blended with Si\,{\sc iv} $\lambda$\,4116 or He\,{\sc ii} $\lambda$\,5412 possibly affected by blends with diffuse interstellar bands), or turned out to be rather insensitive to a small variation in the effective temperature (such as He\,{\sc i} $\lambda$\,4026 which reaches its maximum intensity at spectral type O9 and is also blended with the weak but non-zero He\,{\sc ii} $\lambda$\,4026 line), we finally adjusted the effective temperature on the He\,{\sc i} $\lambda$\,4922 line and inferred a value of $30\,900\pm 1000$\,K. With this effective temperature, the He\,{\sc i} $\lambda\lambda$\,4026, 4388, 4713, and 4922 lines as well as the He\,{\sc ii} $\lambda$\,4200 are perfectly adjusted. A possible explanation for the difficulties encountered when attempting to determine the effective temperature could be the fact that we are applying a 1-D model atmosphere code that assumes a spherical geometry to a spectrum that forms in the atmosphere of a distorted star in a binary system. Gravity darkening then leads to a highly non-uniform temperature distribution \citep{Palate12}, which is not accounted for in the model atmosphere code.

The surface gravity was obtained by adjusting the wings of the Balmer lines H$\beta$, H$\gamma$, H$\delta,$ and H\,{\sc i} $\lambda\lambda$\,3835, 3890. We excluded H$\varepsilon$ because this line is not correctly reproduced by the disentangling code owing to the blend with the interstellar Ca\,{\sc ii} line. We derived $\log g_\text{spectro} = 3.60\pm0.10$.\\

\begin{figure*}[htbp]
\begin{center}
\includegraphics[clip=true,trim=0.5cm 4.5cm 1cm 1cm,width=\linewidth]{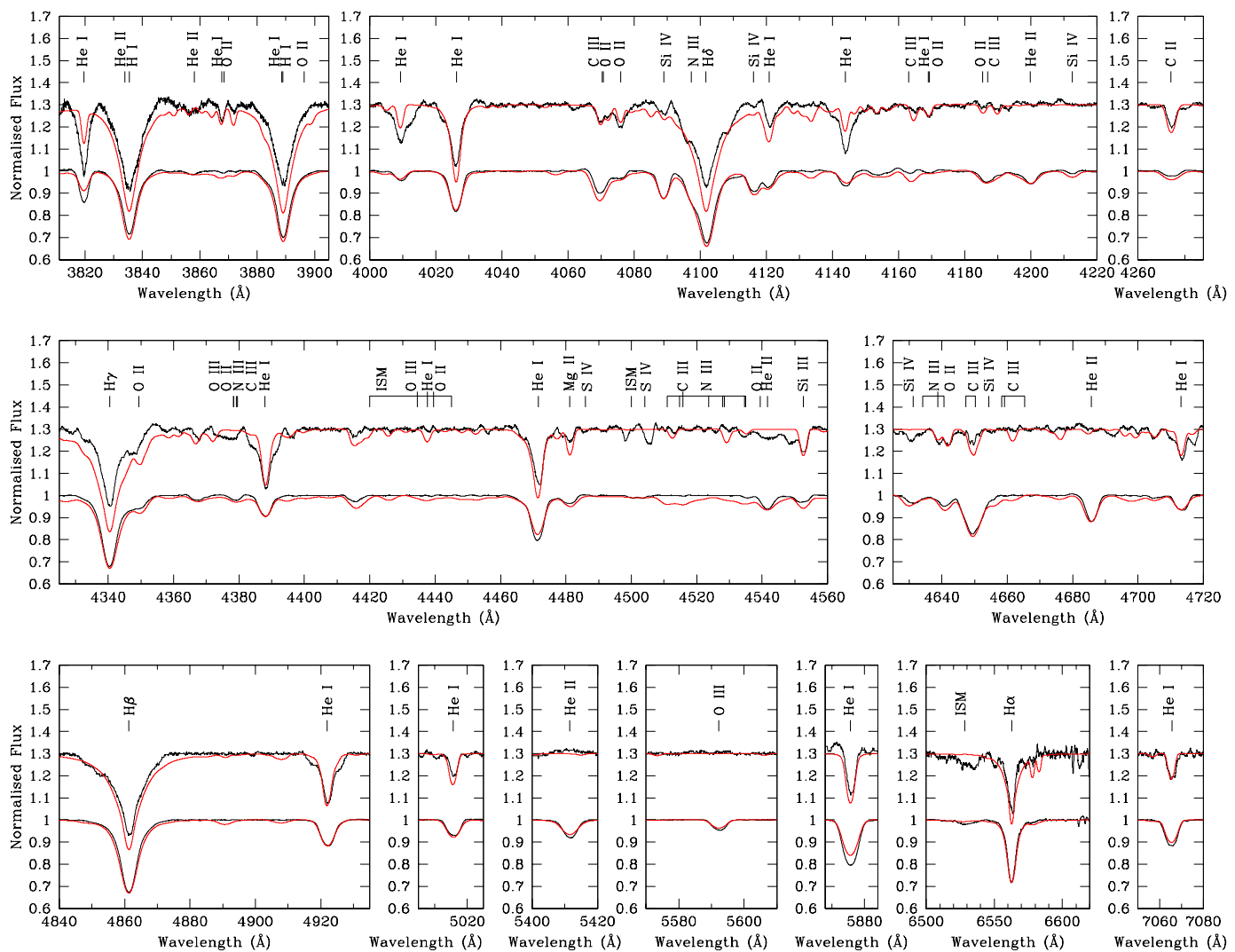}
\end{center}
\caption{Normalised disentangled spectra (in black) of the primary and secondary star of HD\,152219 (note that the spectrum of the secondary star is shifted by +0.3 in the $y$-axis for convenience) together with the respective best-fit {\tt CMFGEN} model atmosphere (in red). \label{Fig:CMFGEN}}
\end{figure*}

The surface chemical abundances of all elements, including helium, carbon, nitrogen, and oxygen, were set to solar as taken from \citet{Asplund}. Indeed, HD\,152219 is a typical case were it is impossible to certify that the chemical abundances of the elements differ from solar; the number of parameters we have to adjust exceeds the number of lines we can use for this purpose. For instance, to adjust the oxygen abundance, the O\,{\sc iii} $\lambda$\,5592 line is the only one free of blends that could be used. To perfectly reproduce this line, an oxygen abundance nearly twice solar would be required, which is highly unlikely. Indeed, for such an oxygen abundance the synthetic {\tt CMFGEN} spectrum predicts a series of O\,{\sc iii} absorption lines ($\lambda\lambda$\,4368, 4396, 4448, 4454, and 4458) that are not present in the observed spectra (neither before nor after disentangling). Moreover, this would also artificially reduce the carbon abundance, inferred from the C\,{\sc iii} $\lambda$\,4070 line. This C\,{\sc iii} line is the only suitable carbon line which is not affected by subtle formation processes controlled by a number of unconstrained far-UV lines. However, it is heavily blended with an oxygen line. As a result, and since HD\,152219 is a relatively unevolved system, we consider that adopting solar abundances for the chemical elements is the best option.

Regarding the wind parameters, the clumping parameters were fixed: The volume filling factor $f_1$ was set to 0.1, and the $f_2$ parameter that determines the wind velocity (and thus the position) where clumping starts was set to 100\,km\,s$^{-1}$, as suggested by \citet{hillier13}. For the sake of completeness, we varied $f_1$ from 0.05 to 0.2 and $f_2$ from 80 to 120\,km\,s$^{-1}$, and we did not observe any significant difference. 
Likewise, the $\beta$ parameter of the velocity law was fixed to the value of 0.9 as suggested by \citet{Muijres} for an O9\,V type star. Again for the sake of completeness, we tested a value of 1.0 for $\beta$ but we did not observe any significant difference in the resulting spectrum.

In principle, the wind terminal velocity could be derived from the H$\alpha$ line. However, because of the degeneracy between the wind parameters, we decided to fix the value of $v_\infty$ to 1930\,km\,s$^{-1}$ as determined in Sect.\,\ref{subsect:vinf}. Regarding the mass-loss rate, the main diagnostic lines in the optical domain are H$\alpha$ and He\,{\sc ii} $\lambda$\,4686. We inferred a value of $\dot{M} = 6\times10^{-8}\,M_\odot/\text{yr}$ based on these lines. \\

\begin{table}[h!]
\caption{Stellar and wind parameters of the best-fit {\tt CMFGEN} model atmosphere derived from the separated spectra of HD\,152219.}
\centering
\begin{tabular}{l l l}
\hline\hline
\vspace*{-3mm}\\
Parameter & \multicolumn{2}{c}{Value} \\
& Primary & Secondary \\
\hline
\vspace*{-3mm}\\
$T_\text{eff}$ (K) & $30\,900\pm1000$ & $23\,500\pm1000$  \\ 
$\log g_\text{spectro}$ (cgs) & $3.60\pm0.10$ & 3.92 (fixed) \\
$v_\text{macro}~(\text{km\,s}^{-1})$ & $120\pm10$ & $50\pm10$ \\ 
$v_\text{micro}~(\text{km\,s}^{-1})$ & $15\pm3$ & 10 (fixed) \\ 
$\dot{M}~(M_\odot\,\text{yr}^{-1})$  & $6.0\times10^{-8}$ &$1.0\times10^{-10}$ \\
$\dot{M}_\text{unclumped}~(M_\odot\,\text{yr}^{-1})^a$  & $1.9\times10^{-7}$ & $3.2\times10^{-10}$ \\
$v_\infty~(\text{km\,s}^{-1})$ & 1930 (fixed)  & $2250\pm50$ \\
$f_1$ & 0.1 (fixed) & 0.1 (fixed)  \\ 
$f_2~(\text{km\,s}^{-1})$ & 100 (fixed) & 100 (fixed)\\
$\beta$ & 0.90 (fixed) & 1.1 (fixed) \\
\vspace*{-4mm}\\
$R_\text{spectro}\,(R_{\odot})$ & $9.7^{+1.8}_{-1.5}$ & $3.7^{+0.8}_{-0.7}$ \\
\vspace*{-3mm}\\
$M_\text{spectro}\,(M_\odot)$ & $17.2^{+7.1}_{-6.3}$ & $4.4^{+2.1}_{-1.9}$ \\
\vspace*{-3mm}\\
\hline
\end{tabular}
\tablefoot{ $^a$$\dot{M}_\text{unclumped}=\dot{M}/\sqrt{f_1}$.}
\label{Table:CMFGEN}
\end{table} 

\subsubsection{Secondary star} 
As the O\,{\sc iii} $\lambda$\,5592 line is not present in the spectrum of the secondary star, we had to rely on He\,{\sc i} and Balmer lines to estimate the macroturbulence. Hence, we determined $v_\text{macro}$, the effective temperature and the surface gravity simultaneously, in an iterative manner, in order to get the best-fit of the spectrum. We inferred a value of  $v_\text{macro}=50\pm 10$\,km\,s$^{-1}$. 

As for the primary star, we adjusted the effective temperature mainly based on the He\,{\sc i} $\lambda$\,4922 line, and found a value of $23\,500 \pm 1000$\,K. With this effective temperature, He\,{\sc i} $\lambda\lambda$\,4026, 4471, 5016, and 5876 are overestimated whilst He\,{\sc i} $\lambda\lambda$\,4388, 4713, and 7065 are underestimated. 

In principle, the surface gravity is obtained by adjusting the wings of the Balmer lines. However, in the present case, we found that the wings, as well as the depths,  of the Balmer lines are systematically overestimated. At least for some of the Balmer lines (e.g.\ H$\delta$ and H$\gamma$), this discrepancy very likely reflects residual normalisation errors which affect the disentangled spectra and are amplified by the large brightness ratio that is applied to renormalise the secondary spectrum. Hence, we decided to adopt a $\log g_\text{spectro} = 3.92\pm0.10$ as inferred by \citet{San09}. With this value, the wings of most hydrogen lines are overestimated.  \\

We fixed the microturbulence velocity $v_\text{micro}$ to 10\,km\,s$^{-1}$, a reference value for such types of stars as suggested by \citet{cazorla17}. For the same reasons as for the primary star, we set the surface chemical abundances to solar as taken from \citet{Asplund}. \\

Regarding the wind parameters, the clumping parameters $f_1$ and $f_2$ were fixed as for the primary star. We fixed the $\beta$ parameter of the velocity law to 1.1 as representative of early B-type stars \citep[see, e.g.][]{lefever10}, and did not find any significant difference when varying the value of this parameter. 

We adjusted the mass-loss rate based on the H$\gamma$ and H$\alpha$ lines and inferred a value of $\dot{M} = 1.0\times10^{-10}\,M_\odot/\text{yr}$. The wind terminal velocity was then adjusted based on the strength of the H$\alpha$ line. We found a value of  $v_\infty = 2250\pm50$\,km\,s$^{-1}$. 

The stellar and wind parameters of the best-fit {\tt CMFGEN} model are summarised in Table\,\ref{Table:CMFGEN}. The normalised disentangled spectra of both components of HD\,152219 are illustrated in Fig.\,\ref{Fig:CMFGEN} along with the best-fit {\tt CMFGEN} adjustment.    \\

\subsubsection{Spectroscopic radii and masses\label{subsubsect:RML}}
The bolometric magnitudes of the stars $M_{\text{bol},1}=-7.46^{+0.31}_{-0.37}$ and $M_{\text{bol},2}=-4.17^{+0.38}_{-0.43}$ were computed assuming that the bolometric correction depends only on the effective temperature through the relation 
\begin{equation}
BC = -6.89\log(T_{\text{eff}}) + 28.07
\end{equation}
\citep{MP}. We further obtained bolometric luminosities $L_{\text{bol},1}$ of $7.74^{+2.66}_{-2.18}\times 10^4 L_\odot$ for the primary star and $L_{\text{bol},2}$ of $3.72^{+1.48}_{-1.29}\times 10^3 L_\odot$ for the secondary star. From the relation between bolometric luminosity, effective temperature, and radius, we inferred spectroscopic radii of $R_{\text{spectro},1}=9.7^{+1.8}_{-1.5} \,R_\odot$ for the primary star and $R_{\text{spectro},2}=3.7^{+0.8}_{-0.7} \,R_\odot$ for the secondary star. \\
We corrected the surface gravities determined with {\tt CMFGEN} to account to first order for the impact of the centrifugal force and of the radiation pressure. In this way, we obtained $\log{g_C}$ of $3.70 \pm 0.08$ and $3.95 \pm 0.09$ for the primary and secondary stars, respectively. From there, we then inferred spectroscopic masses of $M_{\text{spectro},1}=17.2^{+7.1}_{-6.3}\,M_\odot$ for the primary star and $M_{\text{spectro},2}=4.4^{+2.1}_{-1.9}\,M_\odot$ for the secondary star. In Sect.\,\ref{sect:photom}, we will compare the spectroscopic radii and masses with model-independent determinations based on the RV curves and the photometric light curve. Beside the difficulties outlined above to determine the surface gravity of the secondary, it is important to recall that {\tt CMFGEN} assumes that the stars are spherical, static, and isolated. However, because of the binarity, these three assumptions are not verified. The effective temperature is thus not homogeneous all over the surface and the computed temperature has to be considered as a mean value over the star. In addition, because of the companion, the surface gravity $|\vec{\nabla}(\Omega)|$ is not homogeneous over the stellar surface and the local surface gravity is lower than that of an isolated star of same mass and radius \citep{Palate12}.
      
\section{Apsidal motion \label{sect:omegadot}}
Our total dataset of RVs consists of 16 primary RVs taken from \citet{Hil74}, 1 from \citet{Con77}, 4 from \citet{Lev83}, 3 from \citet{Per90}, 8 from \citet{Gar01}, 8 from \citet{Sti01}, and the 6 CAT+CES RV points from \citet{San06}. These data were complemented by our 93 RV points obtained as part of the disentangling process of the FEROS observations of HD\,152219. The third datapoint from the \citet{Per90} dataset (HJD 2\,439\,958.773) was discarded as it was found to be offset by about 80\,km\,s$^{-1}$ from the most-likely orbital solution \citep[see also][]{San06}. Likewise, we found that the four last RV points (HJD 2\,450\,593.600 -- 2\,450\,598.792) given by \citet{Gar01} were discrepant. These points were thus also discarded. We therefore end up with a series of 134 primary RV data points spanning about 38 years.

\citet{San06} and \citet{San09} reported significantly different RV amplitudes for different lines of the primary star. For instance, \citet{San09} found a difference of 15\% between the smallest primary RV amplitude (107.7\,km\,s$^{-1}$ for the He\,{\sc i} $\lambda$\,4922 line) and the largest value (124.4\,km\,s$^{-1}$ for the He\,{\sc ii} $\lambda$\,4686 line). In contrast, the RVs inferred from our disentangling method applied to different wavelength domains did not show significant differences. We therefore suspect that the amplitude differences reported by \citet{San06} and \citet{San09} arise from the double Gaussian fitting method used by these authors to establish the RVs.

For those literature references that explicitly quote errors on the RVs, we adopted the latter values. For the {\it IUE} data from \citet{Sti01} and for the CAT/CES data of \citet{San06}, we adopted errors of 5\,km\,s$^{-1}$ as representative of the measurement error on these values based on the spectral resolution and the method used to derive the RVs. Moreover these values are very close to the dispersions of the RV points over the best orbital solutions. For the RVs derived from the spectral disentangling, we estimated the errors from the dispersion of the RVs determined from the various spectral domains. 

Since the secondary RVs are only available for the most recent data and are subject to larger errors than the primary RVs, we did not use them in the determination of the rate of apsidal motion. For each time of observation $t$, we adjusted the primary RV data with the following relation
\begin{equation}
\label{eq:RV}
RV_{\rm P}(t) = \gamma_{\rm P} + K_{\rm P}\,[\cos{(\phi(t)+\omega(t))} + e\,\cos{\omega(t)}],
\end{equation}
where
$\gamma_{\rm P}$, $K_{\rm P}$, $e$, and $\omega(t)$ are respectively the apparent systemic velocity, the semi-amplitude of the RV curve, the eccentricity and the argument of periastron of the primary orbit. The true anomaly $\phi(t)$ is evaluated through Kepler's equation which involves $e$ and the anomalistic orbital period $P_{\rm orb}$. We explicitly accounted for secular variations of the argument of periastron, by adopting
\begin{equation}
\label{eq:w}
\omega(t) = \omega_0 + \dot{\omega}\,(t - T_0),
\end{equation}
where $\omega_0$ is the value of $\omega$ at the epoch $T_0$ of periastron passage and $\dot{\omega}$ is the rate of secular apsidal motion.
As the RVs of different spectral lines potentially yield slightly different apparent systemic velocities \citep[see Table\,3 of][]{San06}, and since literature RVs were obtained from diverse sets of lines, we adjusted the systemic velocity of each subset of our total dataset so as to minimise the sum of the residuals of the data about the curve given by Eq.\,\eqref{eq:RV} for each combination of the six model parameters. For our best-fit solution, the systemic velocity amounts to $-24.3\pm 1.6$, $-42.9\pm 3.8$, $-13.0\pm 2.9$, $-35.0\pm 3.5$, $-24.6\pm 2.1$, $-42.7\pm 1.8$, $-6.7\pm2.0$, and $-24.5\pm0.1$\,km\,s$^{-1}$ for the \citet{Hil74}, \citet{Con77}, \citet{Lev83}, \citet{Per90}, \citet{Gar01}, \citet{Sti01}, \citet{San06}, and our subset, respectively.

\begin{figure*}[htb]
\centering
\hspace{-7cm}\includegraphics[clip=true,trim=0cm 2.3cm 1cm 7cm, width=0.615\linewidth]{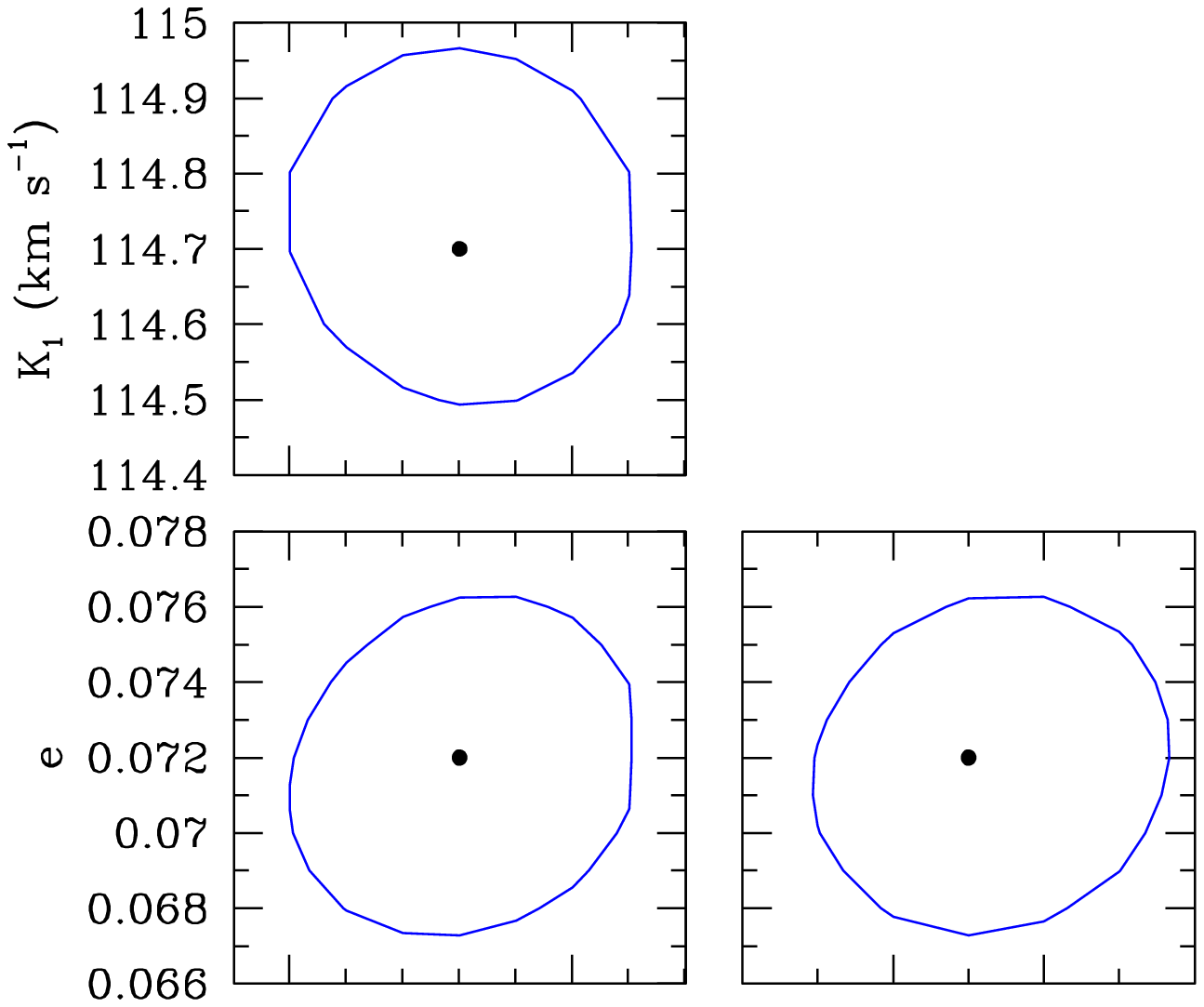}\\
\vspace{-0.85cm}
\includegraphics[clip=true,trim=0cm 1cm 1cm 1.4cm, width=0.615\linewidth]{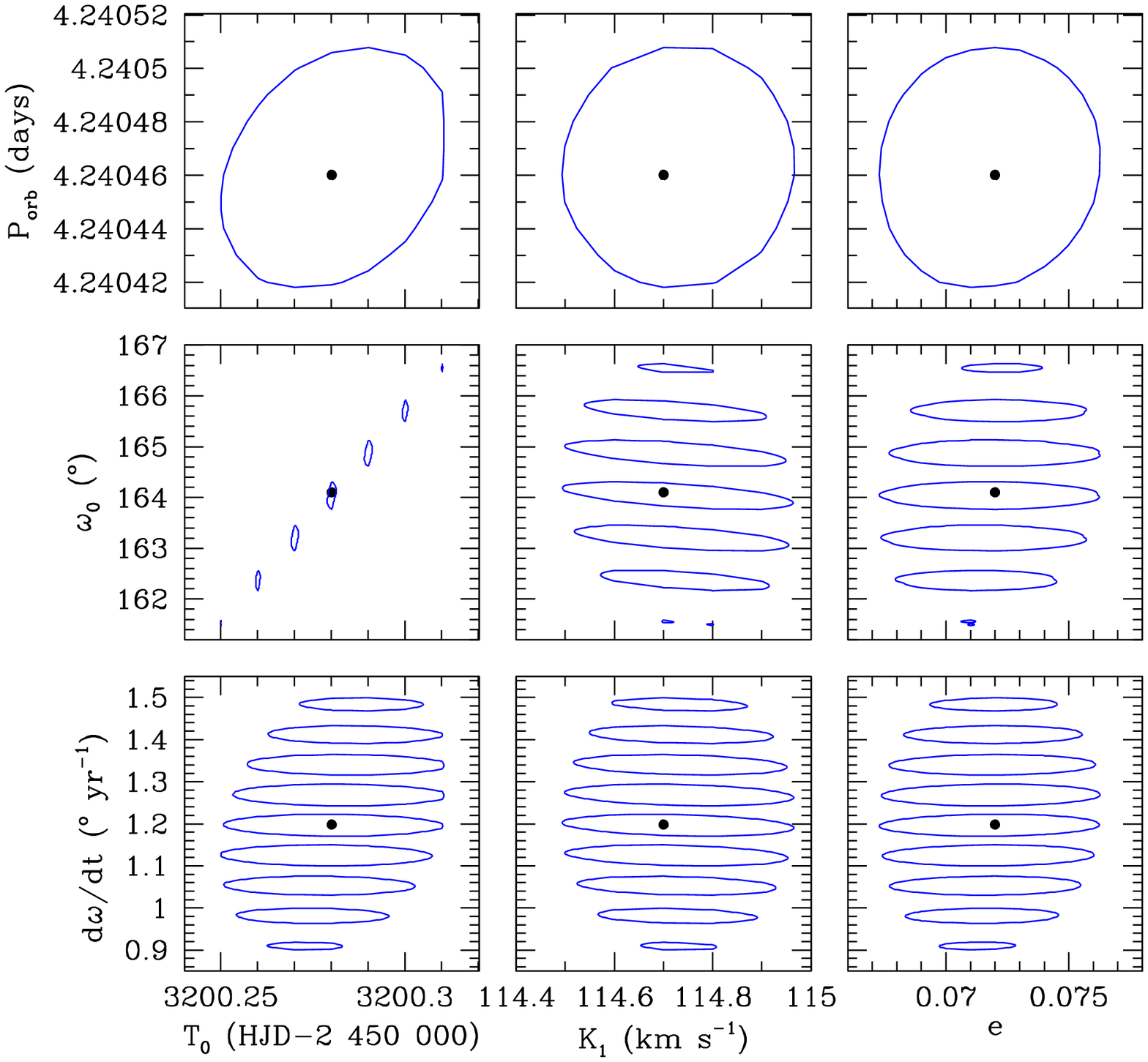}
\includegraphics[clip=true,trim=2.8cm 1cm 5.77cm 0cm, width=0.375\linewidth]{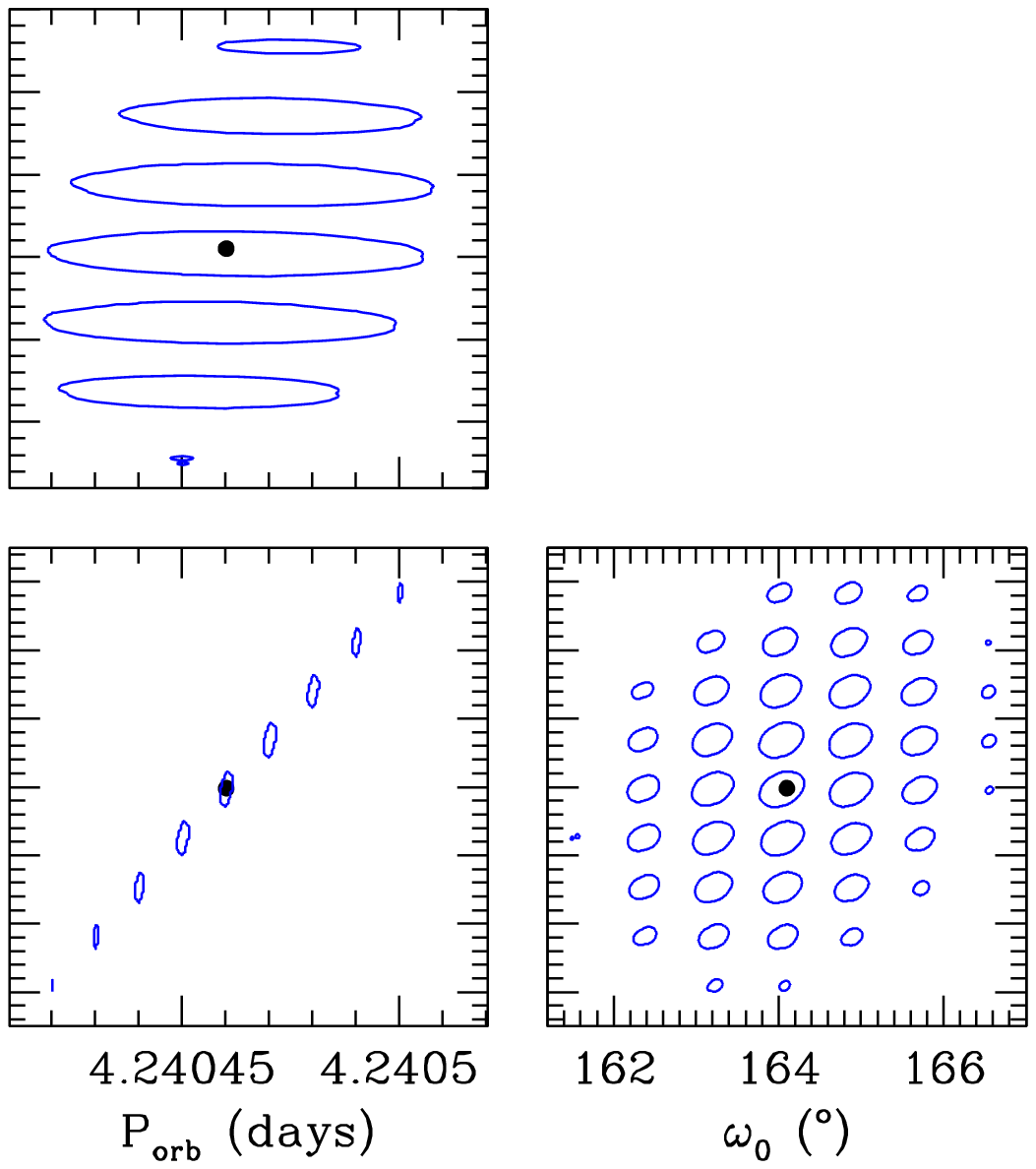}
\caption{Confidence contours for the best-fit parameters obtained from the adjustment of the full set of 134 primary RV data of HD\,152219 with Eqs.\,\eqref{eq:RV} and \eqref{eq:w}. The best-fit solution is shown in each panel by the black filled dot. The corresponding $1\,\sigma$ confidence level is shown by the blue contour. \label{contoursomega}}
\end{figure*}

\begin{figure*}[p]
\centering
\includegraphics[clip=true,trim=0cm 6cm 1cm 0cm,width=0.85\linewidth]{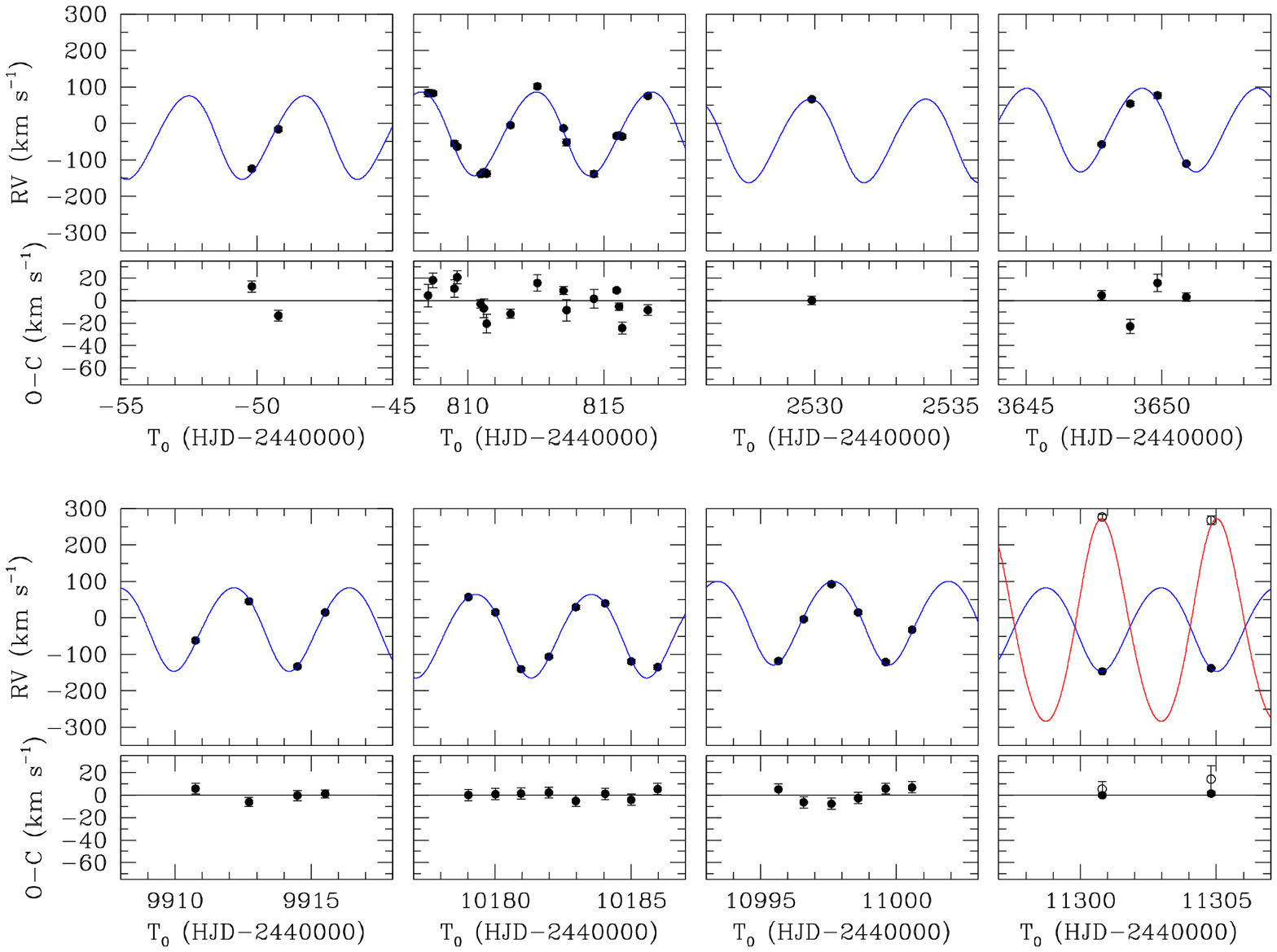}\\
\vspace{0.4cm}
\includegraphics[clip=true,trim=0cm 6cm 1cm 0cm,width=0.85\linewidth]{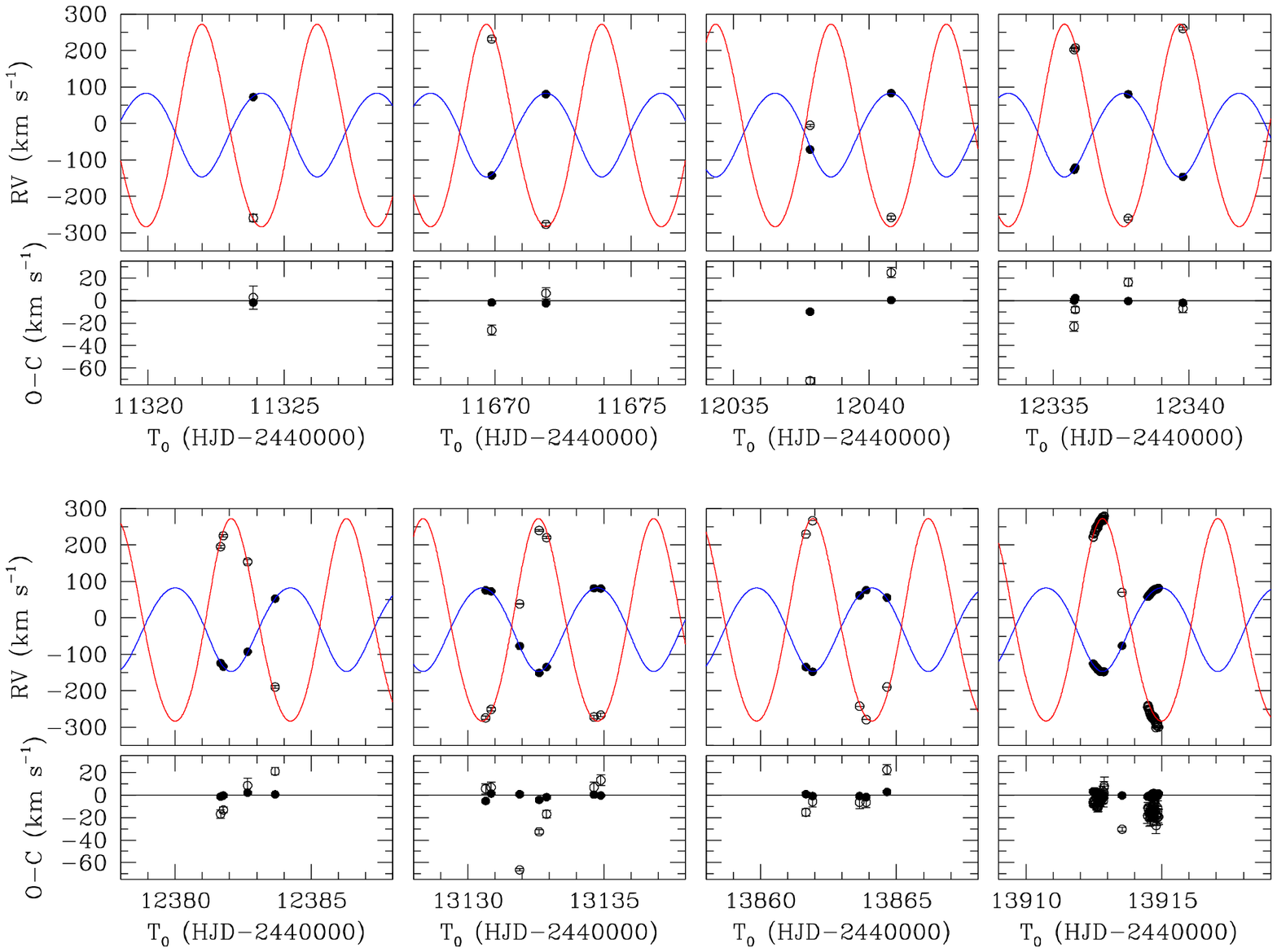}
\caption{Comparison between the measured RVs of the primary (filled dots) and secondary (open dots, when available) with the RVs expected from relations \eqref{eq:RV} and \eqref{eq:w} with the best-fit parameters given in Table\,\ref{bestfitTable}. \label{fitRV}}
\end{figure*}

To find the values of the six free parameters ($K_1$, $P_{\rm orb}$, $e$, $T_0$, $\omega_0$, and $\dot{\omega}$) that provide the best-fit to the whole set of RV data, we scanned the parameter space in a systematic way. Figure\,\ref{contoursomega} illustrates the projections of the 6-D parameter space onto 2-D planes and the best-fit values are given in Table\,\ref{bestfitTable}. The 1$\sigma$ confidence contours are computed in each two-dimensional plane adopting $\Delta\chi^2=2.30$ as appropriate for two free parameters. Figure\,\ref{fitRV} illustrates the best fit of the RV data at all epochs.

\begin{table}[htb]
\caption{Best-fit orbital parameters of HD\,152219 obtained via Eqs.\,\eqref{eq:RV} and \eqref{eq:w} and their 1$\sigma$ uncertainties.}
\centering
\begin{tabular}{l l}
\hline\hline
Parameter & Primary RVs \\
\hline
\vspace*{-3mm}\\
$P_{\rm orb}$\,(d) & $4.24046^{+0.00005}_{-0.00004}$ \\
\vspace*{-3mm}\\
$e$              & $0.072^{+0.004}_{-0.005}$ \\
\vspace*{-3mm}\\
$\dot{\omega}$ ($^{\circ}$\,yr$^{-1}$) & $1.198\pm 0.300$ \\
\vspace*{-3mm}\\
$\omega_0$ ($^{\circ}$) & $164.1 \pm 2.6$ \\
\vspace*{-3mm}\\
$T_0$ (HJD$-$2\,450\,000) & $3200.28 \pm 0.03$ \\
\vspace*{-3mm}\\
$K_{\rm P}$\,(km\,s$^{-1}$) & $114.7^{+0.3}_{-0.2}$ \\
\vspace*{-3mm}\\
\textbf{$\chi_\nu^2$} & 2.72 \\
\vspace*{-3mm}\\
\hline
\vspace*{-3mm}\\
$q = M_\text{S}/M_\text{P}$   & $0.413 \pm 0.004$  \\
\vspace*{-3mm}\\
$K_{\rm S}$\,(km\,s$^{-1}$) & $277.7 \pm 2.9$ \\
\vspace*{-3mm}\\
$a_{\rm P}\,\sin{i}$\,($R_{\odot}$) & $9.59 \pm 0.03$ \\
\vspace*{-3mm}\\
$a_{\rm S}\,\sin{i}$\,($R_{\odot}$) &  $23.22 \pm 0.24$ \\
\vspace*{-3mm}\\
$M_{\rm P}\,\sin^3{i}$\,($M_{\odot}$) & $18.64 \pm 0.47$ \\
\vspace*{-3mm}\\
$M_{\rm S}\,\sin^3{i}$\,($M_{\odot}$) & $7.70 \pm 0.12$\\
\vspace*{-3mm}\\
\hline
\end{tabular}
\label{bestfitTable}
\end{table} 

The value of $\dot{\omega}$ that best fits our RV time series is $(1.198 \pm0.300)^{\circ}$\,yr$^{-1}$. \citet{Mayer08} derived a rate of apsidal motion of $(1.64 \pm 0.18)^{\circ}$\,yr$^{-1}$ based on the then available literature RVs and ASAS-3 photometry. Within the errorbars the two values overlap. 

The primary and secondary radial velocities of an SB2 binary are related to each other through a linear relation
\begin{equation}\label{massratio}
RV_{\rm P}(t) = -q\,RV_{\rm S}(t) + B,
\end{equation}
where $q$ is the mass ratio $(q = \frac{M_{\rm S}}{M_{\rm P}}$, $M_\text{P}$ and $M_\text{S}$ being the mass of the primary and secondary star, respectively) and $B$ is a linear combination of the apparent systemic velocities of the two stars ($B = \gamma_{\rm P} + q\,\gamma_{\rm S}$). Applying this linear regression to the RVs obtained in the disentangling process, we found $q = \frac{M_{\rm S}}{M_{\rm P}} = 0.413 \pm 0.004$. We used this result to build an SB2 orbital solution for HD\,152219. Except for two more discrepant points (at HJD 2\,452\,037.8 and 2\,453\,131.9), the secondary RVs were adjusted with $\chi^2_\nu = 13.1$. Our best-fit parameters and their $1\sigma$ errors are listed in Table\,\ref{bestfitTable}. We note that the mass ratio we found is higher than the value ($q = 0.395 \pm 0.003$) inferred by \citet{San06} from the He\,{\sc i} lines. We further note that our best-fit eccentricity is lower than their value ($e = 0.082 \pm 0.011$) but still compatible within the error bars, whilst their semi-amplitudes of the primary and secondary RV curves are respectively smaller ($K_{\rm P} = 110.7 \pm 0.7$\,km\,s$^{-1}$) and larger ($K_{\rm S} = 279.9 \pm 1.7$\,km\,s$^{-1}$), the latter being compatible within the error bars. Again these differences are likely due to the two-Gaussian fit used by \citet{San06} to establish the RVs as well as to the fact that their solution did not account for the apsidal motion. \\

We further used Eq.\,\eqref{massratio} to convert the RVs of both stars into equivalent RVs of the primary star with
\begin{equation}
RV_\text{eq}(t) = (RV_\text{P}-qRV_\text{S}(t)+B)/2,
\end{equation}
whenever secondary RVs were available. For each time of observation $t$, we adjusted the equivalent RV data explicitly accounting for apsidal motion. The best fit is achieved for $\dot\omega = (1.69 \pm 0.23)^\circ$\,yr$^{-1}$ and has $\chi^2_\nu = 4.61$. This best-fit solution is not as good as the one based on the primary RVs only, as judged by the mean rms residual value, which is three times larger than the one obtained for the primary RVs fit. In addition, those results predict values for $\omega$ at the times of the TESS observations that are not compatible, within the error bars, with the values obtained from the fit of the TESS data. For those reasons, we kept the solution based on the adjustment of the primary RVs only as our best-fit solution.

\section{Light curve analysis\label{sect:photom}}
We analysed the light curves of HD\,152219 with the \texttt{Nightfall} binary star code (version 1.92) developed by R. Wichmann, M. Kuster, and P. Risse\footnote{The code is available at the URL: http://www.hs.uni-hamburg.de/DE/Ins/Per/Wichmann/Nightfall.html} \citep{Wichmann}. The shape of the stars is described by the Roche potential scaled with the instantaneous separation between the stars. In the absence of any surface spots, the eight model parameters are the mass ratio $q$, the effective temperatures $T_\text{eff,P}$ and $T_\text{eff,S}$ of the primary and secondary stars, the Roche lobe filling factors $f_\text{P}$ and $f_\text{S}$ of the primary and secondary stars, the eccentricity of the orbit $e$, the orbital inclination $i$, and the argument of periastron $\omega$. The Roche lobe filling factor is defined as the ratio between the polar radius of the star and the associated polar radius of the Roche lobe at periastron. In the case of HD\,152219, an additional parameter comes from the possible contribution of the third light discussed in Sect.\,\ref{subsect:photometry}. A quadratic limb-darkening law was adopted with the coefficients in the $I$ band coming from \citet{claret00}. In view of the stellar effective temperatures, the gravity darkening exponent was set to 0.25 as appropriate for massive stars with radiative envelopes. The code accounts for reflection effects by taking into account the mutual irradiation of the stellar surface elements of both stars \citep{Hen92}.

Following the radial velocities analysis (see Table\,\ref{bestfitTable} and Sect.\,\ref{sect:omegadot}), we fixed $q$ to 0.413 and $e$ to $0.072$. We fixed the effective temperature of the primary star to 30\,900\,K as derived from the \texttt{CMFGEN} analysis (see Table\,\ref{Table:CMFGEN} and Sect.\,\ref{subsect:cmfgen}). In a first attempt to reproduce the TESS-12 light curve of the binary system, we also fixed the effective temperature of the secondary star to 23\,500\,K as found in the \texttt{CMFGEN} analysis. However, the solutions obtained in this way were unable to reproduce the depth of the primary eclipse. Considering both the difficulties encountered in the spectroscopic analysis to determine the effective temperature of the secondary star and the high uncertainty on the inferred value, we decided to leave $T_\text{eff,S}$ free in the light curve analysis. 
For each adjustment, we therefore had five free parameters: $T_\text{eff,S}$, $f_\text{P}$, $f_\text{S}$, $i$, and $\omega$. The results are given for the TESS-12 data in Table\,\ref{tab:nightfall}, together with the corresponding reduced $\chi_\nu^2$. 
We note that the secondary/primary light ratio in the V-band of 0.084, estimated by the \texttt{Nightfall} code, is compatible with the light ratio inferred from spectroscopy (see Sect.\,\ref{subsect:spectralclass}) within the error bars.

\begin{table}[h]
\caption{Parameters of the best-fit \texttt{Nightfall} model for the TESS-12 light curve.}
\label{tab:nightfall}
\centering
\begin{tabular}{l l}
\hline\hline
\vspace*{-0.3cm} \\
Parameter & Value \\
\vspace*{-0.3cm} \\
\hline
\vspace*{-0.3cm} \\
$I_3$ & 0 \\
\vspace*{-0.3cm} \\
$T_\text{eff,S}$ (K)&  $21\,697^{+370}_{-340}$  \\
\vspace*{-0.3cm} \\
$f_\text{P}$ & $0.711 \pm 0.009$ \\
\vspace*{-0.3cm} \\
$f_\text{S}$ & $0.418 \pm 0.005$\\
\vspace*{-0.3cm} \\
$i$ $(^\circ)$  & $89.58^{+0.42}_{-2.28}$ \\
\vspace*{-0.3cm} \\
$\omega$ $(^\circ)$  & $192.4^{+3.4}_{-9.4}$ \\
\vspace*{-0.3cm} \\
$\chi_\nu^2$ &0.376 \\
 \vspace*{-0.3cm} \\
\hline
\end{tabular}
\end{table}

\begin{figure}[h]
\centering
\includegraphics[clip=true,trim=0 220 40 40,width=\linewidth]{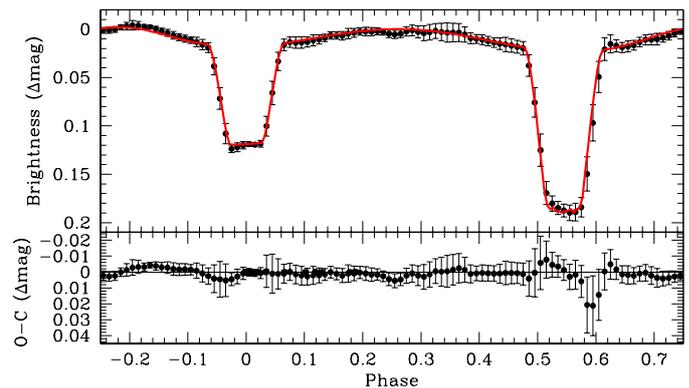}
\caption{\textit{Upper panel:} best-fit \texttt{Nightfall} solution of the TESS-12 light curve of HD\,152219. \textit{Lower panel:} residuals over the best-fit solution. \label{fig:nightfall}}
\end{figure}

We adopted the parameters of the best-fit to the TESS-12 data as our reference photometric solution. The best fit is illustrated in Fig.\,\ref{fig:nightfall}.

\begin{figure*}[h]
\centering
\hspace{-11.8cm}
\includegraphics[clip=true,trim=0 64 330 360,width=0.328\linewidth]{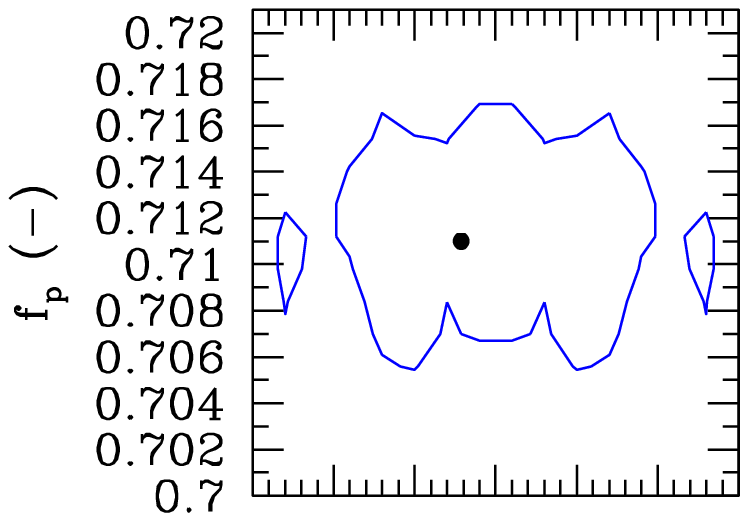}\\
\includegraphics[clip=true,trim=0 20 25 40,width=0.74\linewidth]{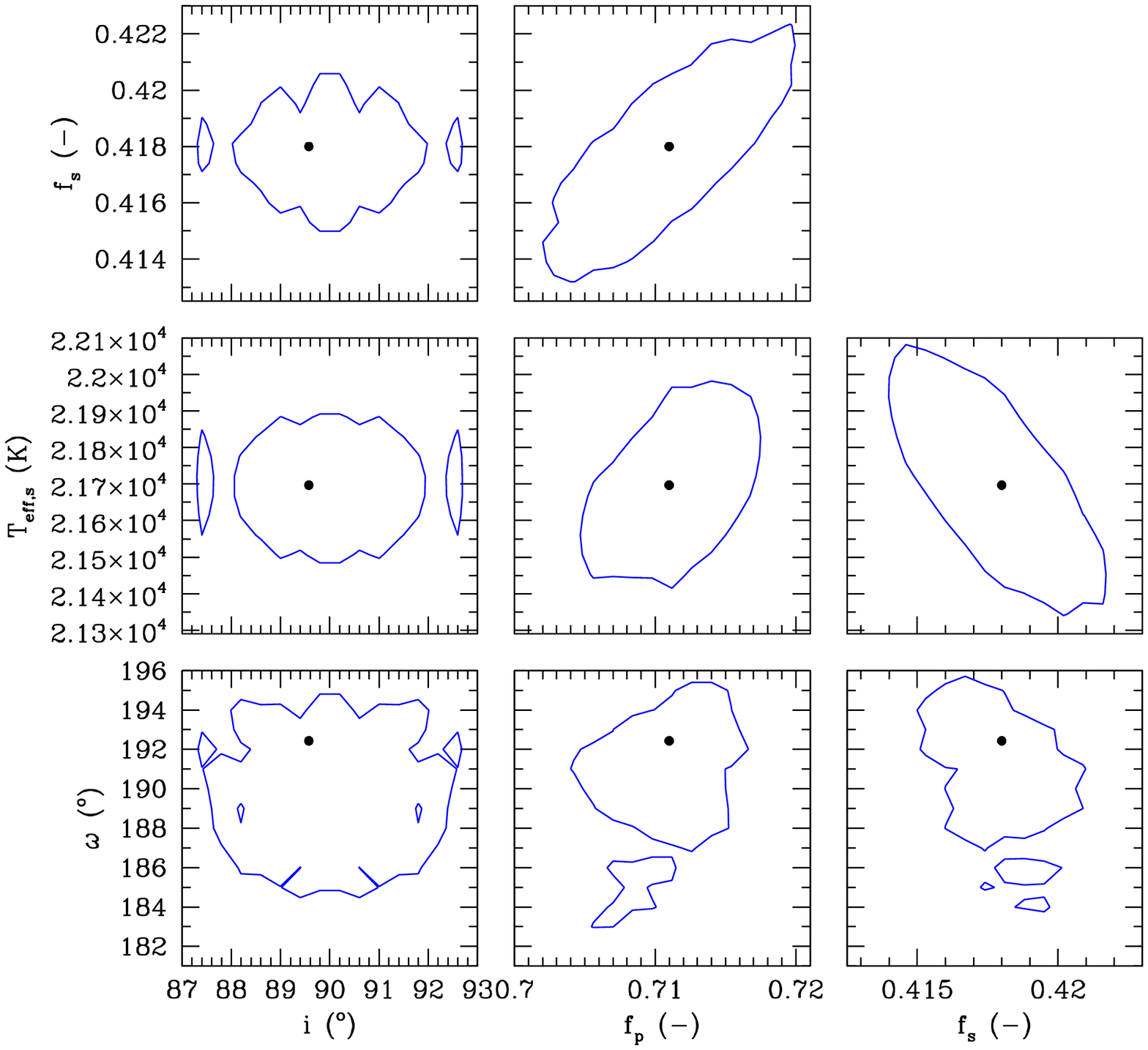}
\includegraphics[clip=true,trim=80 20 330 40,width=0.221\linewidth]{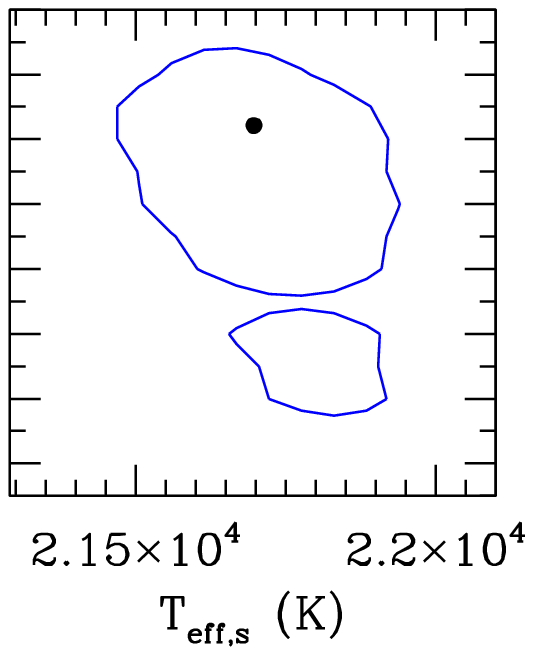}
\caption{Confidence contours for the best-fit parameters obtained from the adjustment of the TESS-12 light curve of HD\,152219 with the \texttt{Nightfall} code. The best-fit solution (i.e. with a third light contribution $I_3 = 0$) is shown in each panel by the black dot, while the corresponding $1\sigma$ confidence level is shown by the blue contour. \label{fig:nightfall_contours}}
\end{figure*}

We scanned the parameter space of the five free parameters to estimate the error bars. For this purpose, we used the $\chi^2$ mapping functionality implemented in \texttt{Nightfall}. This function scans two parameters at a time, fixing the other three to their best-fit value. We then computed the $1\sigma$ confidence contours in each two-dimensional plane adopting $\Delta \chi^2 = 2.30$ as appropriate for two free parameters. The confidence contours obtained in this way are plotted in dark blue in Fig.\,\ref{fig:nightfall_contours}. In this way, adopting the largest contours on the parameters to account for correlations, we obtained the values given in Table\,\ref{tab:nightfall}. 
For a near circular orbit seen under an inclination of $90^\circ$, the depths of the eclipses are mostly set by the ratio of the effective temperatures of the two stars. The primary star's effective temperature, $T_\text{eff,P}$, determined in Sect.\,\ref{subsect:cmfgen} is subject to uncertainties that hence propagate in the determination of $T_\text{eff,S}$. To quantify this dependence, we computed two light curves fixing all parameters to their best-fit value, setting $T_\text{eff,P}$ to its value $\pm 1 \sigma$ (i.e.\ 31\,900 and 29\,900\,K, respectively), and leaving $T_\text{eff,S}$ as the only free parameter. We respectively obtained $T_\text{eff, S} = 22\,320$ and 21\,047\,K, values which differ from the best-fit temperature by +623 and -650\,K, respectively. Considering the results from the {\tt CMFGEN} modelling (Sect.\,\ref{subsect:cmfgen}) and the light curve analysis, we hence adopt an error bar of $\pm 1000$\,K for $T_\text{eff,S}$ to be conservative.

The combination of the minimum semi-major axis and masses inferred from the radial velocity analysis (see Table\,\ref{bestfitTable}) on the one hand and the best-fit inclination value on the other hand yields a semimajor axis of $32.81 \pm 0.24\,R_\odot$ for the system and absolute masses for the primary and secondary stars $M_1 = 18.64 \pm 0.47$\,$M_\odot$ and $M_2 = 7.70\pm 0.12$\,$M_\odot$. As expected from the discussion in Sect.\,\ref{subsubsect:RML}, these masses are higher than the spectroscopic masses even though the primary masses are compatible within the error bars.  From this solution, we inferred values for the primary and secondary stellar radii of $R_1 = 9.40^{+0.14}_{-0.15}\,R_\odot$ and $R_2 = 3.69\pm 0.06\,R_\odot$. Those radii are compatible within the error bars with the spectroscopic ones. This leads to photometric values of the surface gravities of the primary and secondary stars $\log g_1 = 3.76 \pm 0.02$ and $\log g_2 = 4.19\pm 0.01$. As expected, these values are higher than those derived from the spectroscopic analysis (see discussion in Sect.\,\ref{subsect:cmfgen}). From the photometric stellar radii, the primary effective temperature derived in Sect.\,\ref{subsect:cmfgen} and the secondary effective temperature derived here above, we infer bolometric luminosities $L_\text{bol,1} = (7.26 \pm 0.97) \times 10^4\,L_\odot$ and $L_\text{bol,2} = (2.73\pm 0.51)\times 10^3\,L_\odot$ for the primary and secondary stars, respectively. These values are compatible with those derived from the spectroscopic analysis (see Sect.\,\ref{subsubsect:RML}). If we further assume the stellar rotational axes to be aligned with the normal to the orbital plane (but see Sect.\,\ref{sect:discussion}), the combination of the orbital inclination, the stellar radii, and the projected rotational velocities of the stars derived in Sect.\,\ref{subsect:vsini} yields rotational periods for the primary and secondary stars $P_\text{rot,1} = 2.86 \pm 0.18$\,days and $P_\text{rot,2} = 1.97\pm 0.19$\,days. The ratio between rotational angular velocity and instantaneous orbital angular velocity at periastron amounts to $1.28 \pm 0.08$ and $1.86 \pm 0.18$ for the primary and secondary stars, respectively.

\citet{Mayer08} inferred an orbital inclination of $i = 72\pm 3^{\circ}$ from their analysis of the ASAS-3 photometric data. We note that our best-fits values are significantly larger, close to $90^{\circ}$. The total primary eclipse puts a stringent constraint on the inclination of the system. Hence, we checked that their inclination value of $72^{\circ}$, adopting for the other parameters our best-fit values, is not consistent with the total eclipse. To reproduce a total primary eclipse, an inclination of at least $80^\circ$ is required.

\begin{figure}[h]
\centering
\includegraphics[clip=true,trim=0 220 40 40,width=\linewidth]{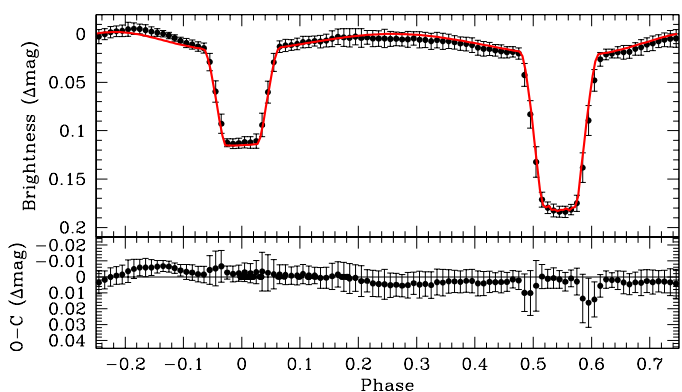}
\caption{\textit{Upper panel:} best-fit \texttt{Nightfall} solution of the TESS-39 light curve of HD\,152219. \textit{Lower panel:} residuals over the best-fit solution. \label{fig:lightcurve_TESS39}}
\end{figure}

The light curves of a binary system can in principle be used to determine the apsidal motion rate of the system \citep[e.g. ][]{zasche}. However, in the case of HD\,152219, the total time span of photometric observations is much smaller than the total time span of spectroscopic observations. Hence, we rather decided to use the photometric observations to perform a consistency check of the apsidal motion determined based on the spectroscopic data. To be consistent with the results obtained for the TESS-12 light curve, we adjusted the TESS-39 light curve keeping all parameters fixed to the values inferred for the TESS-12 except for $\omega$. We found $\omega=192.41$ and $\chi_\nu^2 = 0.500$ for the best-fit adjustment. We observed that the depths of the eclipses were not perfectly reproduced: the theoretical depths were deeper than observed. This suggests that some third light contribution is present in the TESS-39 data and we hence adjusted the light curve keeping the third light contribution as a second free parameter. The best-fit adjustment was achieved for $\omega =192.480^\circ$ and $I_3=0.032$, has $\chi^2_\nu = 0.395$, and is illustrated in Fig.\,\ref{fig:lightcurve_TESS39}. Adopting $1\sigma$ contours for the two-parameter space, we obtained $\omega=(192.5^{+3.8}_{-6.1})^\circ$ and $I_3=0.032^{+0.015}_{-0.014}$. 

\begin{figure}[h]
\centering
\includegraphics[clip=true,trim=0 0 0 0,width=\linewidth]{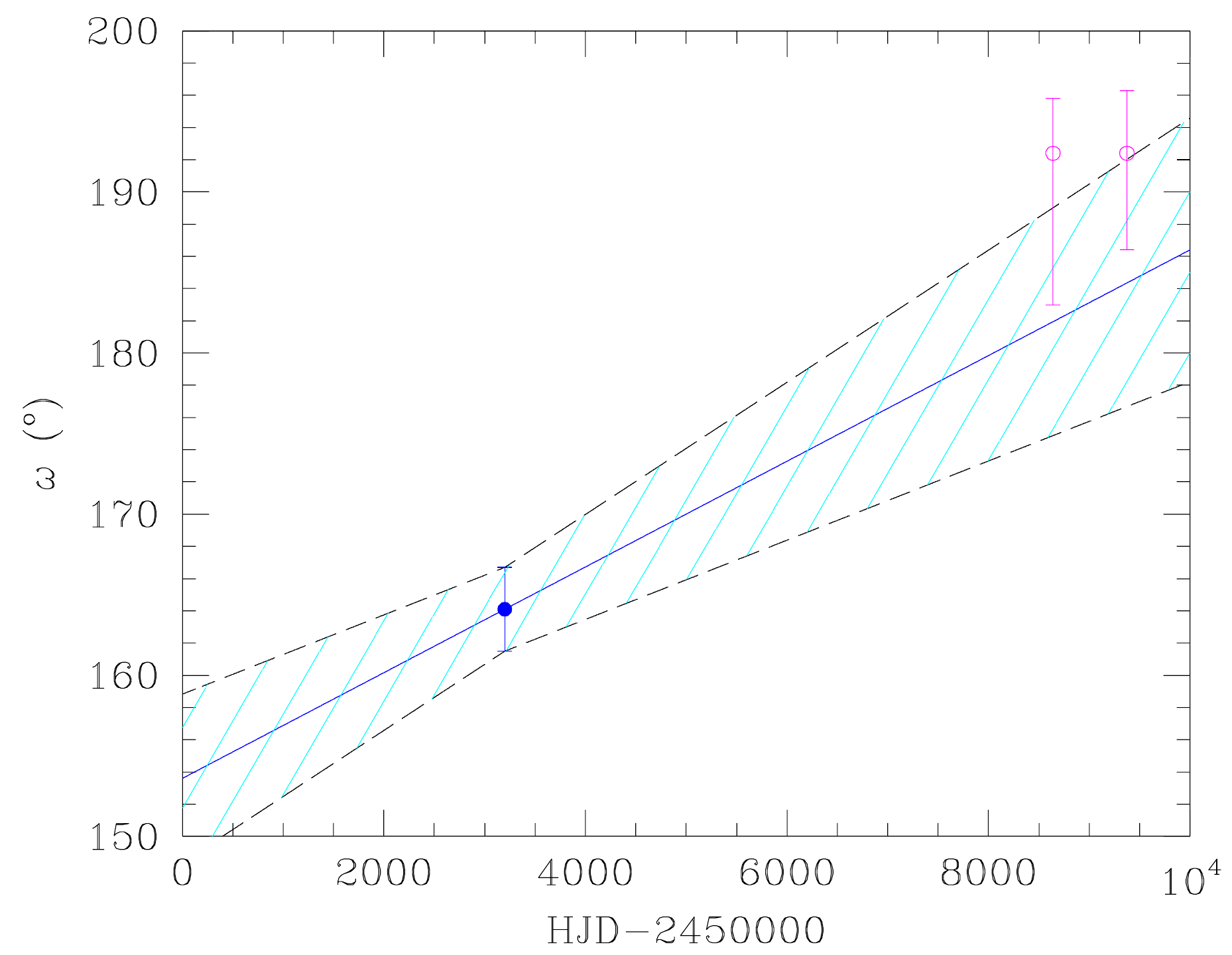}
\caption{Values of $\omega$ as a function of time inferred from the photometric light curves and the RVs. The pink symbols correspond to the data of the fits of the TESS-12 and TESS-39 photometry. The blue dot indicates the $\omega_0$ value obtained from the global fit of all RV data. The solid blue line corresponds to our best-fit value of $\dot\omega$ inferred from the RVs, and the dashed cyan zone corresponds to the range of values according to the $1\sigma$ uncertainties on $\omega_0$ and $\dot\omega$. \label{fig:omega_vs_time}}
\end{figure}

The resulting values of $\omega$ are provided in Table\,\ref{tab:omega_photom} together with the reduced $\chi^2_\nu$. Figure\,\ref{fig:omega_vs_time} illustrates the variations of $\omega$ with time. Both TESS-12 and TESS-39 data agree with the result inferred from the global fit of the RV data within the error bars. 
Unfortunately, the ASAS-3 data are too sparsely sampled to determine meaningful epoch-dependent values of $\omega$. 

\begin{table}[h]
\caption{Best-fit values of $\omega$ from photometry.}
\label{tab:omega_photom}
\centering
\begin{tabular}{l r r r r r}
\hline\hline
\vspace*{-0.3cm} \\
Epoch & $N$ & $\Delta t$ & Instr. & $\omega$ & $\chi_\nu^2$\\ 
(HJD) & & (d) & & ($^\circ$) \\
\vspace*{-0.3cm} \\
\hline
\vspace*{-0.3cm} \\
2\,458\,641.4 & 11\,944 & 23.0 & TESS-12& 192.4 & 0.376\\
2\,459\,374.4 & 3411 & 25.2 & TESS-39 & 192.5& 0.395\\
\hline
\end{tabular}
\end{table}

\begin{figure}[h]
\centering
\includegraphics[clip=true,trim=0 0 0 0,width=\linewidth]{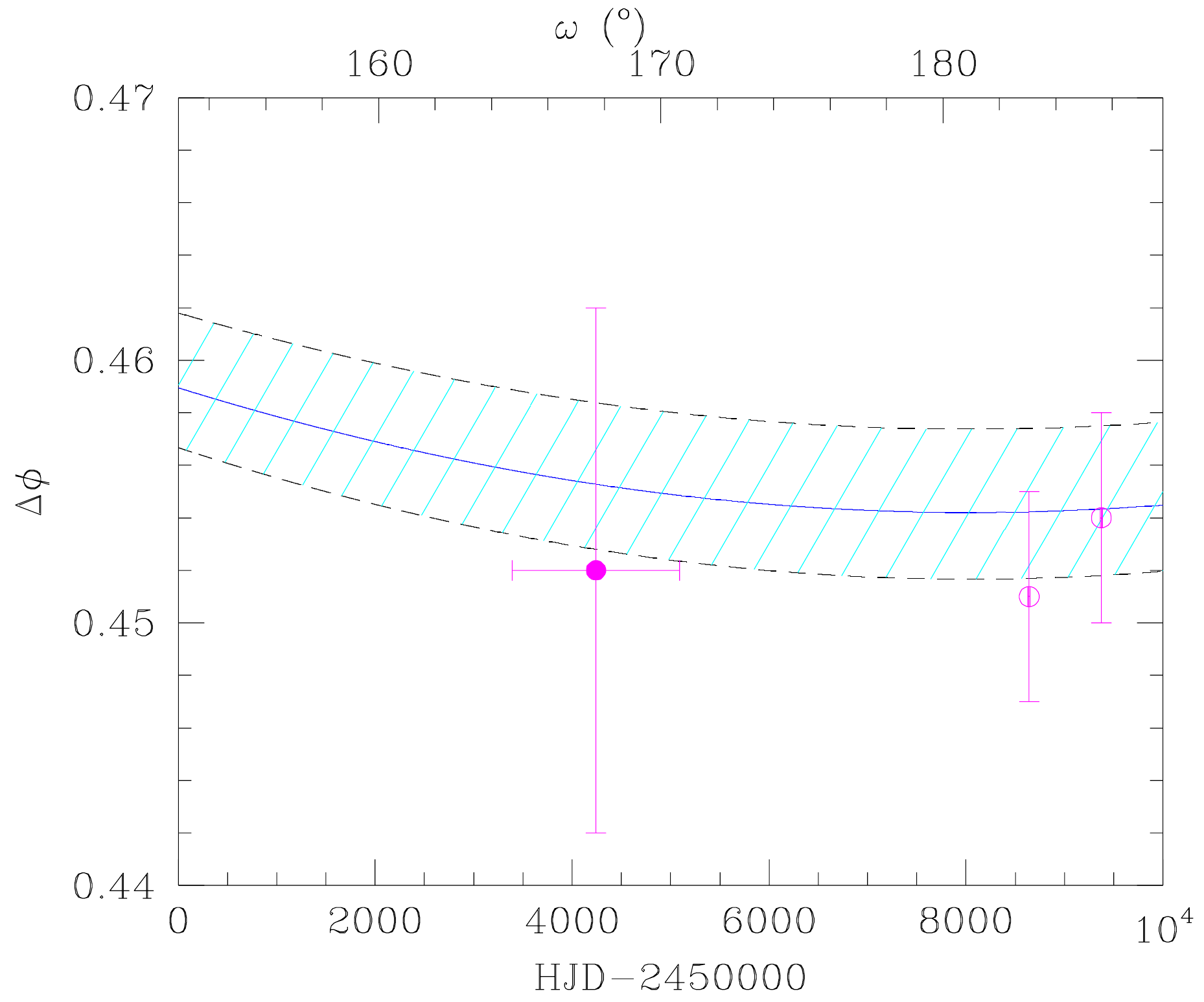}
\caption{Values of the phase difference $\Delta\phi$ between the primary and secondary eclipses as a function of time and $\omega$ inferred from the photometric light curves and the RVs. The pink filled symbol and the pink open symbols correspond to the data of the fits of the ASAS-3 photometry, and the TESS-12 and TESS-39 photometry, respectively.  The solid blue line corresponds to our best-fit value of $\Delta\phi$ inferred from the RVs, and the hatched cyan zone corresponds to the range of values according to the $1\sigma$ uncertainties on $e$.  \label{fig:timing1}}
\end{figure}

We computed the phase difference between the primary and secondary eclipses. To do so, we adjusted a second-order polynomial to the eclipses for the TESS-12, TESS-39, and combined ASAS-3 data and found the values of $0.451 \pm 0.004$, $0.454 \pm 0.004$, and $0.452 \pm 0.010$, respectively (see Fig.\,\ref{fig:timing1}). These values were confirmed by computing the first order moment of the eclipses and from the times of minima determined by the {\tt Nightfall} code. We adopted conservative error bars to take into account the difficulty to determine the minima of the secondary eclipse. We then computed the phase difference as a function of $\omega$, defined as
\begin{equation}
\Delta \phi = \frac{t_2-t_1}{P_\text{orb}} = \frac{\Psi-\sin\Psi}{2\pi},
\end{equation}
where $t_1$ and $t_2$ are the primary and secondary minima, respectively,  
\begin{equation}
\Psi = \pi + 2\arctan{\left(\frac{e\cos\omega}{\sqrt{1-e^2}}\right)},
\end{equation}
and adopting $e$ from the RV analysis. The resulting curve is plotted in Fig.\,\ref{fig:timing1}. We observe that within the errorbars, the observational values of $\Delta \phi$ agree with the curve showing the expected values from the RV analysis. Adopting the periastron length obtained from the RV analysis at the time of the TESS-12 observations, we find an eccentricity of $0.070^{+0.006}_{-0.005}$, that is compatible within the errorbars with an eccentricity of 0.072, therefore confirming the coherence between the RV and photometric analysis.

\section{Internal structure constant $k_2$ \label{sect:k2}}
The apsidal motion rate of a binary system is made of two contributions: 
\begin{equation}
\dot\omega = \dot\omega_\text{N} + \dot\omega_\text{GR}.
\end{equation}
The classical Newtonian contribution (N) is the dominant term in the majority of non-degenerate binary systems, except in a few cases where the general relativistic correction (GR) is the dominant term \citep[see e.g.][and reference therein]{baroch21}.
    
The general relativity contribution to the rate of apsidal motion depends on the total mass of the binary, the orbital period, and the eccentricity of the system through the following expression 
\begin{equation}
\label{eqn:omegadotGR}
\dot\omega_\text{GR} = \left(\frac{2\pi}{P_\text{orb}}\right)^{5/3}\frac{3(G(M_1+M_2))^{2/3}}{c^2 (1-e^2)},
\end{equation} 
when only the quadratic corrections are taken into account \citep{shakura}. In this expression, $G$ and $c$ stand for the gravitational constant and the speed of light, respectively. Using the eccentricity and the orbital period derived from the radial velocity analysis in Sect.\,\ref{sect:omegadot} as well as the masses derived from the photometric analysis in Sect.\,\ref{sect:photom}, we inferred a value for $\dot\omega_\text{GR}$ of $(0.159\pm 0.002)^\circ$\,yr$^{-1}$ for HD\,152219. As a consequence, the observational contribution of the Newtonian term $\dot\omega_\text{N}$ amounts to $(1.039\pm 0.300)^\circ$\,yr$^{-1}$.

The Newtonian contribution takes the form adopted from \citet{sterne},
\begin{equation}
\label{eqn:omegadotN}
\begin{aligned}
\dot\omega_\text{N} = \frac{2\pi}{P_\text{orb}} \Bigg[&15f(e)\left\{k_{2,1}q \left(\frac{R_1}{a}\right)^5 + \frac{k_{2,2}}{q} \left(\frac{R_2}{a}\right)^5\right\} \\
& \begin{aligned} + g(e) \Bigg\{ &k_{2,1} (1+q) \left(\frac{R_1}{a}\right)^5 \left(\frac{P_\text{orb}}{P_\text{rot,1}}\right)^2 \\ 
& + k_{2,2}\, \frac{1+q}{q} \left(\frac{R_2}{a}\right)^5 \left(\frac{P_\text{orb}}{P_\text{rot,2}}\right)^2 \Bigg\}  \Bigg],
\end{aligned}
\end{aligned}
\end{equation}
if the stellar rotation axes are orthogonal to the orbital plane, and where only the contributions arising from the second-order harmonic distortions of the potential are considered. In this expression, $f$ and $g$ are functions of the eccentricity of the system given by
\begin{equation}
\label{eqn:fg}
\left\{
\begin{aligned}
& f(e) = \frac{1+\frac{3e^2}{2}+\frac{e^4}{8}}{(1-e^2)^5}, \\ 
& g(e) = \frac{1}{(1-e^2)^2}.
\end{aligned}
\right. 
\end{equation}
The Newtonian term is itself the sum of the effects induced by stellar rotation and tidal deformation which in the present case amount to about 33\% and 67\% of the total Newtonian term, respectively. The main unknowns in Eq.\,\eqref{eqn:omegadotN} are $k_{2,1}$ and $k_{2,2}$, the internal structure constants of the primary and secondary stars. Also known as the apsidal motion constant, the internal structure constant $k_2$ of a star is expressed as 
\begin{equation}
k_2 = \frac{3-\eta_2(R_*)}{4+2\eta_2(R_*)},
\end{equation} 
where
\begin{equation}
\eta_2(R_{*}) =  \left.\frac{d\ln\epsilon_2}{d\ln r}\right|_{r=R_{*}}
\end{equation}
is the logarithmic derivative of the surface harmonic of the distorted star evaluated at the stellar surface ($r = R_*$), expressed in terms of the ellipticity $\epsilon_2$. $\eta_2(R_*)$ is the solution of the Clairaut-Radau differential equation
\begin{equation}
r \frac{d\eta_2(r)}{dr} + \eta_2^2(r) - \eta_2(r) + 6 \frac{\rho(r)}{\bar\rho(r)} \left(\eta_2(r)+1\right) - 6 = 0
\label{eqn:Radau}
\end{equation}
with the boundary condition $\eta_2(0) = 0$~\citep{Hejlesen}. The terms $\rho(r)$ and $\bar\rho(r)$ are the density at distance $r$ from the centre of the star and the mean density within the sphere of radius $r$, $r$ being the current radius at which the equation is evaluated. Qualitatively, the internal structure constant is a measure of the internal mass distribution inside the star, that is to say the density contrast between the stellar core and external layers. An homogeneous sphere of constant density has $k_2$ equal to 0.75 whilst a massive star having a very dense core and a diluted atmosphere can see its $k_2$ take a value as low as $10^{-4}$ \citep[see e.g.][]{rosu20b}. During stellar evolution, the external layers become more and more diluted compared to the stellar core, hence $k_2$ decreases with time, making this quantity a good indicator of stellar evolution. 

In the case of HD\,152219, it is impossible to observationally constrain the individual values of the apsidal motion constants of both stars, as Eq.\,\eqref{eqn:omegadotN} is an underdetermined system. Nevertheless, we can define a weighted-average apsidal motion constant $\bar{k}_2$ for the binary system. For this purpose, we first rewrite Eq.\,\eqref{eqn:omegadotN} in the form
\begin{equation}
\label{eqn:omegadotN2}
\dot\omega_\text{N} = c_1 k_{2,1} + c_2 k_{2,2},
\end{equation} 
where $c_1$ and $c_2$ are two functions of known stellar parameters which expressions are straightforward from Eq.\,\eqref{eqn:omegadotN}. Dividing Eq.\,\eqref{eqn:omegadotN2} by $(c_1+c_2)$, we get
\begin{equation}
\label{eqn:k2bar}
\bar{k}_2 = \frac{c_1 k_{2,1} + c_2 k_{2,2}}{c_1 + c_2} = \frac{\dot\omega_\text{N}}{c_1+c_2}. 
\end{equation}
All terms appearing in the right-hand side of Eq.\,\eqref{eqn:k2bar} are known from observations, hence we get $\bar{k}_2 = 0.00173\pm0.00052$. As the secondary star is smaller and less massive than the primary star, its $k_2$-value exceeds that of the primary. Therefore, we have $k_{2,1}< \bar{k}_2 < k_{2,2}$. Observationally, we have $\frac{c_1}{c_1+c_2} = 0.95$ and $\frac{c_2}{c_1+c_2}=0.05$, meaning that the $k_{2,1}$ has a much higher weight in the calculation of $\bar{k}_2$\footnote{Note however that this does not imply that the product $\frac{c_1}{c_1+c_2} k_{2,1}$ contributes 95\% to $\bar{k}_2$.}. We further note that - and here we anticipate the analysis performed in Sect.\,\ref{subsect:clesprim} - if we make the hypothesis that $k_{2,1} = \bar{k}_2$, and if $k_{2,2}$ appears to be equal to $n* k_{2,1}$, then the ensuing error beyond the initial hypothesis amounts to $(n-1)*5$\%.

\section{Stellar structure and evolution models\label{sect:cles}}
We computed stellar evolution models with the Code Li\'egeois d'\'Evolution Stellaire\footnote{The \texttt{Cl\'es} code is developed and maintained by Richard Scuflaire at the STAR Institute at the University of Li\`ege.} \citep[\texttt{Cl\'es}, ][]{scuflaire08}. An overview of the main features of \texttt{Cl\'es} in the present context is presented in \citet[Sect.\,2.1]{rosu20b}; we refer to this paper for further information and recall here only the most important points. For massive stars, \texttt{Cl\'es} allows to build stellar structure and evolution models from the Hayashi track to the asymptotic giant branch phase. The zero-age of a model is defined at the beginning of the pre-main sequence phase. The code adopts \texttt{OPAL} opacities from \citet{iglesias96} and solar chemical composition from \citet{Asplund}. The FreeEOS equation of state and the rates of nuclear reactions are implemented following, respectively, \citet[a short description can be found in \citet{cassisi03}]{Irwin12} and \citet{adelberger11}, whilst the mixing length theory is adopted to parameterise the mixing in convective regions \citep{cox68}. The stellar atmosphere is a radiative gray model atmospheres computed in the Eddington approximation, which is treated separately and added as a boundary condition. By default, internal mixing for massive stars is restricted to the convective core, but the user can introduce overshooting and/or turbulent mixing in the models \citep[see e.g.][]{rosu20b}. When overshooting is introduced in a model, the boundary of the central mixed region is displaced from its initial position given by the Ledoux criterion over a distance $\alpha_\text{ov} H_p$. The temperature gradient is the radiative gradient in the overshooting region.
Turbulent mixing can be introduced in the models through the inclusion of a pure diffusion term that reduces the abundance gradient of the considered element. It takes the following form
\begin{equation}
-D_T \frac{\partial \ln X_i}{\partial r},
\end{equation}
where $r$ is the radial coordinate, $X_i$ is the mass fraction of element $i$ at location $r$, and 
\begin{equation}
D_T = D_\text{turb} \left(\frac{\rho}{\rho_0}\right)^n + D_\text{ct}
\end{equation}
is the turbulent diffusion coefficient (positive and expressed in cm$^2$\,s$^{-1}$). In this expression, $n$, $D_\text{turb}$, and $D_\text{ct}$ are three parameters chosen by the user. In the models, we did not include any microscopic diffusion. In the present analysis, we adopted a constant $D_T$ throughout the star, that is to say we set $n=0$.

The \texttt{Cl\'es} code does not include rotational mixing. Nonetheless, \citet{rosu20b} have shown that, to first order, the impact of rotational mixing on the internal structure can be simulated by means of the turbulent diffusion. Finally, the code is designed for single star evolution and does not account for the impact of binarity on the internal structure. However, \citet{rosu20b} have also shown that for an un-evolved binary system, the impact of binarity on the internal structures of the stars is quite modest and only slightly changes the age at which the star reaches a given stage of central condensation for a given external radius. Finally, the mass-loss rate induced by stellar winds is implemented through the \citet{vink01} recipe, to which the user can apply a multiplicative scaling factor $\xi$. \\

We built two sets of stellar evolution models for the primary and secondary stars of HD\,152219. Whilst the primary star shows clear evidence of mass-loss through stellar wind, the mass-loss rate of the secondary star is much smaller. Hence, the initial stellar mass and the mass-loss prescription are only important for the primary star. In accordance with \citet{Asplund}, we adopted as reference values the hydrogen fraction $X=0.715$ and the metallicity $Z=0.015$. These values are used to compute the models, unless otherwise stated (see Appendix\,\ref{subsect:XZ}). The overshooting parameter is poorly constrained: \citet{claret19} showed based on their \texttt{MESA} models that $\alpha_\text{ov}$ saturates around a value of 0.20 for stars more massive that 2\,$M_\odot$ but their sample stopped at 4.5\,$M_\odot$. More recently, \citet{rosu20b} explored values of $\alpha_\text{ov}$ up to 0.40 in their \texttt{Cl\'es} models for the binary system HD\,152248 located in the same cluster as our present target and showed evidence for the need of high $\alpha_\text{ov}$ values. \citet{tkachenko20} reached the same conclusion from the analysis of their sample of stars based on \texttt{MESA} models. \citet{martinet21} suggested to use a value of $\alpha_\text{ov} = 0.20$ for their \texttt{GENEC} rotating models for stars having a mass of 9\,$M_\odot$ or higher. 
We emphasise here that different analyses using different stellar evolution codes with different prescriptions of the mixing processes reach the same qualitative conclusion that strong internal mixing is required.

For each model, the internal structure constant is computed solving the differential equation Eq.\,\eqref{eqn:Radau} using a \texttt{Fortran} routine in which a fourth-order Runge-Kutta method with step doubling is implemented. This code, validated against polytropic models, has already been applied to the massive binaries HD\,152218 \citep{rauw16} and HD\,152248 \citep{rosu20b}. This $k_2$ value is then corrected by the quantity
\begin{equation}
\label{eqn:k2Claret}
\delta(\log k_2) = -0.87\lambda_\text{s},
\end{equation}
with
\begin{equation}
\lambda_\text{s}=\frac{2\Omega_\text{s}^2R_*^3}{3GM_*}
\end{equation}
to account for the centrifugal deformation due to stellar rotation, following \citet{Claret99}. In this expression, $\Omega_\text{s}$ stands for the observationally-determined rotational angular velocity, computed from the $v\,\sin{i}$ obtained in Sect.\,\ref{subsect:vsini}, and the stellar radius and the inclination determined in Sect.\,\ref{sect:photom}, while $R_*$ and $M_*$ stand for the radius and the mass of the corresponding model. In the following, unless explicitely specified, $k_{2,1}$ and $k_{2,2}$ are always corrected for the rotation of the star by Eq.\,\eqref{eqn:k2Claret}.

We used the \texttt{min-Cl\'es} \texttt{Fortran} routine\footnote{The \texttt{min-Cl\'es} code is developed and maintained by Martin Farnir at the STAR Institute at the University of Li\`ege.}, in which a Levenberg-Marquardt minimisation method \citep{press92} is implemented, to search for best-fit models. The routine determines the combination of the free parameters of the model allowing to best reproduce the set of observationally-determined present-day stellar properties. The initial masses and the age of the binary system are two major free parameters, also complemented by $D_T$ in most cases. The set of currently observed properties consists of the masses, radii, and effective temperatures of the stars, as well as the weighted-average internal structure constant of the system: these parameters are summarised in Table\,\ref{table:constraints}. However, in \texttt{min-Cl\'es}, only the first three parameters can directly be constrained in the models. Nonetheless, individual $k_2$ values can be constrained in the models (we will return to this in Sect.\,\ref{subsect:clesprim}). The \texttt{min-Cl\'es} routine assumes symmetric uncertainties, hence we adopted error bars on each parameter given by the maximum value of the upper and lower bounds.

\begin{table}[h]
\caption{Set of observationally-determined properties of the binary system HD\,152219 used for the \texttt{Cl\'es} analysis, for which we adopted symmetric uncertainties.\label{table:constraints}}
\centering
\begin{tabular}{l l l}
\hline\hline
\vspace{-3mm}\\
Parameter & \multicolumn{2}{c}{Value} \\
& Primary & Secondary \\
\hline
\vspace{-3mm}\\
$M$ ($M_\odot$) & $18.64 \pm 0.47$ & $7.70 \pm 0.12$  \\
\vspace{-3mm}\\
$R$ ($R_\odot$) & $9.40 \pm 0.15$ & $3.69\pm 0.06$ \\
\vspace{-3mm}\\
$T_\text{eff}$ (K) & $30\,900 \pm 1000$ & $21\,697\pm 1000$ \\
\vspace{-3mm}\\
$\bar{k}_2$ & \multicolumn{2}{c}{$0.00173 \pm 0.00052$} \\
\vspace{-3mm}\\
\hline
\end{tabular}
\end{table}

\subsection{Preliminary analysis}
In a first attempt to obtain stellar evolution models representative of both stars, we built two \texttt{Cl\'es} models simultaneously, adopting the six corresponding constraints ($M, R$, and $T_\text{eff}$ of both stars), and leaving the initial masses of the two stars and the age of the binary system as the only three free parameters. For both stars, we adopted $\xi=1$, $\alpha_\text{ov}=0.30$, and no turbulent diffusion. The best-fit solution gives an age of 7.8\,Myr, but none of the stellar parameters is correctly reproduced. Adding more free parameters would certainly help the convergence of the models but, due to degeneracies, might also lead to spurious solutions. Hence, we decided to proceed differently and to analyse both stars separately, keeping in mind that the ages of both stars have to match somehow.

\begin{figure}[h!]
\centering
\includegraphics[clip=true,trim=0.5cm 0cm 5.5cm 0.5cm,width=0.49\linewidth]{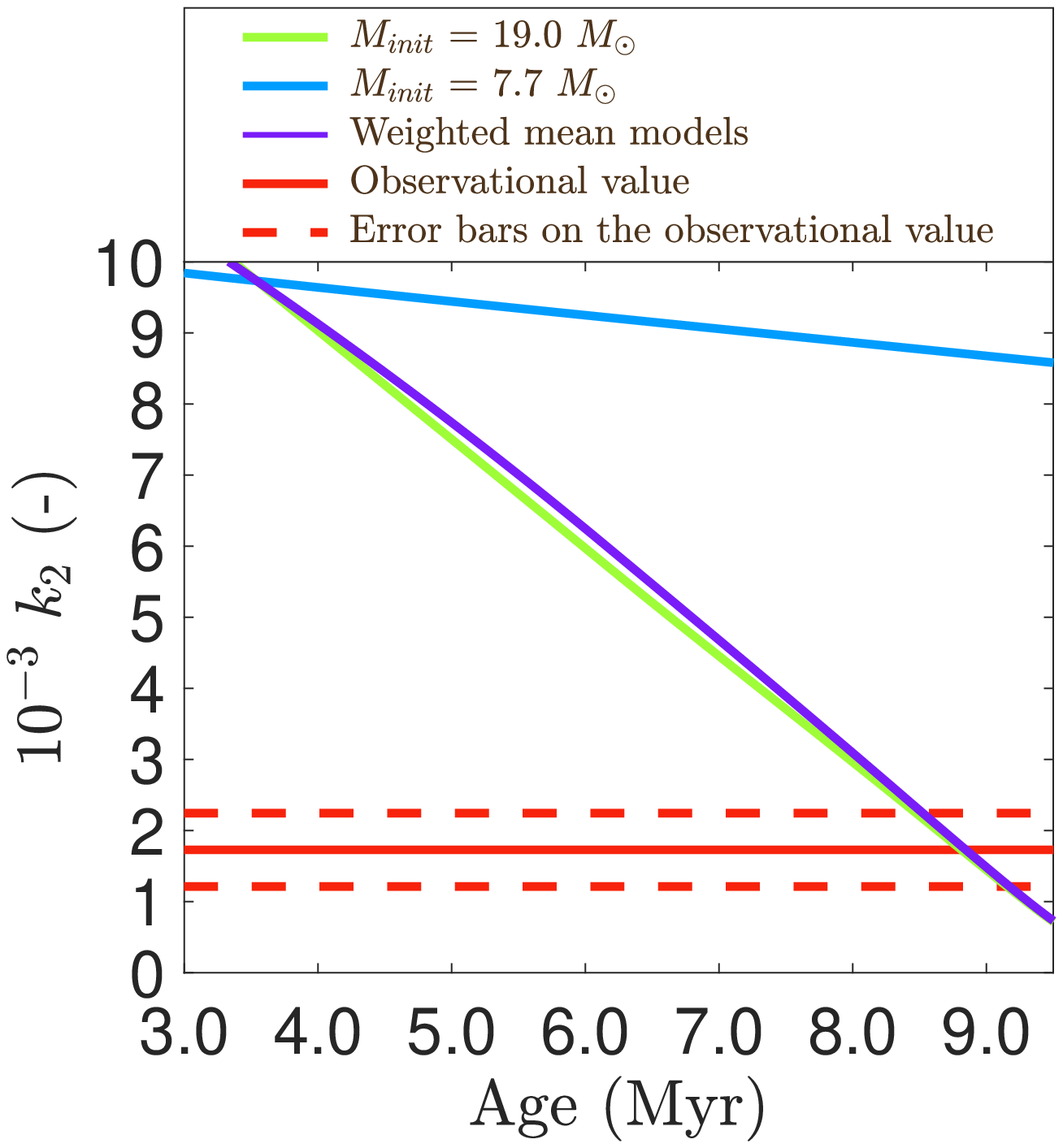}
\includegraphics[clip=true,trim=0.5cm 0cm 5.5cm 0.5cm,width=0.49\linewidth]{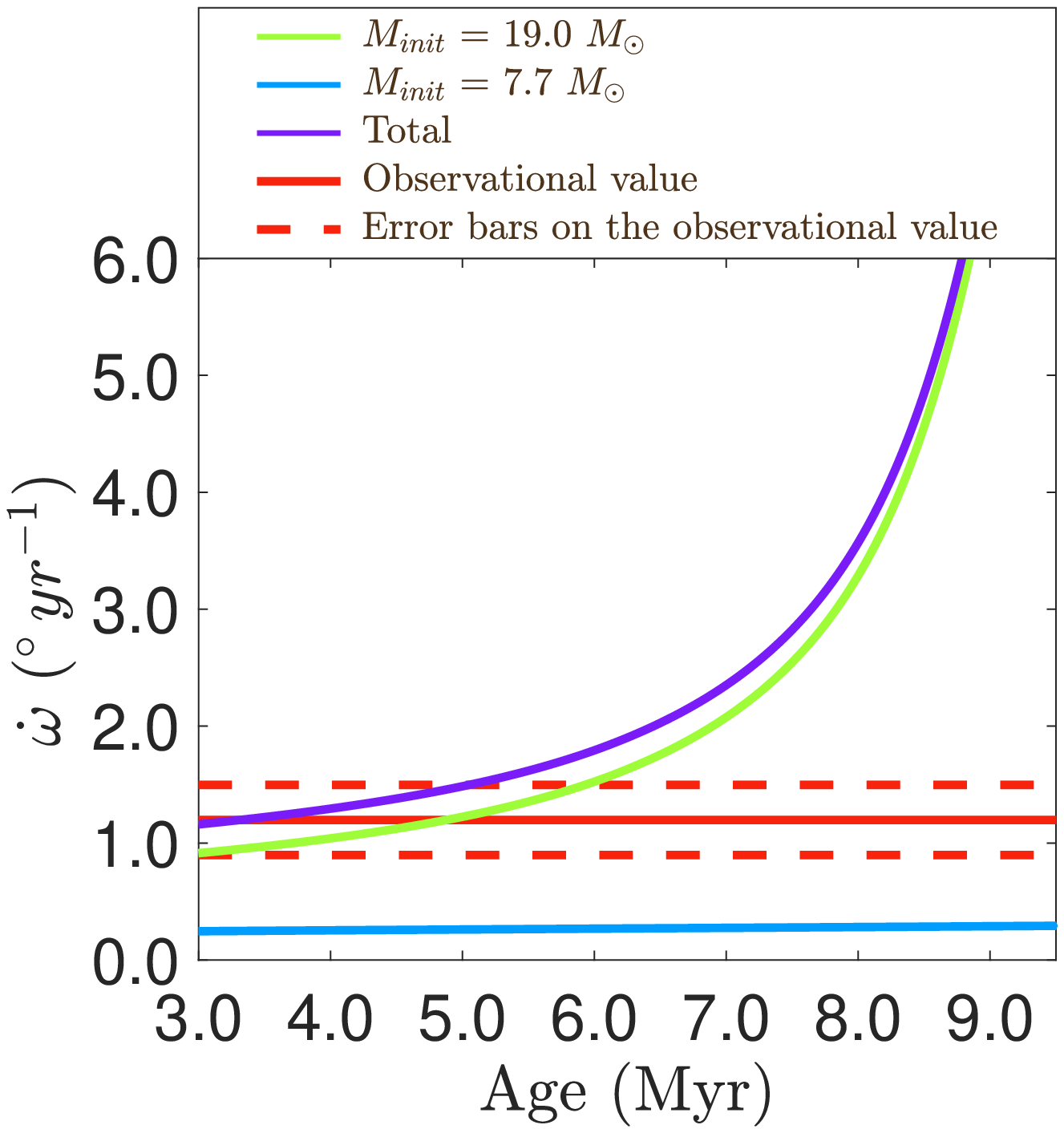}
\caption{\textit{Left panel:} Evolution as a function of stellar age of the internal structure constant (to which the empirical correction Eq.\,\eqref{eqn:k2Claret} has been applied) for \texttt{Cl\'es} models with initial mass of 19.0\,$M_\odot$ and $\xi=1$ ($k_{2,1}$, green) and initial mass of 7.7\,$M_\odot$ and $\xi=0.1$ ($k_{2,2}$, blue), both models have $\alpha_\text{ov}=0.30$ and no turbulent diffusion. The weighted-average mean of the $k_2$-values, computed using Eq.\,\eqref{eqn:k2bar}, is also depicted ($\bar{k}_2$, purple). \textit{Right panel:} Evolution as a function of stellar age of the total apsidal motion rate (purple) and of the contributions of the two stars (green and blue). The observational value of the corresponding parameter and its error bars are represented by the solid red line and the dashed red horizontal lines, respectively.\label{fig:evolution_k2omega}}
\centering
\includegraphics[clip=true,trim=3cm 2cm 10.5cm 5.5cm,width=\linewidth]{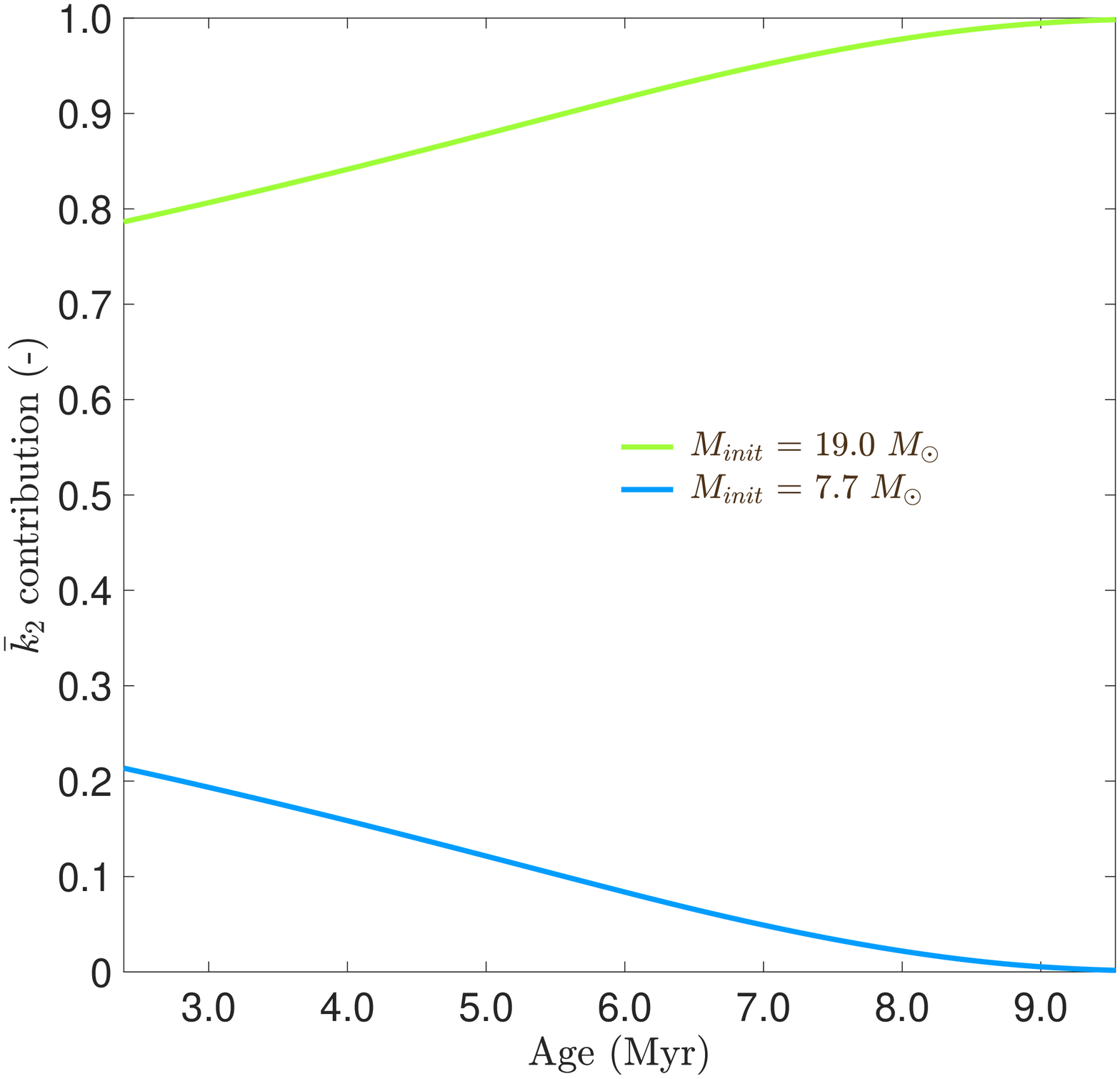}
\caption{Evolution as a function of stellar age of the contributions $\frac{c_1}{c_1+c_2}$ and $\frac{c_2}{c_1+c_2}$ to the weighted-average mean $\bar{k}_2$ of the internal structure constants $k_{2,1}$ and $k_{2,2}$ (to which the empirical correction Eq.\,\eqref{eqn:k2Claret} has been applied) for the models presented in Fig.\,\ref{fig:evolution_k2omega}. \label{fig:k2_contributions}}
\end{figure}

\begin{table*}[h!tb]
\caption{Parameters of some best-fit {\tt Cl\'es} models discussed in Sect.\,\ref{sect:cles}.}
\label{tab:mincles}
\begin{tabular}{l c c c c c c c c c c c c}
\hline\hline
\vspace*{-0.3cm} \\
Model &  Age & $M_{\rm init}$ & $M$ & $R$ & $T_{\rm eff}$ &    $k_{2,\text{un.}}$ & $k_2$    & $10^{-7}$\, $\dot{\text{M}}$  & $\xi$ & $10^6$\,$D_T$ & $\alpha_\text{ov}$ & $\chi^2$ \\
& (Myr) & ($M_{\odot}$) & ($M_{\odot}$)  & ($R_{\odot}$)  & (K)  & ($10^{-3}$)& ($10^{-3}$)  & ($M_{\odot}$\,yr$^{-1}$)&  & (cm$^2$\,s$^{-1}$) & & \\
\vspace*{-0.3cm} \\
\hline
\vspace*{-0.3cm} \\
Model I & $7.30$ & $19.05$ & $18.74$ & $9.39$ & $30\,383$ &  4.3686 &  3.9670 & 0.66  & $1$ & $0.0^\#$ & $0.30$ & 19.11\\
Series II(1) & $7.88$ & $19.00$ & $18.64$ & $9.40$ & $30\,900$ &  3.8505 & 3.4936 & 0.82 &  $1$ & $2.12^*$& $0.20$ & 11.68\\
Series II(2) & $7.86$ & $19.00$ & $18.64$ & $9.40$ & $30\,900$ &  3.8467 & 3.4900 & 0.82 &  $1$ & $1.68^*$& $0.25$ &  11.63\\
Series II(3) & $7.84$ & $19.00$ & $18.64$ & $9.40$ & $30\,900$ &  3.8435 & 3.4871 & 0.82 &  $1$ & $1.29^*$& $0.30$ &  11.59\\
Series II(4) & $7.83$ & $19.00$ & $18.64$ & $9.40$ & $30\,900$ &  3.8405 & 3.4844 & 0.82 &  $1$ & $0.95^*$& $0.35$ &  11.56\\
Series II(5) & $7.81$ & $19.00$ & $18.64$ & $9.40$ & $30\,900$ & 3.8379  & 3.4821 & 0.82 &  $1$ & $0.65^*$& $0.40$ & 11.53\\
Series III(1) & $7.99$ & $18.82$ & $18.64$ & $9.40$ & $30\,900$ &  3.8507 & 3.4936 & 0.41 &  $0.5$ &$2.18^*$& $0.20$ & 11.68\\
Series III(2) & $7.97$ & $18.82$ & $18.64$ & $9.40$ & $30\,900$ &  3.8463 & 3.4896  & 0.41 &  $0.5$ &$1.75^*$& $0.25$ &  11.63\\
Series III(3) & $7.95$ & $18.82$ & $18.64$ & $9.40$ & $30\,900$ &  3.8437 & 3.4873 & 0.41 &  $0.5$ &$1.35^*$& $0.30$ &  11.60\\
Series III(4) & $7.94$ & $18.82$ & $18.64$ & $9.40$ & $30\,900$ &  3.8410 & 3.4849 & 0.41 &  $0.5$ &$1.00^*$& $0.35$ &  11.57\\
Series III(5) & $7.92$ & $18.82$ & $18.64$ & $9.40$ & $30\,900$ &  3.8384 & 3.4825 & 0.41 &  $0.5$ &$0.67^*$& $0.40$ &  11.53\\
Series IV(1) & $8.05$ & $18.73$ & $18.64$ & $9.40$ & $30\,900$ &  3.8511 & 3.4940 & 0.20 &  $0.25$ &$2.22^*$& $0.20$ &  11.69\\
Series IV(2) & $8.03$ & $18.73$ & $18.64$ & $9.40$ & $30\,900$ & 3.8463 & 3.4897  & 0.20 &  $0.25$ &$ 1.78^*$& $0.25$ & 11.63\\
Series IV(3) & $8.01$ & $18.73$ & $18.64$ & $9.40$ & $30\,900$ &  3.8434 & 3.4870 & 0.20 &  $0.25$ &$1.38^*$& $0.30$ & 11.59\\
Series IV(4) & $7.99$ & $18.73$ & $18.64$ & $9.40$ & $30\,900$ &  3.8413 & 3.4852 & 0.20 &  $0.25$ &$1.03^*$& $0.35$ & 11.57\\
Series IV(5) & $7.98$ & $18.73$ & $18.64$ & $9.40$ & $30\,900$ &  3.8385 & 3.4826 & 0.20 &  $0.25$ &$0.73^*$& $0.40$ & 11.54\\
Model V & $9.49$ & $18.91$ & $18.34$ &$9.49$ & $32\,234$ &  2.7753 & 2.5068& 1.44 & $1$ & $9.92^*$ & $0.20$ & 4.85\\
Model VI & $9.57$ & $18.91$ & $18.32$ &$9.51$ & $32\,298$ & 2.7070 & 2.4432 & 1.49 & $1$ & $9.67^*$ & $0.25$ & 4.90\\
Model VII& $9.56$ & $18.94$ & $18.34$ &$9.47$ & $32\,419$ &  2.7089 & 2.4486 & 1.52 & $1$ & $9.37^*$ & $0.30$ & 4.87\\
Model VIII & $9.50$ & $18.91$ & $18.33$ &$9.49$ & $32\,305$ &  2.7179 & 2.4551 & 1.48 & $1$ & $8.20^*$ & $0.35$ & 4.73\\
Model IX & $9.54$ & $18.86$ & $18.28$ &$9.49$ & $32\,266$ &  2.6912 & 2.4298 &1.46 & $1$ & $7.53^*$ & $0.40$ & 4.69\\
\hline 
\vspace*{-0.3cm} \\
Model I$_\text{S}$ &  $10.90$ & $7.70$ & $7.70$ &$3.69$ & $21\,760$ & 8.5115  & 8.2581 & 0.001 & $0.1$ & $0.0^*$ & $0.10$ & 0.01\\  
Model II$_\text{S}$ &  $11.15$ & $7.70$ & $7.70$ &$3.69$ & $21\,789$ & 8.4740  & 8.2217 & 0.001 & $0.1$ & $0.0^*$ & $0.15$ & 0.01\\  
Model III$_\text{S}$ &  $11.40$ & $7.70$ & $7.70$ &$3.69$ & $21\,818$ & 8.4379  & 8.1866 & 0.001 & $0.1$ & $0.0^*$ & $0.20$ & 0.02\\  
Model IV$_\text{S}$ &  $11.66$ & $7.70$ & $7.70$ &$3.69$ & $21\,847$ & 8.4029  & 8.1527 & 0.001 & $0.1$ & $0.0^*$ & $0.25$ & 0.03\\  
Model V$_\text{S}$ &  $11.89$ & $7.70$ & $7.70$ &$3.69$ & $21\,875$ &    8.3694 & 8.1201& 0.001 & $0.1$ & $0.0^*$ & $0.30$ & 0.04\\  
Model VI$_\text{S}$ &  $12.18$ & $7.70$ & $7.70$ &$3.69$ & $21\,903$ & 8.3366  & 8.0883 & 0.001 & $0.1$ & $0.0^*$ & $0.35$ & 0.05\\  
Model VII$_\text{S}$ &  $12.45$ & $7.70$ & $7.70$ &$3.69$ & $21\,930$ & 8.3050  &  8.0576& 0.001 & $0.1$ & $0.0^*$ & $0.40$ & 0.06\\  
\vspace*{-0.3cm} \\
\hline
\end{tabular}
\tablefoot{Columns\,1 and 2 give the name of the model and its current age. Column\,3 lists the initial mass of the corresponding evolutionary sequence. Columns\,4, 5, and 6 give the mass, radius, and effective temperature of the model. Columns\,7 and 8 yield the $k_2$ of the model respectively before and after applying the empirical correction for the effect of rotation of \citet[][Eq.\,\eqref{eqn:k2Claret}]{Claret99}. Column\,9 lists the mass-loss rate of the model. Columns\,10, 11, and 12 give the mass-loss rate scaling factor, the turbulent diffusion, and the overshooting parameter of the model. Column\,13 quotes the $\chi^2$ of the model. \\
Models without and with subscript $\text{S}$ correspond to the models of the primary star and secondary star, respectively.\\For $D_T$, the *-symbol denotes it is a free parameter of the model, while the \#-symbol denotes  it is a fixed parameter of the model. \\All $\chi^2$ for the primary star have been computed based on $M, R, T_\text{eff}$, and $k_{2,1}$, but see the caveats of \citet{andrae10a} and \citet{andrae10b}. All $\chi^2$ for the secondary star have been computed based on $M, R$, and $T_\text{eff}$ only.}
\end{table*}

As an illustration, we built two stellar evolutionary sequences, the first one with initial mass of 19.0\,$M_\odot$ and $\xi =1$ which should represent the primary star, and the second one with initial mass of 7.7\,$M_\odot$ and $\xi=0.1$ which should represent the secondary star. Both sequences assume $\alpha_\text{ov}=0.30$ and no turbulent diffusion. Whilst the stellar parameters of the primary star all cross the observational values, within the error bars, between 7 and 8\,Myr, the stellar parameters of the secondary star cross the observational values, within the error bars, beyond 10\,Myr. This preliminary analysis already shows the difficulty to build consistent stellar evolutionary models for both stars but also coherent in terms of age. Yet, the importance of these two evolutionary sequences rather lies in Fig.\,\ref{fig:evolution_k2omega}: The left panel shows the evolution as a function of the age of the internal structure constants of both evolutionary sequences together with the weighted-average mean of the $k_2$, while the right panel shows the evolution as a function of the age of the apsidal motion contributions of both stars taken separately as well as the total apsidal motion of the binary system. 
This second panel clearly shows that the primary star contributes for the most part to the binary apsidal motion, as expected. The evolution of $\bar{k}_2$ and $k_{2,1}$ overlap nearly perfectly, suggesting that $k_{2,1}$ dominates in $\bar{k}_2$. To quantify this assertion, we show in Fig.\,\ref{fig:k2_contributions} the evolution as a function of the age of the coefficients $\frac{c_1}{c_1+c_2}$ and $\frac{c_2}{c_1+c_2}$ appearing in Eq.\,\eqref{eqn:k2bar} for these two evolutionary sequences. The contribution of $k_{2,1}$ to $\bar{k}_2$ increases with time and, assuming an age older than 5\,Myr, amounts to minimum 90\%. These arguments, together with the fact that the stellar parameters of the primary star are better constrained than those of the secondary star, motivate us to first build stellar evolutions models for the primary star.

\subsection{Primary star\label{subsect:clesprim}}
We built stellar evolution models for the primary star assuming the three constraints $M, R$, and $T_\text{eff}$, and adopting at least the initial mass of the star and its age as free parameters. The first model (Model I) is built fixing $\xi=1$ and $\alpha_\text{ov}=0.30$, and assuming no turbulent diffusion. The only free parameters are thus the initial mass and the current age of the star. This model gives $M_\text{init}=19.05\,M_\odot$ and an age of 7.30\,Myr, but has a slightly smaller effective temperature than observed, even though still within the errorbars (see Table\,\ref{tab:mincles}). We note that the $\chi^2$ is computed based on $M, R, T_\text{eff}$, and $k_{2,1}$. Adopting $\alpha_\text{ov}=0.0$ gives even worse results in terms of $M, R$, and $T_\text{eff}$. Without any mixing, the global parameters of the star cannot be reproduced accurately.

\begin{figure}
\centering
\includegraphics[clip=true,trim=90 60 110 30,width=1\linewidth]{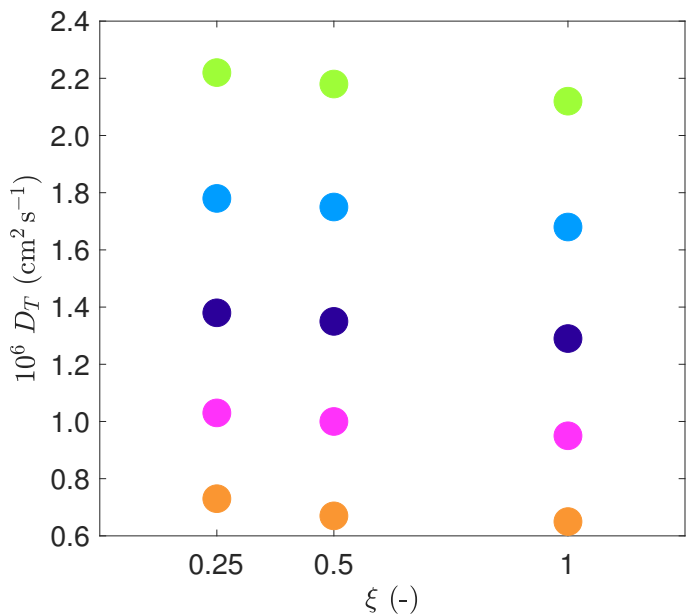}
\caption{Degeneracy between the various processes in the stellar interior for the best-fit \texttt{min-Cl\'es} models: turbulent diffusion coefficient $D_T$ as a function of the mass-loss rate scaling parameter $\xi$ for different values of the overshooting parameter $\alpha_\text{ov}$. The colours stand for $\alpha_\text{ov}=0.20$ (green), 0.25 (blue), 0.30 (purple), 0.35 (pink), and 0.40 (orange). \label{fig:alphaovxiDt}}
\end{figure}

Hence, we next investigated the influence of $\xi, \alpha_\text{ov}$, and $D_T$ on the best-fit models\footnote{The \citet{vink01} recipe has been found to often overpredict the mass-loss rate of O-type stars \citep[and references therein]{sundqvist19}. For the sake of completeness, we decided to investigate the influence of this parameter on the best-fit models.}. In three series of models, Series II to IV, we fixed $\xi$ to 1, 0.5, and 0.25, respectively. For each Series, we built five models with different values of $\alpha_\text{ov}$, namely 0.20, 0.25, 0.30, 0.35, and 0.40, while leaving $D_T$ as an additional free parameter. The results are summarised in Table\,\ref{tab:mincles}. Series II, III, and IV models have an initial mass of 19.00, 18.82, and $18.73\,M_\odot$, respectively, and an age between 7.81 and 7.88\,Myr, 7.92 and 7.99\,Myr, and 7.98 and 8.05\,Myr, respectively. The age is lower for larger values of $\alpha_\text{ov}$ and $\xi$. All these 15 models correctly reproduce the three constraints $M, R$, and $T_\text{eff}$, meaning that models of equal quality, as far as the adjustment of the three stellar parameters is concerned, can be obtained for different triplets of $D_T, \xi$, and $\alpha_\text{ov}$. This $D_T - \xi-\alpha_\text{ov}$ degeneracy (already highlighted for HD\,152248 in \citet{rosu20b}) is illustrated in Fig.\,\ref{fig:alphaovxiDt}. First, for a given value of $\xi$, the higher the value of $\alpha_\text{ov}$, the lower the value of $D_T$. These two parameters affect the stellar structure and evolution the same way. They increase the mixing inside the stars, hence increasing the amount of hydrogen transported to the core. As a consequence, if one of these two parameters is increased, the other one decreases accordingly, and vice-versa. Second, for a given value of $\alpha_\text{ov}$, the higher the value of $\xi$, the lower the value of $D_T$. This is expected as a higher value of $\xi$ implies a higher mass-loss rate and thus a higher initial mass. When the star reaches the state where its mass equals the observational mass, its convective core is larger than for a star of lower $\xi$. The larger convective core in turn implies stronger mixing, thereby reducing the need for turbulent mixing. 

The biggest differences between Series II, III, and IV lie in the different initial masses and ages of the models. For the reasons aforementioned, the higher the $\xi$ value, the higher the initial mass of the stars and, hence, the lower its age when the star reaches the state defined by the observational parameters. 
Finally, all these models have a $k_2$ value $\sim$\,2 times higher than the $\bar{k}_2$ value. There is no significant difference in the $k_2$ value for models having the same $\alpha_\text{ov}$ but a different $\xi$. However, for the models having the same $\xi$, $k_2$ decreases when $\alpha_\text{ov}$ increases notably due to a more pronounced density variation at a radius close to the junction between the convective core and the radiative envelope in the overshooting region. 

\begin{figure}[h]
\centering
\includegraphics[clip=true,trim=70 70 300 80,width=\linewidth]{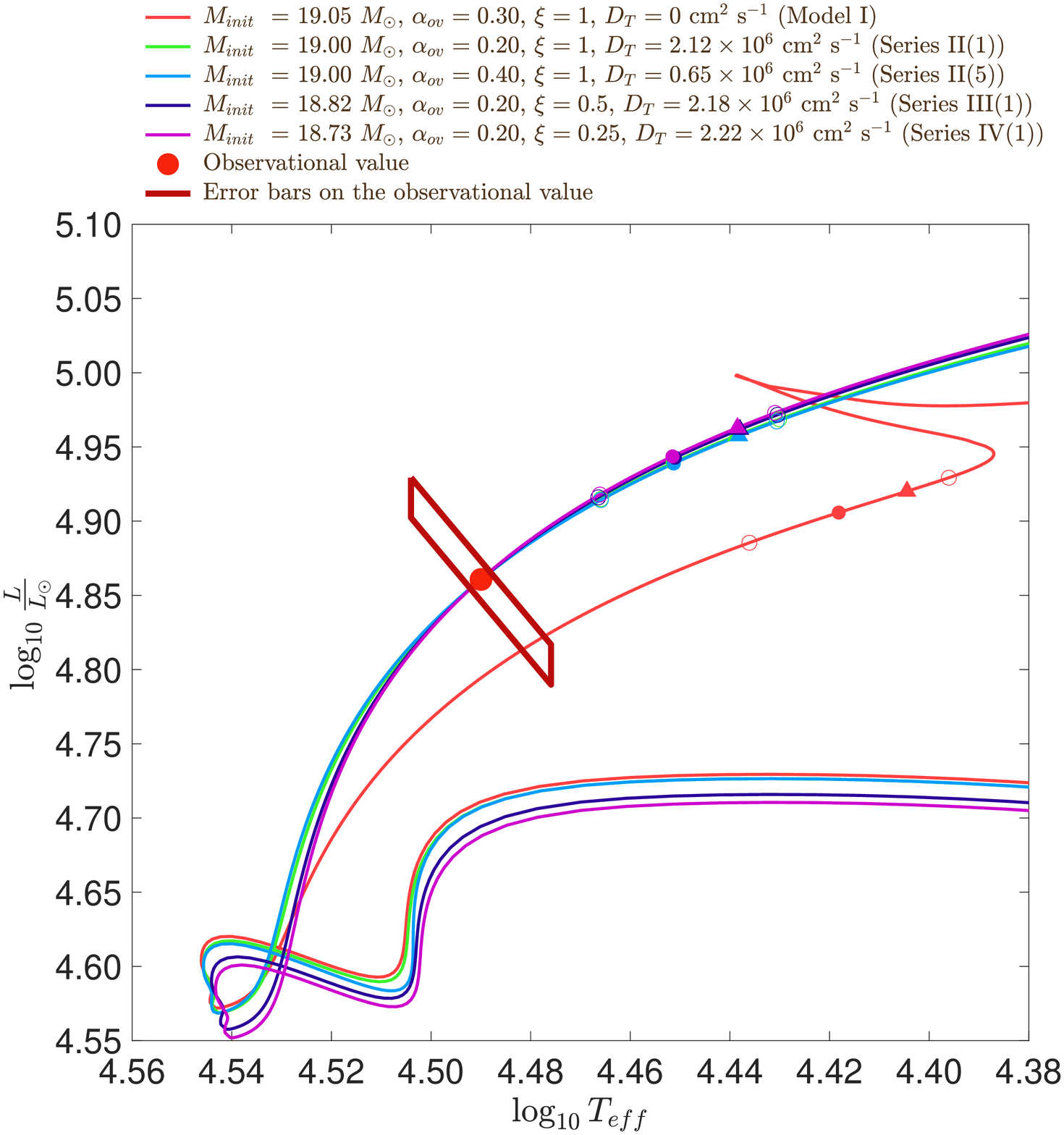}
\caption{Hertzsprung-Russell diagram: evolutionary tracks of \texttt{Cl\'es} models corresponding to the following best-fit models: Model I (coral), Series II(1) (green), Series II(5) (light blue),  Series III(1) (dark blue), and Series IV(1) (plum). The dots and triangles over-plotted on the corresponding tracks correspond to the models for which $k_{2,1}$ is equal to $\bar{k}_2$ and 0.00139, respectively, while the open circles correspond to the models for which $k_{2,1}$ is equal to $\bar{k}_2\pm \Delta \bar{k}_2$. The observational value is represented by the red point, and its error bars are represented by the dark red parallelogram. \label{fig:HRdiag1}}
\end{figure}

The degeneracy between the best-fit models is illustrated in the Hertzsprung-Russell diagram in Fig.\,\ref{fig:HRdiag1} where four evolutionary sequences are presented, together with Model I for comparison and the observational box defined by the observational radius and effective temperature and their respective errors. The evolutionary sequences are built based on the initial mass, $\xi$, $\alpha_\text{ov}$, and $D_T$ adopted from the Series II(1), II(5), III(1), and IV(1). The sequences corresponding to Series II(1) and II(5) are very similar, while the sequences corresponding to Series III(1) and IV(1) follow slightly different tracks. Nonetheless, all four sequences cross the observational box at the same point and follow very close paths near this point. The five dots over-plotted on each track represent the model having $k_{2,1} = \bar{k}_2$, which is, as already stated in Sect.\,\ref{sect:k2}, the maximum value $k_{2,1}$ can take. These five models lie well beyond the observational box, confirming that the best-fit models are too homogeneous. The filled triangles over-plotted on these tracks represent the models having  $k_{2,1}= 0 .00139$ (see Sect.\,\ref{subsect:clessecondary} for the justification of this value). As expected, these models lie even further away from the observational box than those with $k_{2,1}=\bar{k}_2$.

All Series predict mass-loss rates lower than observed. Since the observational $\dot{M}$ is poorly constrained, one cannot totally rule out the possibility of a $\xi$-value different from 1. Yet, without any additional information, we decided, from now on, to restrict our analysis to models having $\xi=1$.

We conclude that in order to find a model which reproduces $M, R$, and $T_\text{eff}$ within their error bars, and which also satisfies the constraint that $k_{2,1} < \bar{k}_2$ within the error bars of $\bar{k}_2$, the value of $D_T$ should be increased for a given value of $\alpha_\text{ov}$. We computed five models (Models V to IX) with the value of $\alpha_\text{ov}$ fixed to 0.20, 0.25, 0.30, 0.35, and 0.40, respectively, leaving the age, initial mass, and $D_T$ as free parameters. Compared to previously computed models, one additional constraint is enforced: $k_{2,1} = \bar{k}_2$ within the error bars. The resulting models are reported in Table\,\ref{tab:mincles}.

The mass and radius of all these models are comprised within the error bars on these parameters. However, none of these models has an effective temperature and $k_2$-value compatible with observations: both parameters are overestimated. 
This highlights the difficulty to reconcile the observational and theoretical $k_2$-values, especially since, as already discussed in Sect.\,\ref{sect:k2}, $\bar{k}_2$ is only the upper-bound value on $k_{2,1}$. Requesting $k_{2,1}$ to be much lower than $\bar{k}_2$ would lead to an even more pronounced discrepancy.

\begin{figure}[h]
\centering
\includegraphics[clip=true,trim=70 70 300 120,width=\linewidth]{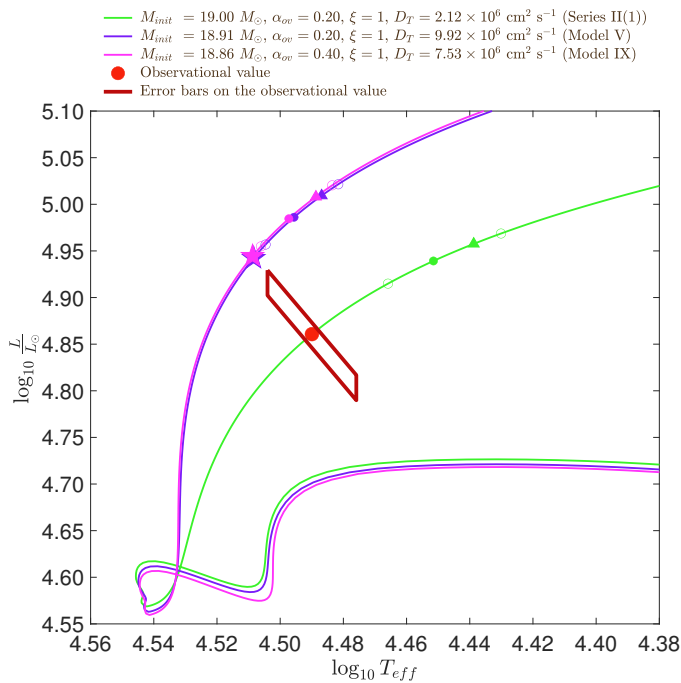}
\caption{Hertzsprung-Russell diagram: evolutionary tracks of \texttt{Cl\'es} models corresponding to the following best-fit models: Series II(1) (green), Model V (purple), and Model IX (pink). The dots and triangles over-plotted on the corresponding tracks correspond to the models for which $k_{2,1}$ is equal to $\bar{k}_2$ and 0.00139, respectively, while the open circles correspond to the models for which $k_{2,1}$ is equal to $\bar{k}_2\pm \Delta \bar{k}_2$. The stars on the purple and pink tracks correspond to the best-fit models Model V and Model IX, respectively. The observational value is represented by the red point, and its error bars are represented by the dark red parallelogram. \label{fig:HRdiag2}}
\end{figure}

Even though the mass and radius can be correctly reproduced within the error bars, it seems impossible to find a model reproducing both the effective temperature and the internal structure constant simultaneously. This issue is highlighted in the Hertzsprung-Russell diagram in Fig.\,\ref{fig:HRdiag2}: three evolutionary sequences are presented, together with the observational box. The evolutionary sequences are built based on the initial mass, $\xi$, $\alpha_\text{ov}$, and $D_T$ adopted from the Series II(1), Model V, and Model IX. As expected, the tracks corresponding to Models V and IX do not cross the observational box. The three dots over-plotted on each track represent the model having $k_{2,1} = \bar{k}_2$, while the filled triangles over-plotted on these tracks represent the models having $k_{2,1}= 0 .00139$ (see Sect.\,\ref{subsect:clessecondary} for the justification of this value). These six models lie well beyond the observational box, confirming that the best-fit models still have a too small density contrast between the core and the external layers. The two stars over-plotted on the tracks of Models V and IX correspond to the best-fit models given in Table\,\ref{tab:mincles} (note that the one corresponding to Series II(1) perfectly matches the observational value).

\subsection{Secondary star\label{subsect:clessecondary}} 
We built stellar evolution models for the secondary star assuming the three constraints $M, R,$ and $T_\text{eff}$ and adopting the initial mass, the age and the turbulent diffusion as free parameters. We built seven models (Models I$_\text{S}$ to VII$_\text{S}$) fixing $\xi=0.1$ and adopting values for $\alpha_\text{ov}$ of 0.10, 0.15, 0.20, 0.25, 0.30, 0.35, and 0.40.  The resulting best-fit models are summarised in Table\,\ref{tab:mincles}. 
All models reproduce perfectly the mass and radius, but slightly overestimate the effective temperature even though still within the error bars. All these models have an initial mass of $7.70$\,$M_\odot$ and an age ranging between 10.9 and 12.5\,Myr. We observe that $D_T$ converges towards a value of 0\,cm$^{2}$\,s$^{-1}$ for all models. Assuming Models IX and I$_\text{S}$ correctly reproduce the primary and secondary star, respectively, we get an apsidal motion rate of $1.8793^\circ\,\text{yr}^{-1}$ for the binary system. This value is 57\% higher than the observational value and is therefore incompatible with the observational results. The second problem here lies in the fact that the age of these models is incompatible with the range of ages derived for the primary star in Sect.\,\ref{subsect:clesprim}. The internal structure constant of these models, corrected for the rotation of the star through Eq.\,\eqref{eqn:k2Claret} amounts to $\sim$\,$8\times 10^{-3}$. Assuming these models correctly reproduce the secondary star, then $k_{2,2} \sim$\,$4.7\,\bar{k}_2$ and, making use of Eq.\,\eqref{eqn:k2bar}, we deduce that $k_{2,1}$ should take the value of 0.00139. If it appeared that an enhanced mixing would be necessary to model the secondary star, then on the one hand $k_{2,2}$ would be lower and $k_{2,1}$ higher but, on the other hand, the age of the secondary star would be higher, thus reenforcing the age discrepancy between primary and secondary stars.

\subsection{Towards a common age of the stars?\label{subsect:HR}}
In order to examine the age discrepancy, we built series of evolutionary tracks with \texttt{Cl\'es} for the two stars. We assumed $\alpha_\text{ov}= 0.30$ for both stars. For the primary star, we assumed $\xi=1$ and built 20 evolutionary tracks with initial mass ranging from 17 to 21\,$M_\odot$ with a step of 1\,$M_\odot$ and a $D_T$ value of 0, $2\times 10^6$, $6\times 10^6$, and $10\times 10^6$\,cm$^2$\,s$^{-1}$. For the secondary star, we assumed $\xi=0.1$ and built 9 evolutionary tracks with initial mass 7.5, 7.7, and 8.0\,$M_\odot$, and a $D_T$ value of 0, $2\times 10^5$, and $5\times 10^5$\,cm$^2$\,s$^{-1}$.\\

\begin{figure}[h]
\centering
\includegraphics[clip=true,trim=90 70 300 220,width=\linewidth]{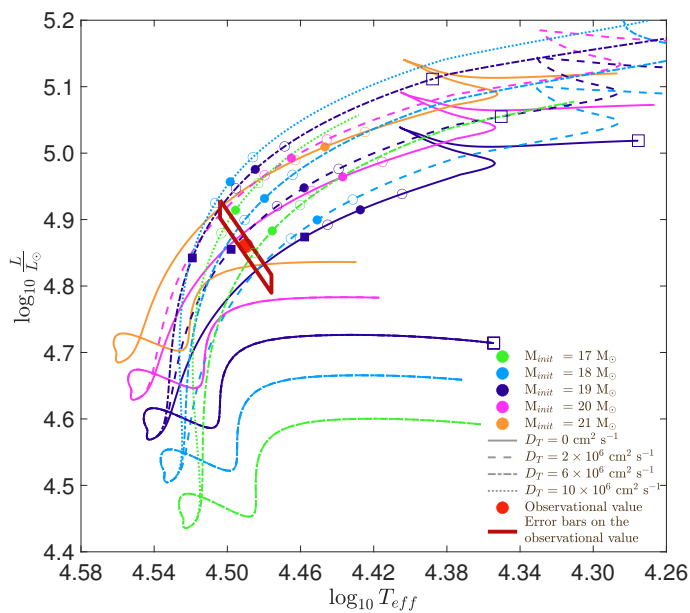}
\caption{Hertzsprung-Russell diagram: evolutionary tracks of \texttt{Cl\'es} models for the primary star of $M_{\rm init}= 17$\,$M_\odot$ (green), 18\,$M_\odot$ (light blue), 19\,$M_\odot$ (dark blue), 20\,$M_\odot$ (pink), and 21\,$M_\odot$ (orange), and $D_T=0$\,cm$^2$\,s$^{-1}$ (solid line), $2\times 10^6$\,cm$^2$\,s$^{-1}$ (dashed line), $6\times 10^6$\,cm$^2$\,s$^{-1}$ (dotted-dashed line), and $10\times 10^6$\,cm$^2$\,s$^{-1}$ (dotted line). All models have $\alpha_\text{ov}=0.30$ and $\xi=1$. The dots over-plotted on the corresponding tracks correspond to the models for which $k_{2,1}$ is equal to $\bar{k}_2$, while the open circles correspond to the models for which $k_{2,1}$ is equal to $\bar{k}_2\pm \Delta \bar{k}_2$. The filled and open squares over-plotted on the tracks having an initial mass of 19\,$M_\odot$ correspond to the models for which the mass equals the observational value within the error bars. The observational value is represented by the red point, and its error bars are represented by the dark red parallelogram. \label{fig:HR_all}}
\end{figure}

Evolutionary tracks for the primary star are presented in the Hertzsprung-Russell diagram in Fig.\,\ref{fig:HR_all}. In this figure, only the evolutionary tracks crossing (or nearly crossing) the observational box are presented. The models with initial mass of 17 and 18\,$M_\odot$ have a too low mass compared to observational, even within the error bars and therefore do not fulfil the constraint on the mass. On the contrary, the models with initial mass of 20 and 21\,$M_\odot$ have too high a value of the mass compared to observational and, at best, fulfil the constraint on the mass at the very end of the main-sequence phase. Only the evolutionary tracks having an initial mass of 19\,$M_\odot$ have a mass compatible with the observation during the major part of the main-sequence phase -- or at least when crossing (or nearly crossing) the observational box. During the main-sequence phase, the turbulent diffusion impacts the evolutionary tracks as follows: the higher the turbulent diffusion in absolute value, the higher the bolometric luminosity of the star for a given effective temperature. Hence, for low-mass models (i.e.\ models with initial mass of 17 and 18\,$M_\odot$), only models with enhanced turbulent mixing cross the observational box, and only those having a $D_T$-value of $6\times 10^6$ and $10\times 10^6$\,cm$^2$\,s$^{-1}$ have a $k_2$-value (almost) compatible with $\bar{k}_2$ within the error bars when crossing the observational box. On the opposite, for the high-mass models (i.e. models with initial mass of 20 and 21\,$M_\odot$), only the models without or with a small $D_T$-value cross the observational box, but none of them has a $k_2$-value compatible with $\bar{k}_2$ within the error bars when crossing the observational box. Regarding the models with an initial mass of 19\,$M_\odot$, only those with $D_T$ equal to 0, $2\times 10^6$, and $6\times 10^6$\,cm$^2$\,s$^{-1}$ cross the observational box, but none of them has a $k_2$-value compatible with $\bar{k}_2$, even though the track with $D_T = 6\times 10^6$\,cm$^2$\,s$^{-1}$ has a $k_2$-value very close to this value. These evolutionary sequences highlight the issue encountered in Sect.\,\ref{subsect:clesprim} about the difficulty to reconcile the mass, radius, effective temperature, and internal structure constant of the star simultaneously. 

\begin{figure}[h]
\centering
\includegraphics[clip=true,trim=90 70 300 220,width=\linewidth]{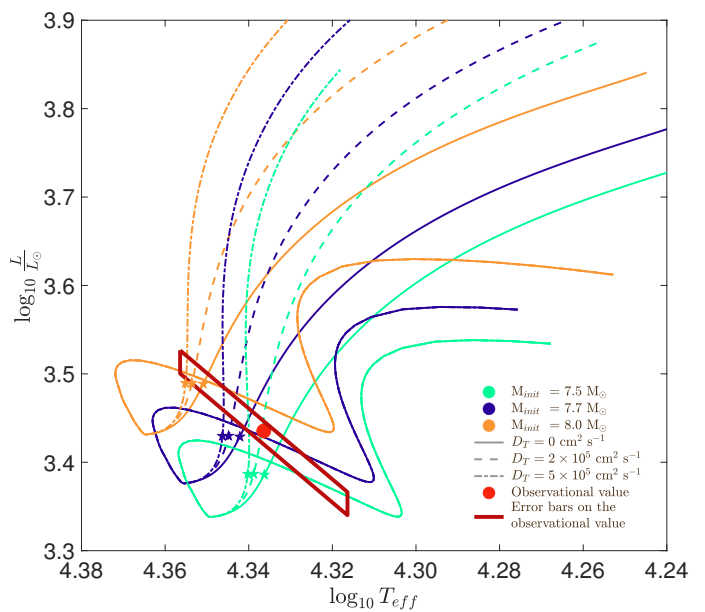}
\caption{Hertzsprung-Russell diagram: evolutionary tracks of \texttt{Cl\'es} models for the secondary star of $M_{\rm init}= 7.5$\,$M_\odot$ (water green), 7.7\,$M_\odot$ (dark blue), and 8.0\,$M_\odot$ (orange), and $D_T=0$\,cm$^2$\,s$^{-1}$ (solid line), $2\times 10^5$\,cm$^2$\,s$^{-1}$ (dashed line), and $5\times 10^5$\,cm$^2$\,s$^{-1}$ (dotted-dashed line). All models have $\alpha_\text{ov}=0.30$ and $\xi=0.1$. The filled 5-branch stars over-plotted on the tracks correspond to the models for which the age is equal to the age of the best-fit models of the primary star (i.e.\ about 9\,Myr). The observational value is represented by the red point, and its error bars are represented by the dark red parallelogram. \label{fig:HR_sec}}
\end{figure}

Evolutionary tracks for the secondary star are presented in the Hertzsprung-Russell diagram in Fig.\,\ref{fig:HR_sec}. All these tracks cross the observational box defined by the radius and effective temperature at some point of their evolution. However, only the models with an initial mass of 7.7\,$M_\odot$ have a mass compatible with the observation when crossing the observational box. Given that the mass-loss rate is negligible, only evolutionary tracks having an initial mass ranging between 7.58 and 7.82\,$M_\odot$ are acceptable. Regarding the evolutionary track of initial mass 7.7\,$M_\odot$ and $D_T = 0$\,cm$^2$\,s$^{-1}$, the model closest to the observational value has a $k_{2,2}$ value approximately equal to 4.7\,$\bar{k}_2$, confirming the value we found in Sect.\,\ref{subsect:clessecondary}. Within the observational box, the age of the star ranges from 10 to 14\,Myr, above the age of the best-fit models of the primary star of approximately 9\,Myr. The filled 5-branch stars over-plotted on the tracks correspond to the models having an age of 9\,Myr. These models are slightly below the observational box meaning that their stellar properties are not compatible with the observational values.  

The influence of a potential non-solar metallicity and helium abundance is investigated in Appendix\,\ref{subsect:XZ}. Besides being highly unlikely, we show that a non-solar metallicity and helium abundance does not solve the discrepancies discussed hereabove.

We inevitably come to the conclusion that we cannot find stellar evolution models for the two stars reproducing the stellar properties properly and having a compatible age at the same time. Whilst stellar evolution models have their intrinsic limitations, notably regarding the way the internal mixing is implemented, we might also have been biased by the observational data. There is no obvious issue arising from the spectroscopic and radial velocity analysis and for example the primary effective temperature and the mass ratio that we have derived should be robust. However, the situation is more complicated for photometry. The zero contribution of the third light to the photometric data of the binary system might have been underestimated. The Roche lobe filling factors of the two stars and the effective temperature of the secondary star have been determined based on the assumed value of the third light contribution, these parameters, and consequently also the radii of both stars, are not as reliable as their error bars might suggest. Yet, by taking $I_3=0$, we have actually the smallest possible radius of the secondary star, and larger radii would lead to even more severe age discrepancies than what we have found here.

\section{Discussion: what if the rotation axis were misaligned?\label{sect:discussion}} 
The highlighted difficulty to reconcile observational and theoretical internal structure constant values suggests there might be some bias in the interpretation of the apsidal motion in terms of the internal structure constant. One possibility could be that higher-order terms in the expression of the apsidal motion rate have been neglected. Yet, these higher-order terms have been shown to have a negligible contribution \citep[][and reference therein]{rosu20b}. Another scenario could arise from the presence of a third body orbiting the binary system. However, to date, there is no signature of a third body in the spectrum of HD\,152219 (see review in Sect.\,\ref{sect:introduction} as well as Sects.\,\ref{sect:specanalysis} and \ref{sect:photom}). The possibility that a ternary star would contribute to the apsidal motion rate without being detected though is a priori conceivable (see e.g. \citet{naoz13} and \citet{borkovits15} for a review of the ensuing apsidal motion equations).  At this stage, there is unfortunately no way to infirm or confirm the presence of a third body in the system. Though, it is highly unlikely that a non-detected third body would contribute for as much as 57\% of the inner binary apsidal motion rate necessary to reconcile theoretical and observational $k_2$-values (see Sect.\,\ref{subsect:clessecondary}).

A third possibility could be a misalignment of the rotation axes of the stars with respect to the normal to the orbital plane. Indeed, the Newtonian contribution to the total apsidal motion rate given by Eq.\,\eqref{eqn:omegadotN} assumes the rotation axes of the stars to be parallel to the normal to the orbital plane. In case of a misalignment, the Newtonian contribution takes the general expression given by Eq.\,(3) of \citet{shakura}: 
\begin{equation}
\label{eqn:omegadotNmisaligned}
\begin{aligned}
\dot\omega_\text{N} = \frac{2\pi}{P_\text{orb}} \Bigg[&15f(e)\left\{k_{2,1}q \left(\frac{R_1}{a}\right)^5 + \frac{k_{2,2}}{q} \left(\frac{R_2}{a}\right)^5\right\} \\
& \begin{aligned} - \frac{g(e)}{(\sin i)^2} \Bigg\{ &k_{2,1} (1+q) \left(\frac{R_1}{a}\right)^5 \left(\frac{P_\text{orb}}{P_\text{rot,1}}\right)^2 \\ 
& \bigg[\cos\alpha_1\left(\cos\alpha_1-\cos\beta_1\cos i\right) \\
& +\frac{1}{2}(\sin i)^2\left(1-5(\cos\alpha_1)^2\right)\bigg] \\
& + k_{2,2}\frac{1+q}{q} \left(\frac{R_2}{a}\right)^5 \left(\frac{P_\text{orb}}{P_\text{rot,2}}\right)^2 \\
& \bigg[\cos\alpha_2\left(\cos\alpha_2-\cos\beta_2\cos i\right) \\
& +\frac{1}{2}(\sin i)^2\left(1-5(\cos\alpha_2)^2\right)\bigg] \Bigg\}  \Bigg].
\end{aligned}
\end{aligned}
\end{equation}
In this expression, $\alpha_1$ (resp. $\alpha_2$) stands for the angle between the primary (resp. secondary) star rotation axis and the normal to the orbital plane, and $\beta_1$ (resp. $\beta_2$) stands for the angle between the primary (resp. secondary) star rotation axis and the line joining the binary centre and the observer. 

We define the function $F_{\alpha,j}$ (with $j=1, 2$ for the primary and secondary star, respectively) as 
\begin{equation}
\label{eq:falpha}
F_{\alpha,j} = \cos\alpha_j\left(\cos\alpha_j-\cos\beta_j\cos i\right)  +\frac{1}{2}\sin^2 i\left(1-5\cos^2\alpha_j\right).
\end{equation}
We can express $\cos\beta_j$ in terms of the inclination and $\alpha_j$ using the cosine relationship in spherical trigonometry:
\begin{equation}
\label{eqn:beta}
\cos\beta_j = \cos i \cos\alpha_j + \sin i \sin\alpha_j\cos\theta,
\end{equation}
where $\theta$ is the azimutal angle of the rotation axes of the stars, such that $\theta=0^\circ$ and $180^\circ$ correspond to the cases where the rotation axes, the line of sight, and the normal to the orbital plane lie in the same plane. Replacing in Eq.\,\eqref{eq:falpha} $\cos\beta_j$ by its expression given by Eq.\,\eqref{eqn:beta} gives us an expression of $F_{\alpha,j}$ which depends upon the inclination, the $\alpha_j$, and $\theta$ only. We can further show that $F_{\alpha,j}$ takes its extremum values for $\theta =0^\circ$ and $180^\circ$. Assuming that $\theta = 180^\circ$, we get $\beta_j = i+\alpha_j$.

\begin{figure}[h]
\centering
\includegraphics[clip=true,trim=100 60 300 200,width=\linewidth]{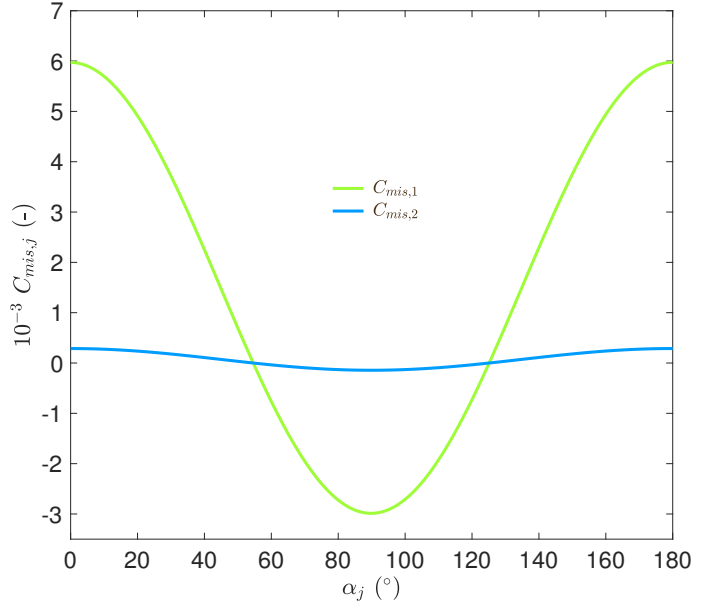}
\caption{Behaviour of $C_{\text{mis},1}$ and $C_{\text{mis},2}$ as a function of $\alpha_1$ and $\alpha_2$, the angle between the primary and secondary stellar rotation axis, respectively, and the normal to the orbital plane.   \label{fig:Cmis}}
\end{figure}

We define the quantities
\begin{equation}
C_{\text{mis},j} = -\frac{1}{(\sin i)^2} \left(1+\frac{M_j}{M_{3-j}}\right)\left(\frac{R_j}{a}\right)^5\left(\frac{P_\text{orb}}{P_{\text{rot},j}}\right)^2 F_{\alpha,j},
\end{equation}
and represent these as a function of the angle $\alpha_j$ in Fig.\,\ref{fig:Cmis} for the specific case of $i=89.58^\circ$. $C_{\text{mis},j}$ takes its minimum and maximum values at $\alpha_j=90^\circ$ and $180^\circ$, respectively.

\begin{figure}[h]
\centering
\includegraphics[clip=true,trim=80 70 300 210,width=\linewidth]{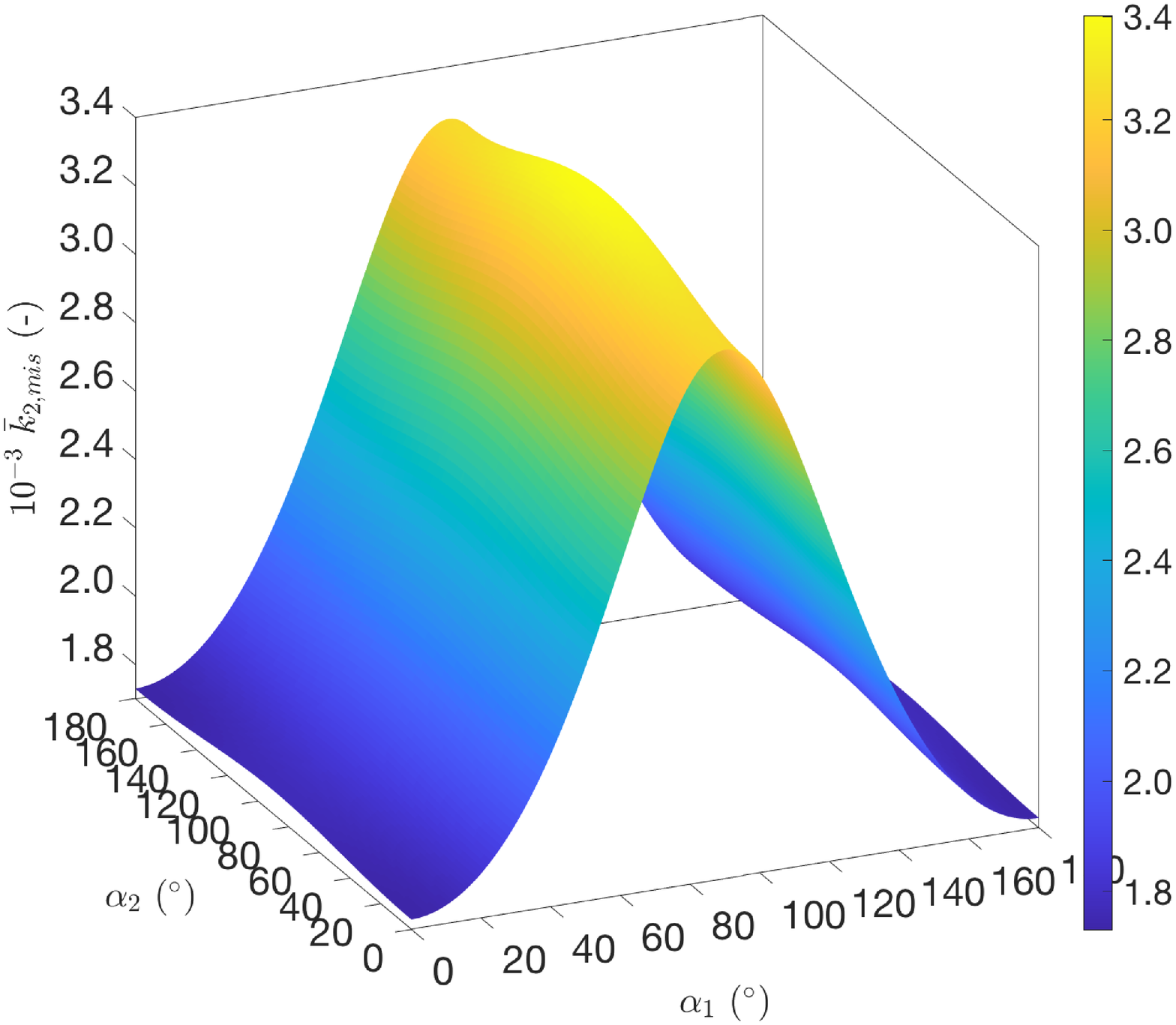}
\includegraphics[clip=true,trim=80 70 300 210,width=\linewidth]{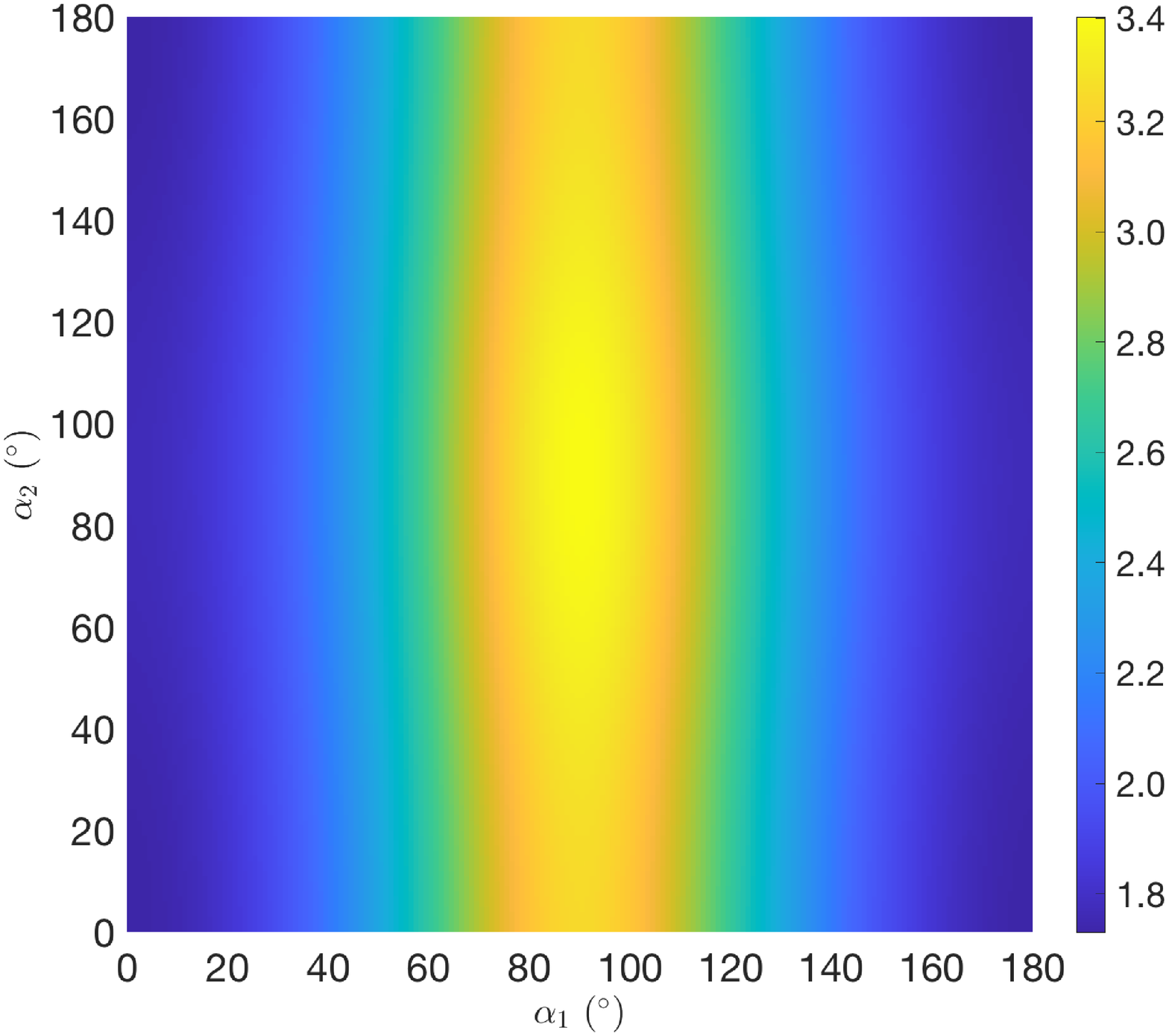}
\caption{Behaviour of $\bar{k}_{2,\text{mis}}$ as a function of $\alpha_1$ and $\alpha_2$, the angle between the primary and secondary stellar rotation axis, respectively, and the normal to the orbital plane. \label{fig:k2mis}}
\end{figure}

We compute $\bar{k}_{2,\text{mis}}$ considering values of $\alpha_j$ ranging from 0 to 180$^\circ$. Figure\,\ref{fig:k2mis} shows the $\bar{k}_{2,\text{mis}}$ value as a function of the two angles $\alpha_1$ and $\alpha_2$. The maximum value of $\bar{k}_{2,\text{mis}}$ is $3.39\times 10^{-3}$ and is reached for $\alpha_1=\alpha_2=90^\circ$. As expected, the $\bar{k}_{2,\text{mis}}$-value mostly change with the primary stellar rotation axis inclination $\alpha_1$ and is only slightly affected by a change in the secondary stellar rotation axis inclination. This reflects the predominant contribution of the primary star to the total rate of apsidal motion. For $\alpha_1$ lower than about 40$^\circ$, there is no significant impact on the $\bar{k}_{2,\text{mis}}$-value.

In Sect.\,\ref{subsect:clesprim}, the best-fit models we obtained for the primary star all had a $k_{2,1}$-value between $\sim$\,2.4 and $2.5\times 10^{-3}$. For these values to be compatible with the observational value, the primary stellar rotation axis should be inclined by an angle of at least $\sim$\,$50^\circ$ with respect to the normal to the orbital plane. Such a high misalignment angle would be surprising in a close binary system, especially since the two stars are about to reach pseudo-synchronisation.

An important question is whether the condition of sub-critical stellar rotation leads to a restriction on the values of $\alpha_1$ and $\alpha_2$. The condition of sub-critical rotation velocity expresses the fact that the stellar rotation rate $\Omega_\text{rot}$ cannot exceed the critical value $\Omega_\text{crit}$ corresponding to the centrifugal force at the stellar equator exactly compensating the effective gravity (i.e.\ the gravity corrected for the effect of radiation pressure). This condition writes 
\begin{equation}
\Omega_{\rm rot} \leq \Omega_{\rm crit} = \sqrt{\frac{G\,M_*}{R_*^3}\,(1 - \Gamma_e)},
\label{eqomega_c}
\end{equation}
where $\Gamma_e$ is the Eddington factor:
\begin{equation}
\Gamma_e = \frac{\kappa_e\,L_*}{4\,\pi\,c\,G\,M_*},
\end{equation}
$\kappa_e$ being the electron scattering opacity. For this parameter, we adopt the value $\kappa_e \simeq 0.34$\,cm$^2$\,g$^{-1}$ \citep[e.g.][]{Sanyal}. Hence, we get $\Gamma_e = 2.6\,10^{-5}\,\frac{L_*\,M_{\odot}}{L_{\odot}\,M_*}$. In terms of the equatorial rotational velocity, this translates into
\begin{equation}
v_{\rm eq} \leq 436.7\,\sqrt{\frac{M_*\,R_{\odot}}{M_{\odot}\,R_*}\,\left(1 - 2.6\,10^{-5}\,\frac{L_*\,M_{\odot}}{L_{\odot}\,M_*}\right)}\,{\rm km\,s}^{-1}.
\end{equation}
For the primary and secondary stars, we get $v_{\text{eq},1} \le 583$\,km\,s$^{-1}$ and $v_{\text{eq},2} \le 628$\,km\,s$^{-1}$, respectively. Using the projected rotational velocities derived in Sect.\,\ref{subsect:vsini}, namely $v_{\text{eq},1} \sin\beta_1 = 166$\,km\,s$^{-1}$ and $v_{\text{eq},2} \sin\beta_2 = 95$\,km\,s$^{-1}$, we get the following constraints on the angles $\beta_j$: $\beta_1 \ge 16.5^\circ$ and $\beta_2 \ge 8.7^\circ$. In order to translate this condition on $\beta_j$ into a condition on $\alpha_j$, we need to assume a value for the azimuthal angle $\theta$. As shown before, the largest impact of $\alpha_j$ on the apsidal motion rate occurs when $\theta$ equals 0$^\circ$ and $180^\circ$. For an azimuthal angle of $0^\circ$, we get the conditions that $\alpha_1> 106.1^\circ$ or $\alpha_1 < 73.0^\circ$, and $\alpha_2> 98.3^\circ$ or $\alpha_2 < 80.9^\circ$, whilst for an azimuthal angle of $180^\circ$, we get the conditions that $\alpha_1> 107.0^\circ$ or $\alpha_1 < 73.9^\circ$, and $\alpha_2> 99.1^\circ$ or $\alpha_2 < 81.7^\circ$. These conditions do not rule out the possibility to have inclination angles of $\sim$\,50$^\circ$ for the stellar rotation axis of the stars. \\

The possibility of a misalignment between the primary stellar rotation axis and the axis of the binary system can in principle be tested via the Rossiter-McLaughlin effect \citep[RM, e.g.][]{hosokawa53, ohta05, gimenez06, albrecht07} during the eclipse of the primary star by the secondary. As the secondary passes in front of the primary, assuming a prograde orbit, it first eclipses those parts of the rotating primary that produce the blue wings of the spectral lines, thereby shifting the line centroids to the red. Likewise, near the end of the eclipse, the secondary occults those parts of the primary that produce the red wings, resulting in a shift of the line centroids to the blue. The signature of a RM effect in HD\,152219 was reported by \citet{San09} from measurements of the He\,{\sc ii} $\lambda$~4686 line which, in this system, is only formed in the spectrum of the primary star.

We have designed a \texttt{Fortran} code to simulate the RM effect for different values of the $\alpha_1$ and $\beta_1$ angles, adopting the photometric value of the orbital inclination and the primary's $v_{\rm eq,1}\,\sin{\beta_1}$. Synthetic line profiles were computed adopting the quadratic limb darkening law of \citet{claret00} and assuming the stars to have spherical shapes with radii as determined from our photometric solution. The centroid of the synthetic profiles was computed by means of the first order moment of the line profile.

\citet{San09} reported maximum RV deviations of about 26\,km\,s$^{-1}$, with the negative and positive RV deviations having nearly identical amplitudes. We found the amplitudes predicted by our code difficult to compare with those reported by \citet{San09}. Indeed, \citet{San09} used Gaussian profile fitting to determine the line RVs, whilst we used the first order moment instead. Our method resulted in an amplitude of the RM effect of at most 19\,km\,s$^{-1}$. Because of the highly non-gaussian shape of the line profiles during the eclipse, the Gaussian fitting results in a different amplitude of the RM effect, which is difficult to reproduce as it also depends on the wavelength range over which the Gaussian fitting is performed. Nevertheless, the fact that the negative and positive RV deviations have nearly identical amplitudes yields a constraint on the value of $\alpha$: $\alpha \le 15^\circ$ and $\alpha \ge 165^\circ$.

Moreover, given the proximity of the system's orbital inclination to $90^{\circ}$, the largest amplitude of the RM effect is expected for an alignment of the rotation and binary axes. In view of the amplitudes reported by \citet{San09}, and keeping in mind the above caveat about these amplitudes, it seems thus unlikely that there exists a very large misalignment between the rotation and binary axes. Whilst this conclusion applies of course only to the primary star, we stress that it's actually the primary that has the largest rotational contribution to the rate of apsidal motion. Hence, a misalignment of the axes seems unlikely to be responsible for the discrepancy between the observational and theoretical values of $\dot{\omega}$.

\section{Conclusion \label{sect:conclusion}}
The eccentric massive binary HD\,152219, belonging to the open cluster NGC\,6231, has been analysed both from an observational and a stellar evolution point of view.

We made use of the huge set of spectroscopic observations of the system to reconstruct the individual spectra of the components by means of a disentangling code. These spectra were subsequently analysed by means of the \texttt{CMFGEN} model atmosphere code to asses the properties of the binary components, notably an effective temperature of $30\,900 \pm 1000$\,K for the primary star. The radial velocities obtained via disentangling were combined with those coming from the literature to establish the rate of apsidal motion along with the orbital period and eccentricity. For these parameters, we respectively found $\dot\omega = \left(1.198 \pm 0.300\right)^\circ$\,yr$^{-1}$, $P_\text{orb}= 4.24046^{+0.00005}_{-0.00004}$\,d and $e=0.072 ^{+0.004}_{-0.005}$. The TESS-12 light curve was analysed by means of the \texttt{Nightfall} code and we notably inferred an effective temperature of $21\,697 \pm 1000$\,K for the secondary star and an inclination of $\left(89.58^{+0.42}_{-2.28}\right)^\circ$ for the system. The Roche lobe filling factors allowed us to infer stellar radii of $9.40^{+0.14}_{-0.15}$\,$R_\odot$ and $3.69\pm0.06$\,$R_\odot$ for the primary and secondary stars, respectively. Combining the results from the radial velocity analysis and light curve analysis, we derived masses of $18.64 \pm 0.47$\,$M_\odot$ and $7.70 \pm 0.12$\,$M_\odot$ for the primary and secondary stars, respectively. We inferred a value of $0.00173 \pm 0.00052$ for the weighted-average mean of the internal structure constants.

We built \texttt{Cl\'es} stellar evolution models using the \texttt{min-Cl\'es} routine to search for  best-fit models of the two stars adopting the mass, radius, and effective temperature as constraints, sometimes complemented by $k_2$ for the primary star. A simultaneous analysis of the two stars gave no coherent results. We considered different prescriptions for the internal mixing occurring inside the stars. We notably performed several tests with different values of the overshooting parameter $\alpha_\text{ov}$, the mass-loss rate scaling factor $\xi$ and the turbulent diffusion $D_T$. The primary star models were shown to have a smaller density contrast than suggested by the observations, that is to say, the theoretical internal structure constant was two times greater than the observational weighted-mean $\bar{k}_2$-value.
To simultaneously satisfy the three previously mentioned constraints together with the constraint that $k_{2,1} < \bar{k}_2$ in the models was impossible: The models were shown to have a strongly enhanced turbulent diffusion but both the effective temperature and $k_2$-value were overestimated. The best-fit models, that is to say the models having the best compromise in terms of $M, R, T_\text{eff}$, and $k_{2,1}$ all have an initial mass of $18.3\pm 0.5$\,$M_\odot$ and an age of $9.5 \pm 0.6$\,Myr. Regarding the secondary star, given that there is no strong constraint on $k_{2,2}$, we only enforced the stellar models to reproduce $M, R$, and $T_\text{eff}$. The best-fit models obtained in this way all have an initial mass of $7.70\pm 0.12$\,$M_\odot$, a $k_{2,2}$-value approximately equal to 4.7\,$\bar{k}_2$, and an age estimate of $11.7 \pm 0.6$\,Myr. Assuming these stellar models are representative of the secondary star, the $k_{2,2}$-value implies that $k_{2,1} = 0.00139$, thereby increasing even more the discrepancy with the stellar evolution models of the primary star. All things considered, the age estimated for the secondary star is incompatible with the age of the primary star and, furthermore, with the age estimate of the cluster. We also tested the impact of non-solar metallicity and helium abundance, and found that both an enhanced metallicity and depleted helium abundance go towards adjustments of the primary star that better reproduce $M, R, T_\text{eff}$, and $k_{2,1}$. However, the corresponding models of the secondary star yield even higher ages, further increasing the inconsistency between primary and secondary ages.

Finally, we investigated the impact of a misalignment of the stellar rotation axis with respect to the normal to the orbital plane. Such a misalignment reduces the contribution of the rotational term to the Newtonian apsidal motion rate, thereby implying an increased value of $\bar{k}_2$ compared to the aligned configuration. In the present case, we found that a severe misalignment angle is required to get a $\bar{k}_2$-value high enough for the stellar models of the primary star to satisfy the constraint on $k_{2,1}$. Such a high misalignment is however ruled out by the observational constraint obtained via the analysis of the Rossiter-McLaughlin effect which sets an upper limit of $15^\circ$ on the misalignment angle. 

The difficulty to reconcile the $k_2$-value of the primary stellar models with the observational constraint on the $\bar{k}_2$-value as well as the internal inconsistency between the ages of the two stars are indications that some physics of the stellar interior are still not completely understood. Obviously, all the tools used in our observational or theoretical analysis have their own limitations. Whilst we are quite confident in our determination of the masses of the stars as well as in the effective temperature of the primary star, we are less confident in the inferred values of the secondary effective temperature and the radii of both stars. 

\begin{acknowledgements}
S. R. acknowledges support from the Fonds de la Recherche Scientifique (F.R.S.- FNRS, Belgium). M. F. acknowledges the support from STFC consolidated grant ST/T000252/1. We thank Dr John Hillier for making his code {\tt CMFGEN} publicly available. We thank Dr Rainer Wichmann for making his code {\tt Nightfall} publicly available as well as for discussions regarding his code. We also thank Dr Eric Gosset for our numerous discussions about the $\chi^2$. We finally thank Dr Ya\"el Naz\'e for her contribution in the extraction of the TESS light curves. This paper makes use of data collected by the TESS mission, whose funding is provided by the NASA Explorer Program. 
\end{acknowledgements}

\begin{appendix} 
\onecolumn
\section{Journal of the spectroscopic observations\label{appendix:spectrotable}}
This appendix provides the journal of spectroscopic observations of HD\,152219 including the radial velocities obtained by means of the disentangling code (Table\,\ref{Table:spectro+RV}).

\begin{table*}[h]
\caption{Journal of the spectroscopic observations of HD\,152219. }
\centering
\label{Table:spectro+RV}
\begin{tabular}{ccrrl}
\hline\hline
\vspace{-3mm}\\
HJD & $\phi$  & $RV_1$ & $RV_2$ & Instrumentation\\
--\,2\,450\,000&  & (km\,s$^{-1}$) & (km\,s$^{-1}$) & \\
\hline
\vspace{-3mm}\\
1300.802 &0.060&$-146.7\pm1.7$&  $277.1 \pm6.4$& ESO 1.5 m + FEROS\\
1304.806 &0.004&$-137.9\pm 2.0$&  $268.2\pm12.0$&ESO 1.5 m + FEROS \\
1323.872 &0.500&$ 72.1\pm 0.8$& 	$-259.7\pm10.1$&ESO 1.5 m + FEROS \\
1669.881 &0.097&$-142.5\pm 1.7$&  $231.2\pm4.7$& ESO 1.5 m + FEROS\\
1671.875 &0.567&$80.2\pm 1.2$& 	$-277.0\pm5.0$& ESO 1.5 m + FEROS\\
2037.824 &0.866&$-71.7\pm 1.3$&   $-5.3\pm3.0$&ESO 1.5 m + FEROS \\
2040.806 &0.569&$82.8\pm 1.0$& 	$-258.2\pm4.7$&ESO 1.5 m + FEROS \\
2335.774 &0.130&$-127.5\pm 1.7$&  $202.5\pm4.1 $&ESO 1.5 m + FEROS \\
2335.814 &0.139&$-121.0\pm 1.2$&  $206.9\pm2.6$&ESO 1.5 m + FEROS \\
2337.765 &0.599&$79.6\pm 0.8$& 	$-260.8\pm3.4$& ESO 1.5 m + FEROS\\
2339.774 &0.073&$-146.9\pm 1.7$&  $260.2\pm3.8$&ESO 1.5 m + FEROS \\
2381.671 &0.953&$-123.6\pm 1.3$&  $195.7\pm3.8$&ESO 1.5 m + FEROS \\
2381.772 &0.977&$-133.4\pm 1.4$&  $225.3\pm2.4$& ESO 1.5 m + FEROS\\
2382.668 &0.188&$-92.5\pm 1.3$&  	$154.0\pm6.3$&ESO 1.5 m + FEROS \\
2383.674 &0.426&$53.0\pm 1.2$& 	$-189.4\pm3.3$&ESO 1.5 m + FEROS \\
3130.655 &0.581&$75.9\pm 1.3$& 	$-274.0\pm4.6$& ESO 2.2 m + FEROS\\
3130.856 &0.628&$73.3\pm 1.2$& 	$-250.6\pm4.4$& ESO 2.2 m + FEROS\\
3131.906 &0.876&$-77.0\pm 1.3$&   $38.3	\pm1.1$&ESO 2.2 m + FEROS \\
3132.624 &0.045&$-151.1\pm 1.3$&  $239.7\pm2.1$&ESO 2.2 m + FEROS \\
3132.897 &0.109&$-134.4\pm 1.4$&  $220.5\pm3.2$&ESO 2.2 m + FEROS \\
3134.636 &0.520&$81.2\pm 0.8$& 	$-272.0\pm4.8$&ESO 2.2 m + FEROS \\
3134.890 &0.579&$80.7\pm 0.8$& 	$-266.3\pm4.6$&ESO 2.2 m + FEROS \\
3861.670 &0.971&$-134.8\pm1.7$ &  $230.0\pm3.4$&ESO 2.2 m + FEROS \\
3861.921 &0.030&$-147.9\pm 1.5$ & $266.9 \pm4.9$&ESO 2.2 m + FEROS \\
3863.643 & 0.436&$61.9\pm1.2$ & $-242.0\pm5.5$&ESO 2.2 m + FEROS \\
3863.886 &0.493&$76.4\pm1.0$ & $-279.1\pm4.6$&ESO 2.2 m + FEROS \\
3864.651 & 0.674&$55.8 \pm 0.9$& $-189.3\pm4.5$ &ESO 2.2 m + FEROS\\
3912.477 &0.952&$-125.3 \pm 1.5$&  $221.6\pm3.0$ &ESO 2.2 m + FEROS\\
3912.488 &0.955&$-126.5 \pm 1.6$& $222.6 \pm2.6$ &ESO 2.2 m + FEROS\\
3912.507 & 0.959&$-129.2 \pm 1.4$& $230.2\pm2.8$ &ESO 2.2 m + FEROS\\
3912.533 &0.966&$-131.9 \pm 1.5$& $235.5\pm2.3$ &ESO 2.2 m + FEROS\\
3912.546 &0.969&$-133.3 \pm 1.5$& $236.4\pm2.5$ &ESO 2.2 m + FEROS\\
3912.577 &0.976&$-134.5 \pm 1.5$& $248.7\pm3.0$ &ESO 2.2 m + FEROS\\
3912.597 &0.981&$-136.4 \pm 1.6$& $243.9\pm3.2$ &ESO 2.2 m + FEROS\\
3912.615 &0.985&$-138.5 \pm 1.6$& $245.4\pm3.0$ &ESO 2.2 m + FEROS\\
3912.625 &0.987&$-139.4 \pm 1.6$& $247.2 \pm3.5$ &ESO 2.2 m + FEROS\\
3912.634 &0.989&$-140.7 \pm 1.7$& $249.4\pm4.1$ &ESO 2.2 m + FEROS\\
3912.643 &0.991&$-141.3 \pm 1.6$& $249.3\pm4.0$ &ESO 2.2 m + FEROS\\
3912.651 &0.993&$-142.0 \pm 1.6$& $251.0\pm3.9$ &ESO 2.2 m + FEROS\\
3912.659 &0.995&$-142.6 \pm 1.6$& $255.6 \pm4.6$ &ESO 2.2 m + FEROS\\
3912.667 &0.997&$-143.3 \pm 1.7$& $257.9\pm4.2$ &ESO 2.2 m + FEROS\\
3912.695 &0.004&$-145.0 \pm 1.6$& $262.6\pm4.3$ &ESO 2.2 m + FEROS\\
3912.702 &0.005&$-145.1 \pm 1.7$& $264.3\pm4.0$ &ESO 2.2 m + FEROS\\
3912.710 &0.007&$-145.3 \pm 1.6$& $264.0\pm4.0$ &ESO 2.2 m + FEROS\\
3912.717 & 0.009&$-145.7 \pm 1.6$& $264.0\pm4.4$ &ESO 2.2 m + FEROS\\
3912.725 &0.011&$-146.3 \pm 1.8$& $264.0\pm4.2$ &ESO 2.2 m + FEROS\\
3912.734 &0.013&$-147.3 \pm 1.8$& $263.9\pm4.5$ &ESO 2.2 m + FEROS\\
3912.743 &0.015&$-147.4 \pm 1.7$& $268.2 \pm4.8$ &ESO 2.2 m + FEROS\\
3912.752 &0.017&$-147.6 \pm 1.7$& $265.1 \pm4.1$ &ESO 2.2 m + FEROS\\
3912.769 &0.021&$-147.4 \pm 1.7$& $267.8\pm4.1$ &ESO 2.2 m + FEROS\\
3912.785 &0.025&$-146.9 \pm 1.7$& $271.8\pm4.5$ &ESO 2.2 m + FEROS\\
\hline 
\end{tabular} 
\end{table*} 
\setcounter{table}{0}
\begin{table*}[h!]
\caption{Continued.}
\centering
\begin{tabular}{ccrrl}
\hline\hline
\vspace{-3mm}\\
HJD & $\phi$  & $RV_1$ & $RV_2$ & Instrumentation\\
--\,2\,450\,000&  & (km\,s$^{-1}$) & (km\,s$^{-1}$) & \\
\hline
\vspace{-3mm}\\
3912.794 &0.027&$-146.8 \pm 1.7$& $271.2\pm4.3$ &ESO 2.2 m + FEROS\\
3912.818 &0.033&$-146.8 \pm 1.8$& $276.5 \pm4.5$ &ESO 2.2 m + FEROS\\
3912.827 &0.035&$-147.1 \pm 1.7$& $270.1\pm4.6$ &ESO 2.2 m + FEROS\\
3912.836 &0.037 &$-147.8 \pm 1.9$& $271.2\pm6.0$ &ESO 2.2 m + FEROS\\
3912.853 &0.041&$-148.5 \pm 1.8$& $267.7\pm5.8 $ &ESO 2.2 m + FEROS\\
3912.862 &0.043&$-148.1 \pm 1.9$& $274.3 \pm5.9$ &ESO 2.2 m + FEROS\\
3912.871 &0.045&$-147.6 \pm 1.8$& $278.1\pm5.3$ &ESO 2.2 m + FEROS\\
3912.880 &0.047&$-146.9 \pm 2.1$& $279.2\pm7.7$ &ESO 2.2 m + FEROS\\
3913.535 &0.202&$-76.4  \pm 1.5$& $70.1\pm1.7$ &ESO 2.2 m + FEROS\\
3914.480 &0.425&$57.9 \pm 1.2$&$-244.8 \pm6.7$ &ESO 2.2 m + FEROS\\
3914.491 & 0.427&$58.4 \pm 1.0$&$-239.9\pm4.9$ &ESO 2.2 m + FEROS\\
3914.502 & 0.430&$59.7 \pm 1.1$&$-243.1\pm5.9$ &ESO 2.2 m + FEROS\\
3914.525 & 0.435&$61.8 \pm 1.1$&$-250.3\pm7.3$ &ESO 2.2 m + FEROS\\
3914.533 & 0.437&$62.2 \pm 1.0$&$-252.4\pm7.2$ &ESO 2.2 m + FEROS\\
3914.541 & 0.439&$63.1 \pm 1.0$&$-255.1 \pm6.6$ &ESO 2.2 m + FEROS\\
3914.558 & 0.443&$63.8 \pm 1.0$&$-262.6\pm7.8$ &ESO 2.2 m + FEROS\\
3914.565 & 0.445&$65.0 \pm 1.0$&$-261.1\pm7.1$ &ESO 2.2 m + FEROS\\
3914.573 & 0.447&$65.9 \pm 0.9$&$-261.6\pm6.2$ &ESO 2.2 m + FEROS\\
3914.590 & 0.451&$67.8 \pm 0.9$&$-268.0 \pm6.5$ &ESO 2.2 m + FEROS\\
3914.597 & 0.452&$68.6 \pm 0.9$&$-266.1 \pm6.3$ &ESO 2.2 m + FEROS\\
3914.615 & 0.457&$69.8 \pm 0.9$&$-267.1 \pm5.7$ &ESO 2.2 m + FEROS\\
3914.629 & 0.460&$71.1 \pm 0.8$&$-270.0\pm6.5$ &ESO 2.2 m + FEROS\\
3914.633 & 0.461&$71.3 \pm 0.8$&$-272.3\pm5.9$ &ESO 2.2 m + FEROS\\
3914.637 & 0.462&$72.0 \pm 0.9$&$-274.0 \pm6.1$ &ESO 2.2 m + FEROS\\
3914.647 & 0.464&$73.1 \pm 0.9$&$-273.2\pm4.5$ &ESO 2.2 m + FEROS\\
3914.653 & 0.465&$73.3 \pm 0.8$&$-275.9\pm4.2$ &ESO 2.2 m + FEROS\\
3914.659 & 0.467&$73.9 \pm 0.8$&$-269.0\pm5.1$ &ESO 2.2 m + FEROS\\
3914.683 & 0.473&$75.3 \pm 0.9$&$-270.0\pm5.7$ &ESO 2.2 m + FEROS\\
3914.690 & 0.474&$75.9 \pm 0.7$&$-273.1\pm4.5$ &ESO 2.2 m + FEROS\\
3914.705 & 0.478&$77.1 \pm 0.8$&$-272.6\pm4.4$ &ESO 2.2 m + FEROS\\
3914.727 & 0.483&$77.8 \pm 0.9$&$-273.5\pm4.5$ &ESO 2.2 m + FEROS\\
3914.737 &0.485& $77.4 \pm 0.9$&$-277.8\pm5.6$ &ESO 2.2 m + FEROS\\
3914.748 & 0.488&$78.2 \pm 0.9$&$-277.5\pm4.1$ &ESO 2.2 m + FEROS\\
3914.759 & 0.490&$78.7 \pm 1.1$&$-285.9\pm7.2$ &ESO 2.2 m + FEROS\\
3914.787 &0.497& $78.4 \pm 0.9$&$-301.4 \pm7.3$ &ESO 2.2 m + FEROS\\
3914.798 & 0.500&$78.7 \pm 1.3$&$-293.2\pm8.3$ &ESO 2.2 m + FEROS\\
3914.809 & 0.502&$78.6 \pm 0.9$&$-293.7\pm5.5$ &ESO 2.2 m + FEROS\\
3914.820 & 0.505&$79.2 \pm 1.1$&$-296.3\pm7.7$ &ESO 2.2 m + FEROS\\
3914.840 & 0.510&$79.6 \pm 1.0$&$-288.9\pm5.7$ &ESO 2.2 m + FEROS\\
3914.851 & 0.512&$80.7 \pm 1.0$&$-294.1\pm5.5$ &ESO 2.2 m + FEROS\\
3914.865 & 0.515&$82.3 \pm 1.2$&$-298.8\pm4.9$ &ESO 2.2 m + FEROS\\
3914.876 & 0.518&$82.6 \pm 1.4$&$-299.6\pm6.9$ &ESO 2.2 m + FEROS\\
\hline
\end{tabular}
\tablefoot{Column\,1 gives the heliocentric Julian date (HJD) of the observations at mid-exposure. Column\,2 gives the observational phase $\phi$ computed with the orbital period determined in Sect.\,\ref{sect:omegadot} (Table\,\ref{bestfitTable}). Columns\,3 and 4 give the radial velocity $RV_1$ and $RV_2$ of the primary and secondary star, respectively. The errors represent $\pm1\sigma$. Column\,5 provides information about the instrumentation.}
\end{table*}

\twocolumn
\section{Stellar evolution models: influence of the metallicity and helium abundance\label{subsect:XZ}}
As a last attempt to reconcile the ages of the primary and secondary stars, we investigated the influence of a potential non-solar metallicity and helium abundance. Both the metallicity and the initial helium abundance affect the internal structure essentially through a change in the opacity. The uncertainties on the opacity hence justify the influence analysis we performed here.

\begin{figure}[h]
\centering
\includegraphics[clip=true,trim=80 70 300 210,width=\linewidth]{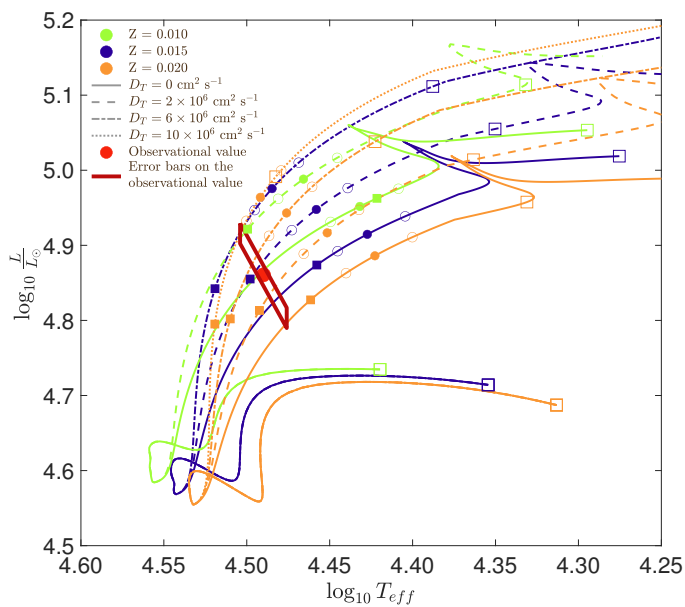}
\caption{Hertzsprung-Russell diagram: evolutionary tracks of \texttt{Cl\'es} models for the primary star having $M_{\rm init}= 19$\,$M_\odot$, $\alpha_\text{ov}=0.30$, $\xi=1$, $X=0.715$, $Z=0.010$ (green), 0.015 (dark blue), and 0.020 (orange), and $D_T=0$\,cm$^2$\,s$^{-1}$ (solid line), $2\times 10^6$\,cm$^2$\,s$^{-1}$ (dashed line), $6\times 10^6$\,cm$^2$\,s$^{-1}$ (dotted-dashed line), and $10\times 10^6$\,cm$^2$\,s$^{-1}$ (dotted line). The dots over-plotted on the corresponding tracks correspond to the models for which $k_{2,1}$ is equal to $\bar{k}_2$, while the open circles correspond to the models for which $k_{2,1}$ is equal to $\bar{k}_2\pm \Delta \bar{k}_2$. The filled and open squares over-plotted on the tracks correspond to the models for which the mass is equal to the observed value within the error bars. The observational value is represented by the red point, and its error bars are represented by the dark red parallelogram. \label{fig:HR_Z}}
\end{figure}

Regarding the metallicity, we built eight evolutionary sequences for the primary star, all having an initial mass of 19\,$M_\odot$, $\alpha_\text{ov} = 0.30$, and $\xi=1$, but different values of $Z$, namely 0.010 and 0.020, and different values of $D_T$, namely 0, $2\times 10^6$, $6\times 10^6$, and $10\times 10^6$\,cm$^2$\,s$^{-1}$. These evolutionary tracks are presented in the Hertzsprung-Russell diagram in Fig.\,\ref{fig:HR_Z} together with the evolutionary tracks for $Z=0.015$ presented in Sect.\,\ref{subsect:HR}. For more clarity, only the tracks crossing at some point of their evolution the observational box are shown. For $Z=0.010$, only the two models with the lowest values of $D_T$ cross the observational box, while for $Z=0.020$, all models cross the observational box. In addition, all tracks have a mass compatible with the observational value when crossing the observational box. Compared to Fig.\,\ref{fig:HR_all}, we observe that a higher metallicity has the same impact on the evolutionary tracks as a lower mass and/or lower $D_T$: the evolutionary track with $Z=0.020$ and the highest value of $D_T$ has now a value of $k_{2,1}$ compatible with $\bar{k}_2$ when it crosses the observational box. 

\begin{figure}[h]
\centering
\includegraphics[clip=true,trim=80 70 300 210,width=\linewidth]{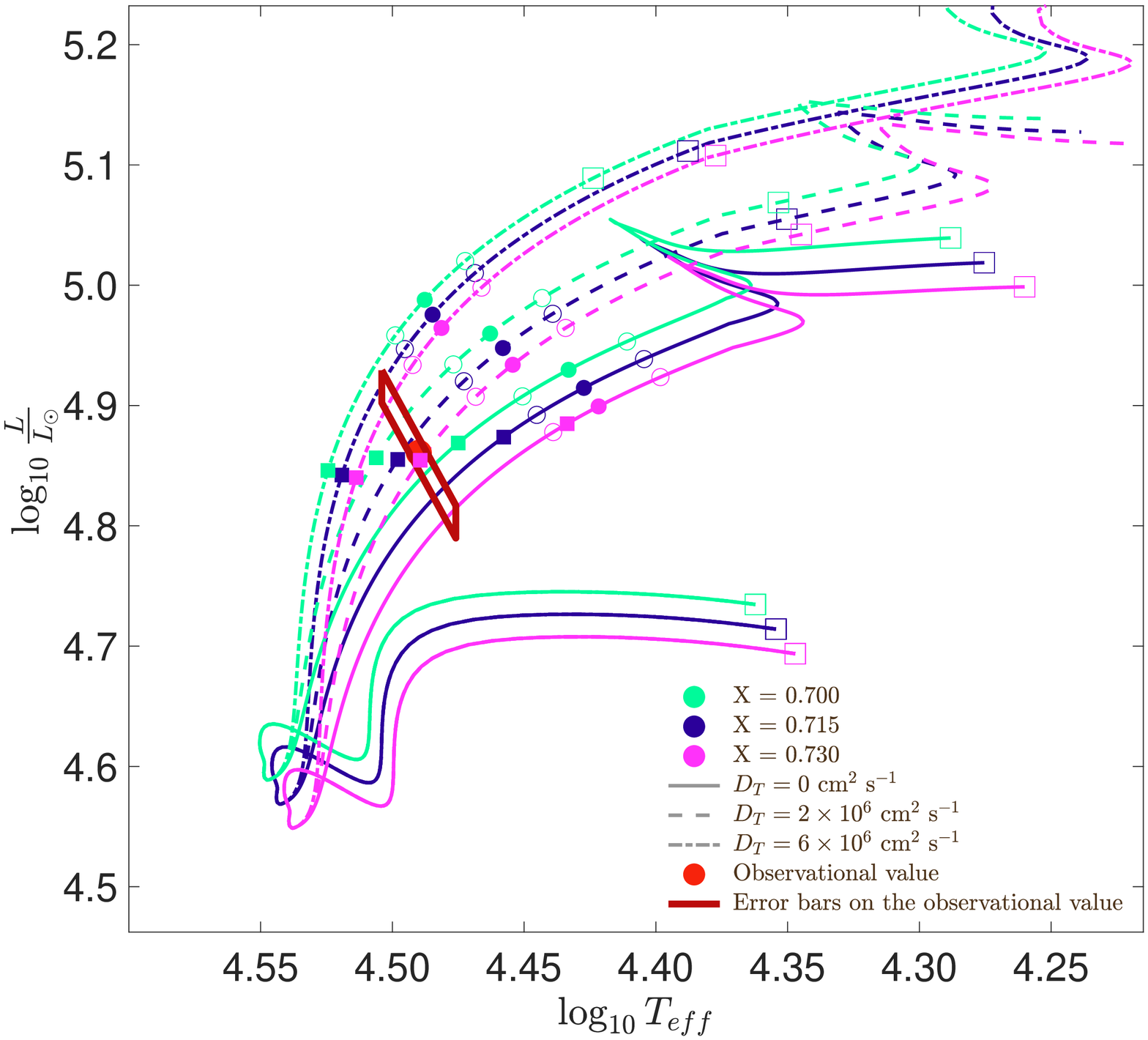}
\caption{Hertzsprung-Russell diagram: evolutionary tracks of \texttt{Cl\'es} models for the primary star having $M_{\rm init}= 19$\,$M_\odot$, $\alpha_\text{ov}=0.30$, $\xi=1$, $Z=0.015$, $X=0.700$ (water green), 0.715 (dark blue), and $0.730$ (pink), and $D_T=0$\,cm$^2$\,s$^{-1}$ (solid line), $2\times 10^6$\,cm$^2$\,s$^{-1}$ (dashed line), $6\times 10^6$\,cm$^2$\,s$^{-1}$ (dotted-dashed line), and $10\times 10^6$\,cm$^2$\,s$^{-1}$ (dotted line). The dots over-plotted on the corresponding tracks correspond to the models for which $k_{2,1}$ is equal to $\bar{k}_2$, while the open circles correspond to the models for which $k_{2,1}$ is equal to $\bar{k}_2\pm \Delta \bar{k}_2$. The filled and open squares over-plotted on the tracks correspond to the models for which the mass is equal to the observed value within the error bars. The observational value is represented by the red point, and its error bars are represented by the dark red parallelogram. \label{fig:HR_X}}
\end{figure}

Regarding the helium abundance, we built eight evolutionary sequences for the primary star, all having an initial mass of 19\,$M_\odot$, $\alpha_\text{ov} = 0.30$, $\xi=1$, and $Z=0.015$, but different values of $X$, namely 0.700 and 0.730, and different values of $D_T$, namely 0, $2\times 10^6$, $6\times 10^6$, and $10\times 10^6$\,cm$^2$\,s$^{-1}$. These evolutionary tracks are presented in the Hertzsprung-Russell diagram in Fig.\,\ref{fig:HR_X} together with the evolutionary tracks having $X=0.715$ presented in Sect.\,\ref{subsect:HR}. For more clarity, only the tracks crossing at some point of their evolution the observational box defined by the radius and effective temperature are plotted. For all values of X, only the three models with the lowest values of $D_T$ cross the observational box at some point of their evolution. In addition, all tracks have a mass compatible with the observed value when crossing the observational box. Compared to Fig.\,\ref{fig:HR_all}, we observe that a higher $X$-value, hence a lower helium abundance, has the same impact on the evolutionary tracks as a lower mass and/or lower $D_T$. The impact of a change in $X$ is however smaller than that of a change in $Z$ and none of the models has a $k_{2,1}$ compatible with $\bar{k}_2$ when crossing the observational box.

\begin{figure}[p]
\centering
\includegraphics[clip=true,trim=80 60 300 60,width=\linewidth]{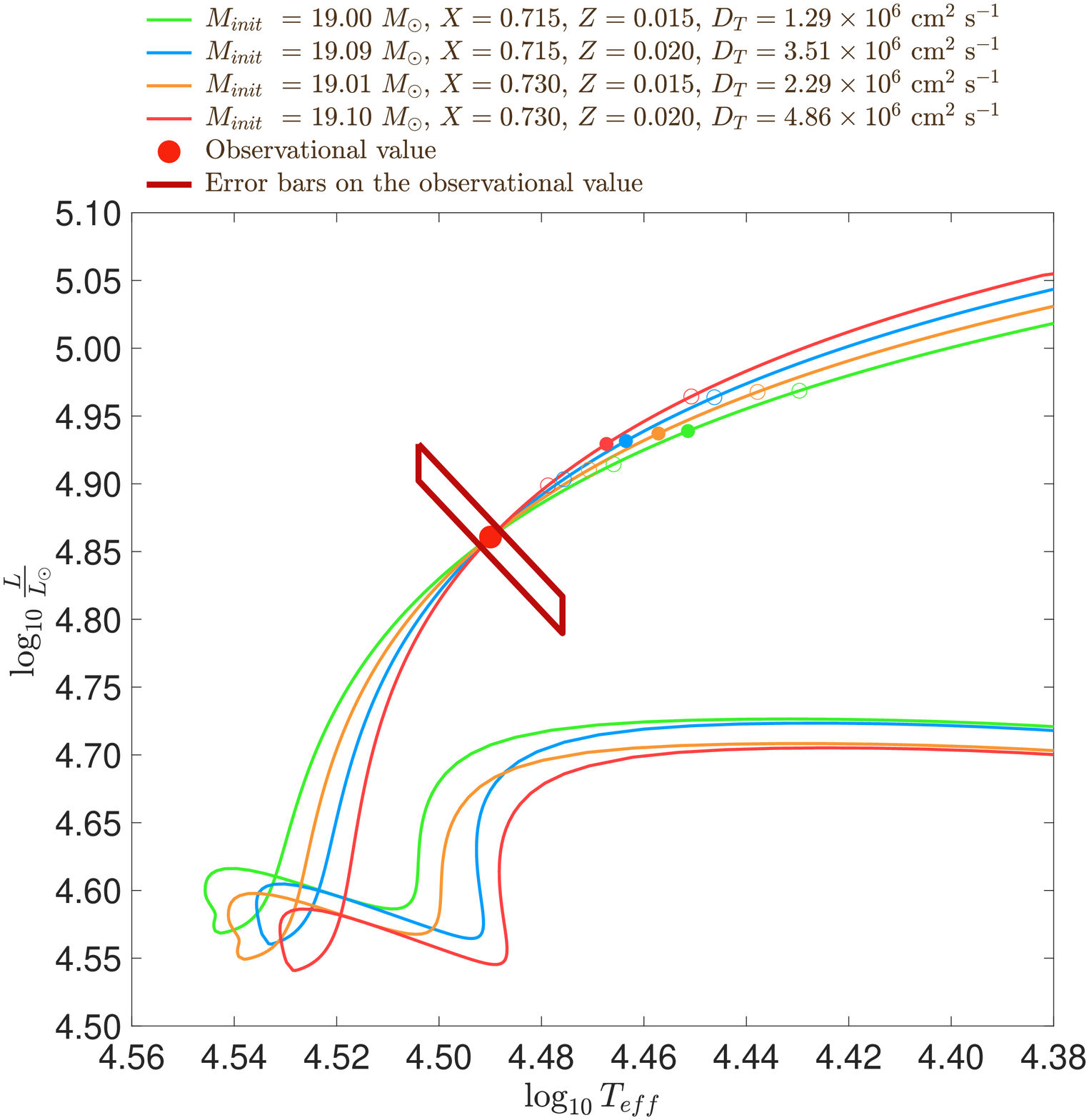}
\includegraphics[clip=true,trim=80 60 300 60,width=\linewidth]{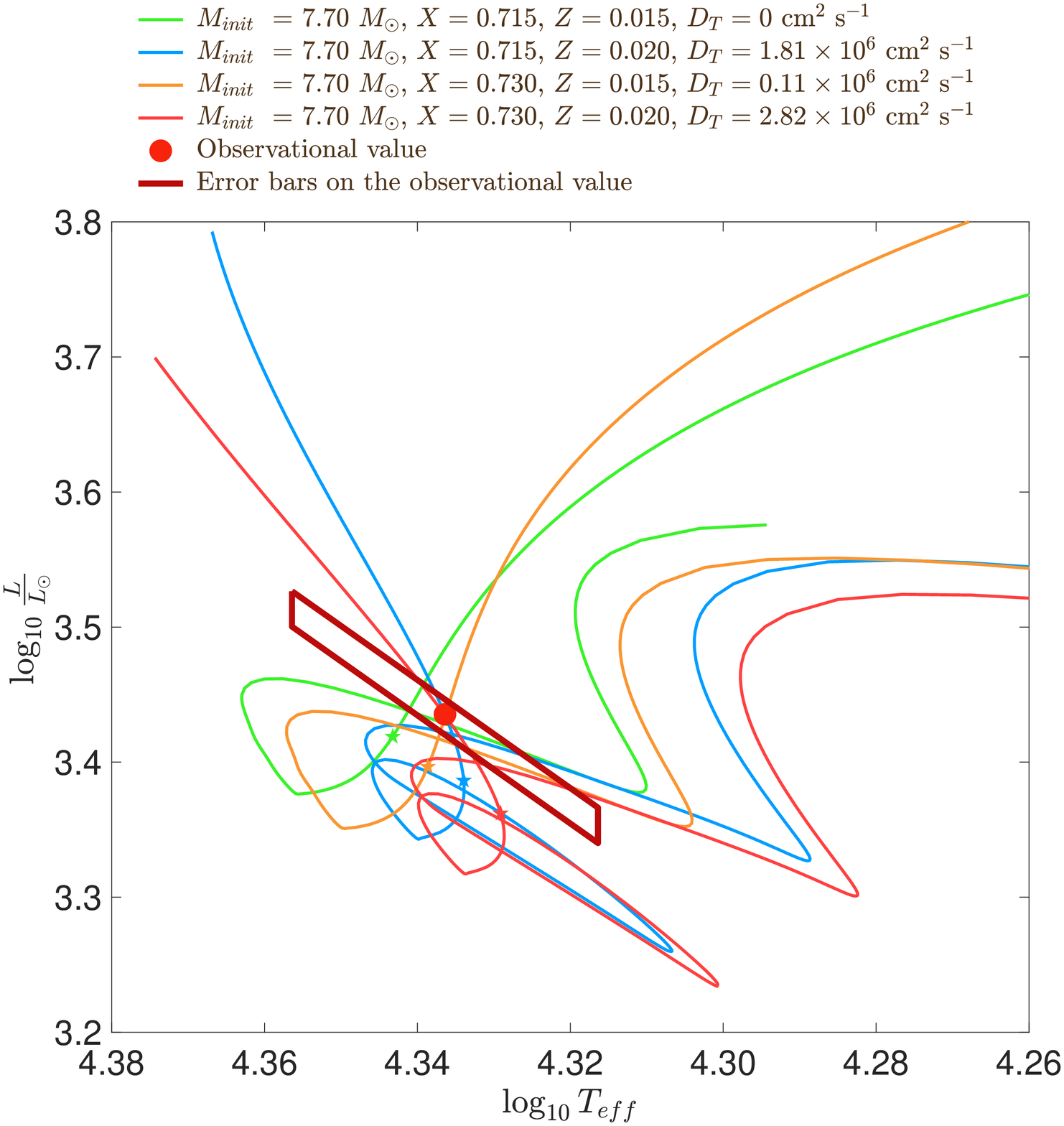}
\caption{Hertzsprung-Russell diagram: evolutionary tracks of \texttt{Cl\'es} models for the primary star (\textit{upper panel)} and secondary star (\textit{lower panel}) corresponding to the best-fit models listed in Table\,\ref{tab:mincles_C} together with Series II(3) for the primary and Model III$_\text{S}$ for the secondary. All models have $\alpha_\text{ov}=0.30$, primary models have $\xi=1$ while secondary models have $\xi=0.1$. The dots over-plotted on the evolutionary tracks of the primary correspond to the models for which $k_{2,1}$ is equal to $\bar{k}_2$, while the open circles correspond to the models for which $k_{2,1}$ is equal to $\bar{k}_2\pm \Delta \bar{k}_2$. The filled 5-branch stars over-plotted on the secondary tracks correspond to the models for which the age is equal to the age of the best-fit model of the primary star of same $X$ and $Z$. The observational value is represented by the red point, and its error bars are represented by the dark red parallelogram. \label{fig:HR_bestfitZX}}
\end{figure}

We then built three best-fit models for the primary star and the secondary star adopting the constraints on $M, R$, and $T_\text{eff}$, fixing $\alpha_\text{ov}=0.30$ for both stars, and $\xi=1$ for the primary star and 0.1 for the secondary star, and leaving the age, initial mass, and $D_T$ as free parameters, for the three following couples of values for $(X,Z)$: (0.715, 0.020), (0.730, 0.015), and (0.730, 0.020). The results are given in Table\,\ref{tab:mincles_C} under the names of Model Z+, X+, and Z+X+, respectively. All six models correctly reproduce the observational mass, radius, and effective temperature of the corresponding star. Regarding the three primary star models, compared to Series II(3), all models have a higher age, a slightly higher initial mass, and a higher turbulent diffusion coefficient $D_T$. Their lower $k_{2,1}$-value compared to Series II(3) make these three models better suited to reproduce the primary star. Regarding the secondary star, compared to Model III$_\text{S}$ the three models have a higher age, the same initial mass, and a higher $D_T$-value, as well as a lower $k_{2,2}$-value. The evolutionary tracks corresponding to these best-fit models are presented in Fig.\,\ref{fig:HR_bestfitZX} together with Series II(3) for the primary and  Model III$_\text{S}$ for the secondary. Compared to the models having a solar metallicity and helium abundance, the age discrepancy between the primary and secondary best-fit models is enhanced when higher metallicity or/and lower helium abundance is/are considered.

\begin{table*}[h!tb]
\caption{Parameters of some best-fit {\tt Cl\'es} models discussed in Sect.\,\ref{sect:cles}.}
\label{tab:mincles_C}
\begin{tabular}{l c c c c c c c c c c c c}
\hline\hline
\vspace*{-0.3cm} \\
Model &  Age & $M_{\rm init}$ & $M$ & $R$ & $T_{\rm eff}$ &    $k_{2,\text{un.}}$ & $k_2$    & $10^{-7}$\, $\dot{\text{M}}$  & $\xi$ & $10^6$\,$D_T$ & $\alpha_\text{ov}$ & $\chi^2$ \\
& (Myr) & ($M_{\odot}$) & ($M_{\odot}$)  & ($R_{\odot}$)  & (K)  & ($10^{-3}$)& ($10^{-3}$)  & ($M_{\odot}$\,yr$^{-1}$)&  & (cm$^2$\,s$^{-1}$) & & \\
\vspace*{-0.3cm} \\
\hline
\vspace*{-0.3cm} \\
Model Z+ & 8.17 & 19.09 & 18.64 & 9.40 & 30\,900 & 3.4777 & 3.1553 & 1.04 & 1 &$3.51$ & 0.30 &  7.63\\
Model X+ & 8.54 & 19.01 & 18.64 & 9.40 & 30\,900 & 3.6527 & 3.3141 & 0.82 & 1 &$2.29$& 0.30 &  9.42\\
Model Z+X+ & 8.91 & 19.10 & 18.64 & 9.40 & 30\,900 & 3.3245 & 3.0162 & 1.04 & 1 &$4.86$& 0.30 &  6.22\\
\hline 
\vspace*{-0.3cm} \\
Model Z+$_\text{S}$ &  $15.36$ & $7.70$ & $7.70$ &$3.69$ & $21\,697$ & 8.0328  & 7.7938 & 0.001 & $0.1$ & $1.81$ & $0.30$ & 0.01 \\  
Model X+$_\text{S}$ &  $14.13$ & $7.70$ & $7.70$ &$3.69$ & $21\,697$ & 8.1705  & 7.9274 & 0.001 & $0.1$ & $0.11$ & $0.30$ & 0.01\\  
Model Z+X+$_\text{S}$ &  $20.81$ & $7.70$ & $7.70$ &$3.69$ & $21\,697$ & 7.9312  & 7.6952 & 0.001 & $0.1$ & $2.82$ & $0.30$ & 0.01\\  
\vspace*{-0.3cm} \\
\hline
\end{tabular}
\tablefoot{Columns\,1 and 2 give the name of the model and its current age. Column\,3 lists the initial mass of the corresponding evolutionary sequence. Columns\,4, 5, and 6 give the mass, radius, and effective temperature of the model. Columns\,7 and 8 yield the $k_2$ of the model respectively before and after applying the empirical correction for the effect of rotation of \citet[][Eq.\,\eqref{eqn:k2Claret}]{Claret99}. Column\,9 lists the mass-loss rate of the model. Columns\,10, 11, and 12 give the mass-loss rate scaling factor, the turbulent diffusion, and the overshooting parameter of the model. Column\,13 quotes the $\chi^2$ of the model.\\
Models without and with subscript $\text{S}$ correspond to the models of the primary star and secondary star, respectively.\\ 
All $\chi^2$ for the primary star have been computed based on $M, R, T_\text{eff}$, and $k_{2,1}$, but see the caveats of \citet{andrae10a} and \citet{andrae10b}. All $\chi^2$ for the secondary star have been computed based on $M, R$, and $T_\text{eff}$ only. The Models Z+ and Z+$_\text{S}$ have $Z=0.020$, the Models X+ and X+$_\text{S}$ have $X=0.730$, and the Models Z+X+ and Z+X+$_\text{S}$ have $Z=0.020$ and $X=0.730$.}
\end{table*}

The increase in $D_T$ with metallicity can be explained as follows \citep[see e.g.][]{rosu20b}. An increase in metallicity, at a given mass, increases the opacity which in turn lowers the effectiveness of the radiative transport, hence decreasing the effective temperature and luminosity. The same holds true for the increase in $D_T$ with the decrease in helium abundance which is the direct consequence of an increase in the opacity with $X$. Since an increase in $D_T$ leads to an increase in the luminosity, it compensates the decrease in $T_\text{eff}$ and $L$ induced by the enhanced metallicity. Given that the turbulent diffusion increases the mass of the convective core, it directly follows that $k_2$ decreases accordingly. In this search for best-fit models with enhanced metallicity, the modification of the various parameters is not a direct consequence of a change of metallicity but rather an indirect consequence resulting from the corresponding change in the turbulent diffusion.  

There is no observational evidence that HD\,152219 would have a non-solar metallicity or helium abundance. Even though a higher metallicity and a subsolar helium abundance would improve the best-fit models of the primary star by allowing to simultaneously satisfy the constraints on $M, R, T_\text{eff}$, and $k_{2,1}$, such a high metallicity seems rather unlikely, especially since \citet{kilian94} found a subsolar CNO mass fraction from the analysis of ten B-type stars in NGC\,6231. At this stage, insufficient information is available to conclude that the metallicity and helium abundance of the star deviate from solar. 
\end{appendix}
\end{document}